\documentclass{article}
\usepackage{arxiv}

\usepackage[utf8]{inputenc} 
\usepackage[T1]{fontenc}    
\usepackage{hyperref}       
\usepackage{url}            
\usepackage{booktabs}       
\usepackage{amsfonts}       
\usepackage{nicefrac}       
\usepackage{microtype}      
\usepackage{lipsum}
\usepackage{graphicx}
\graphicspath{ {./images/} }

\renewcommand{\phi}{\varphi}
\renewcommand{\H}{\mathcal{H}}
\renewcommand{\S}{\mathcal{S}}
\newcommand{\cR}{\mathcal{R}}
\newcommand{\M}{\mathcal{M}}
\newcommand{\Hp}{\mathcal{H}_{\pi}}

\renewcommand{\P}{\mathbb{P}}

\newcommand{\I}{\Gamma}
\newcommand{\A}{\mathcal{A}}
\newcommand{\R}{\mathbb{R}}
\newcommand{\Q}{\mathbb{Q}}
\newcommand{\N}{\mathbb{N}}
\renewcommand{\t}{\widetilde}
\newcommand{\be}{\begin{equation}}
\newtheorem{thm}{Theorem}
\newtheorem{prop}{Proposition}
\newtheorem{lemma}{Lemma}

\title{Theorems motivated by foundations of quantum mechanics and some of their applications}

\author{
 Roberto H. Schonmann \\
  Mathematics Department\\
  University of California at Los Angeles\\
  Los Angeles, CA 90095 \\
  \texttt{rhs@math.ucla.edu} \\
}

\begin{document}
\maketitle
\begin{abstract}
This paper provides theorems 
aimed at shedding light on issues 
in the foundations of quantum mechanics. 
These theorems can 
be used to propose new interpretations to 
the theory, or to better 
understand, evaluate and improve 
current interpretations.
Some of these applications include: (1) 
A proof of the existence of pilot-wave
theories that are fully equivalent to 
standard quantum mechanics in a path-wise
sense. This equivalence 
is stronger than what is 
entailed from the more traditional 
requirements 
of equivariance, or good mixing properties, 
and is necessary to assure 
proper correlations across time and
proper records of the past. (2) A proposal for a minimalistic 
ontology for
non-collapse quantum 
mechanics, in which Born's rule
provides the proper predictions.(3) The observation of a 
close relationship between Born's rule 
and a version of the superposition 
principle. 
\end{abstract}


\section{Introduction}
\label{sec:intro}

Problems in the 
foundations of quantum mechanics involve 
Physics, Philosophy and Mathematics. 
Here we state and prove 
theorems motivated by such issues.
These theorems have implications in the evaluation
and understanding
of various distinct interpretations of 
quantum mechanics, and can also be used to propose
new interpretations. 

A separation between the 
mathematical work presented here 
and the discussion
of its uses to evaluate, or build 
interpretations seems appropriate and 
beneficial. This is so because 
mathematical work can be understood and 
judged based solely 
on its correctness, in an objective way.
In contrast,  
applications to interpretations seem to unavoidably
lead to less precise and less objective ideas. 
For this reason, applications of the 
theorems 
will be 
separated from the purely mathematical 
part of this paper, and presented 
in the final section.

A few words
about the relation between the theorems 
presented here 
and quantum mechanics and its foundations
are nevertheless in order
in this introduction, to help the reader 
understand our motivation and purpose, 
before engaging with the mathematics.

Our mathematical 
setting will correspond to a quantum mechanical universe
described in the Heisenberg picture, in which the
wave function does not evolve, while operators evolve 
in time. We assume 
that no collapse of the wave function ever occurs,
so that the state of this universe is given by an 
unchanging vector $\Psi$ in a Hilbert space, and 
no probabilistic postulates are introduced. 

One of our main goals is 
in producing a coherent 
interpretation of quantum mechanics in which 
there is no collapse of the wave 
function, and no 
special notion of ``observation'' 
or ``experiment'', but in which predictions 
coincide with those of standard 
textbook quantum 
mechanics, based on Born's rule and  
the associated 
collapses of the wave function. Our work 
is in the tradition of 
Everettian approaches, 
\cite{WG}, \cite{SBKW}, \cite{Wallace}, 
\cite{Vaidman}, and 
follows our contribution  
in \cite{Sch1} and \cite{Sch2}.
The current project was in part 
motivated by the 
desire, expressed in
\cite{Sch1} and \cite{Sch2},
of providing an adequate ontology for such a 
theory, which would be compatible with the 
prediction postulate introduced there. 
This is done in
Subsection~\ref{subsec:ontology}, 
based on Theorem~\ref{thm:basic} from 
Section~\ref{sec:basic results}, 
and further elaborated in the following three
subsections, using also results from other 
sections of the paper. 

Readers will 
notice that the right-hand side of the
probability formula 
in part (b) of Theorem~\ref{thm:basic}
corresponds to 
the usual quantum mechanics 
prescription based on 
Born's rule with collapses 
after each observation.
(See also Subsection~\ref{subsec:Born}.)
And the theorems in Sections \ref{sec:F} 
and \ref{sec:Farbitrary} 
indicate, modulo intuitive assumptions 
that include a version of the 
superposition principle, and which 
are explained in
Subsections \ref{subsec:superposition}
and \ref{subsec:Born-superposition},
that this probability should 
indeed correctly predict 
which experiences we can have, 
and which ones are
ruled out, in a universe without wave 
function collapse. 

Readers familiar with the literature 
on foundations
of quantum mechanics will have no 
difficulty in 
seeing various additional ways in which
several of the theorems in this paper 
relate to quantum mechanics and 
issues in its foundations. 

In particular, the setting in which 
these theorems are formulated is clearly related to the ``Consistent, or Decoherent, Histories'' 
approach, 
\cite{Omnes}, \cite{GMH}, \cite{Griffiths}.
In this respect, the main thing to keep
in mind is that no consistency condition will be 
assumed. Rather, we will see that the Hilbert 
space can be decomposed into two orthogonal subspaces, with very distinct properties. 
These properties suggest 
(again, modulo considerations from 
\cite{Sch1}, \cite{Sch2} and 
Subsection~\ref{subsec:superposition} of this 
paper) that the component of $\Psi$ in one of these subspaces does not affect 
our experiences. On the other hand, the other component,
that is therefore responsible 
for our experiences, satisfies 
very strong consistency conditions,
in the form of commutation of 
a large class of 
relevant operators, 
and yields predictions in full equivalence 
with textbook quantum mechanics.
It will be argued, in 
Subsection~\ref{subsec:decoherence},
that this splitting of the Hilbert space
into two orthogonal subspaces 
is associated to the concept 
of environmental decoherence, 
\cite{JZKGKS}, 
\cite{Zurek}, \cite{Omnes}, \cite{GMH}, 
\cite{Schlosshauer},
\cite{Bacciagaluppi}.

Better understanding 
``Pilot-Wave'' theories, 
\cite{DT}, \cite{Bricmont} (Chapter 5), 
\cite{Goldstein}, \cite{Nelson2}, 
\cite{Davidson}, 
\cite{DG}, \cite{SV}, 
\cite{Bell2}, \cite{Vink}, 
\cite{DGTZ}, \cite{Struyve},  
was also one of the original 
motivations for this paper. 
Part (b) of Theorem~\ref{thm:basic} 
can be seen as providing a sort of ``Pilot-Wave''
picture, though in what we will call
``$\I$-space'', 
rather than physical space. 
This can, nevertheless, 
be used to produce pilot-wave theories 
in physical space.
(See Subsections \ref{subsec:pilot-wave}
and \ref{subsec:comparison}.)
And those display
stronger probabilistic agreement with 
quantum mechanics, in a path-wise 
sense, than the 
minimal agreement associated to 
the concept of equivariance
(sometimes strengthened by requiring  
good mixing properties, \cite{Valentini}),
fulfilled by various versions of
pilot-wave theories, including 
Bohmian mechanics.
Full agreement with standard
quantum mechanics
in a path-wise sense is needed
to assure that the path of the process 
produced by the pilot-wave theory displays  
correlations across time in accordance with 
quantum mechanics, and in particular 
yields appropriate records of its own past.
In this way 
problems with some pilot-wave theories, 
pointed out in 
Section 10.2 of \cite{Nelson1}, Section 5 of \cite{Nelson2}, and Section 5 of 
\cite{Bell} are avoided. 

Theorem~\ref{thm:basic}, 
presented in Section~\ref{sec:basic results}, extends substantially 
the mathematical results in \cite{Gutmann} and
\cite{VanWesep}. 
And Theorems \ref{thm:F} 
and \ref{thm:Farbitrary}, 
presented in Sections \ref{sec:F}
and \ref{sec:Farbitrary}, extend 
substantially the mathematical results in 
\cite{Sch2} (and its longer version \cite{Sch1}).
The main difference with respect to these papers 
is that there all the inputs were in the form of 
operators (usually projections) assumed, explicitly
or implicitly, to 
commute with each other, 
while here no such assumption
is made. We will see that this 
leads to the decomposition 
of the Hilbert space into the two 
subspaces mentioned 
above. Commutativity will turn out to be 
restricted 
then to the subspace accessible to our 
experiences.

The purely mathematical sections of this
paper (which are all but the last one) can be read without knowledge of quantum mechanics. 
Requisites for their reading are knowledge 
of the basic theorems about Hilbert spaces, with 
emphasis on projection operators (Chapters 1 
and 2 of \cite{Halmos} 
provide an excellent presentation of this material),
and measure theory (see, e.g., Chapters 11 and 12 of 
\cite{Royden}, Chapter 1 of \cite{Folland}, and 
Chapters 1 and 2 of \cite{Billingsley}). 
Additional knowledge of probability theory and 
the theory of (unbounded) operators in Hilbert 
spaces will be needed occasionally, and are 
covered in the books listed in the bibliography.

Sections \ref{sec:basic results}, \ref{sec:F}
and \ref{sec:Farbitrary}
of the paper contain the core mathematical  
results. Section \ref{sec:pvm} can be read independently 
of these and is included here to provide tools
used in the first two core sections, and also a 
brief introduction to projection valued measures. 
Sections \ref{sec:converse} and \ref{sec:refinements}
contain additional results, related to the material 
in the core sections.
Section~\ref{sec:examples and remarks} contains 
examples, remarks (both of a mathematical 
and philosophical nature) and applications 
of the theorems to issues in the foundations 
of quantum mechanics. 

The fastest way to learn what the content of this 
paper is is to read the material in the beginning of 
each section, from \ref{sec:basic results} to 
\ref{sec:refinements} (possibly skipping 
Section~\ref{sec:pvm}), 
stopping after the statement of the 
first theorem in each section. 
And then reading subsections
\ref{subsec:basic examples}
to 
\ref{subsec:pilot-wave},
\ref{subsec:needof},
and 
\ref{subsec:superposition} to  
\ref{subsec:comparison}.

The logic dependence of the proofs in the various 
sections is as follows. Section~\ref{sec:pvm} is 
independent of the other sections. 
Section \ref{sec:basic results} depends  
at one place on Section~\ref{sec:pvm}. 
Section~\ref{sec:F} depends on 
Section~\ref{sec:basic results} and 
at one place on Section~\ref{sec:pvm}.
Section~\ref{sec:Farbitrary} depends on 
Sections \ref{sec:basic results} and 
\ref{sec:F}. 
Sections \ref{sec:converse} and \ref{sec:refinements}
depend on Section~\ref{sec:basic results}.

\section{Basic results }
\label{sec:basic results}
Let $\H$ be a Hilbert space 
(not necessarily separable)
over the complex field. For 
$\phi, \psi \in \H$ their inner product,   
assumed to be linear in the first argument and
conjugate linear in the second, 
will be denoted by 
$\langle \phi, \psi \rangle$. The norm of $\phi$ will be denoted by $|| \phi ||$. If $\phi \not= 0$, 
we denote by $\hat{\phi} = ||\phi||^{-1} \phi$ its normalized counterpart.
The topological closure of a set $V \in \H$
will be denoted by $\overline{V}$.
By a subspace of $\H$ we will mean a subset of $\H$ that is linearly and topologically closed. Subsets of $\H$ that are linearly closed will be referred to as vector spaces, or linear spaces. 
By a projection $p$ we will mean a self-adjoint 
($\langle \phi , p \psi \rangle = 
\langle p \phi, \psi \rangle$), idempotent ($p^2 = p$)
operator.
Given a family $\{p_\alpha\}$ of projections, 
$\wedge_\alpha p_\alpha$ will denote the 
projection on the intersection of the ranges 
of the $p_\alpha$, called the ``meet'', or the ``infimum'' of these projections (see Section 
30 of \cite{Halmos}). 
If $\{Q_i\}$ is a countable family of bounded 
operators in $\H$, and $Q$ is another bounded 
operator in $\H$, the statement $\sum_i Q_i = Q$
will always mean that 
$\sum_i Q_i \phi = Q \phi$, for all $\phi \in \H$.
The concept of projection valued measures will play a major 
role in this paper. Section~\ref{thm:extension} includes their
definition and a 
brief introduction to their basic properties.
(They are called ``spectral measures'' in \cite{Halmos},
and ``resolutions of the identity in 
\cite{Royden}.)
We will make extensive use of theorems from \cite{Halmos}; when referring to Theorem $x$ in Section $y$ in that text, we will indicate it by Thm.H.$y$.$x$.     

Let $S$ be an arbitrary set. 
In applications to quantum mechanics, 
we will take $S \subset \R$ and think of 
elements of $S$ as moments in time, but in this
paper no structure or constraint 
needs to be assumed on $S$, 
unless when stated otherwise 
(as will happen in Section~\ref{sec:F}, 
where $S$ will often be supposed to be countable, and
in Section~\ref{sec:converse}, where $S$ will 
be supposed to be totally ordered).
Still, we will refer to elements of $S$ 
as ``times''. 
For each $t \in S$, let $\I(t)$ be a countable index set. To each $t \in S$ and $a \in \I(t)$ we associate a projection 
$p^t_a$ in $\H$. We assume that, for each $t \in S$, 
\begin{equation}
\sum_{a \in \I(t)}  \, p^t_a \ = \  I, 
\label{sumt}
\end{equation}
where $I$ is the identity operator, so that 
$\{p^t_a: a\in \I(t)\}$ is a partition of the identity. 
This assumption implies 
(see Thm.H.28.2 and Thm.H.27.4)
the orthogonality condition $p^t_a p^t_b = p^t_b p^t_a = 0$, if $a \not= b$. 
Set
$$
\pi = \{p^t_a : t \in S, a \in \I(t) \}.
$$

The range of $p^t_a$ will be denoted by $\mathcal{H}^t_a$, 
and for $k = 2, 3, ...$ we extend this definition by setting $\H^{t_1,t_2,..., t_k}_{a_1,a_2, ..., a_k} = \cap_{i=1}^k \H^{t_i}_{a_i}$. 
The orthogonality condition mentioned above is equivalent to the 
statement that for each $t_1,...,t_k$, 
\begin{equation}
\H^{t_1, ..., t_k}_{a_1,..., a_k} \perp 
\H^{t_1, ..., t_k}_{b_1,..., b_k}, \ \mbox{  if  } \  (a_1,...,a_k) \not= (b_1,...,b_k).
\label{perp}
\end{equation}
This orthogonality condition allows us to define the direct sum 
$$
\H^{t_1, ...,t_k} \ = \ \bigoplus_{a_1,...,a_k} \H^{t_1, ..., t_k}_{a_1,...,a_k}.
$$
The projection on $\mathcal{H}^{t_1,t_2,..., t_k}_{a_1,a_2, ..., a_k}$ will be denoted by 
$p^{t_1,t_2,..., t_k}_{a_1,a_2, ..., a_k}
= \wedge_{i=1}^{k} p^{t_i}_{a_i}$.
It is clear that for $k < l$,  $\H^{t_1,...,t_k,...,t_l}_{a_1,...,a_k, ..., a_l} \subset \H^{t_1,...,t_k}_{a_1,...,a_k}$ and therefore 
\begin{equation}
 p^{t_1,...,t_k,...,t_l}_{a_1,...,a_k, ..., a_l} \phi =0, \ \mbox{ whenever }  \ p^{t_1,...,t_k}_{a_1,...,a_k} \phi =0.
    \label{monotone}
\end{equation}

We introduce now a measurable space in the following fashion. Define the Cartesian product $\Omega = \times_{t \in S} \I(t)$.
In applications in which we think of elements of 
$S$ as moments in time, $\Omega$ can be 
thought of as the
set of trajectories, or histories,  in ``$\I$ space''.
And let $\Sigma$ be the smallest sigma-algebra 
of subsets of $\Omega$ that contains all the 
sets of the form $\{\omega \in \Omega : 
\omega_t = a\}$, $t \in S$, $a \in \I(t)$.  
Define the functions $X_t$, $t \in S$, from 
$\Omega$ to $\I(t)$ by $X_t(\omega) = \omega_t$. 
We will often use probabilistic terminology and notation, so elements of $\Sigma$ will sometimes be called ``events'', the $X_t$ will sometimes be called ``random variables'' and we abbreviate $\{\omega
\in \Omega : \omega_{t_i} = a_i, i = 1, ..., k\} 
= \{X_{t_i} = a_i, i=1,...,k\}$. 
The events of the form $\{ (X_{t_1}, ..., X_{t_k}) \in G \}$, for some $t_1, ...,t_k$ and $G \subset \I(t_1) \times ... \times \I(t_k)$ form an algebra $\A$, which generates the sigma-algebra $\Sigma$. 
The class of events obtained by countable unions of events in $\A$ 
will be denoted by $\A_{\sigma}$, and
the class of events obtained by countable intersections of events in $\A$ will be denoted by $\A_{\delta}$.

We turn now to the definition of a subspace of $\H$ that will play a central role in this paper. Define $\mathcal{W}$ as the set of operators on $\H$ that are products of finitely many elements of $\pi$. And for $W,V \in \mathcal{W}$, write $W \sim V$ in case $W$ and $V$ are obtained from the same elements of $\pi$, but possibly multiplied in different orders. Given a vector $\phi \in \H$ we say that ``$\pi$ commutes on $\phi$'' if $W \phi = V \phi$, whenever $W \sim V$. And we define
$$
\H_{\pi} \ = \ \{\phi \in \H \, : \, \mbox{$\pi$ commutes on $\phi$} \}. 
$$
It is clear that $\Hp$ is a subspace of $\H$ and that for each $t \in S$ and $a \in \I(t)$, 
\begin{equation}
    p^t_a \Hp \subset \Hp,
    \label{invariant}
\end{equation} 
i.e., $\Hp$ is invariant under the projections in $\pi$. Whenever $\Hp$ is invariant under a projection $p$, we will denote by $\t{p}$ the restriction of $p$ to $\Hp$.
We will denote by $p_{\pi}$ the projection on $\Hp$. And for $\phi \in \H$ we define
$\phi_\pi = p_\pi \phi$.

Define also 
\begin{eqnarray*}
\Hp' & =   & \left\{ \phi \in \H \, : \, \mbox{$p^{t_1, ...,t_k}_{a_1,...,a_k} \phi = 
p^{t_1}_{a_1} ... p^{t_k}_{a_k} \phi$ \ \ for all \ \ $t_1, ..., t_k$ and $a_1, ..., a_k$ } \right\},
\\ 
\Hp'' & =  &  \left\{ \phi \in \H \, : \, \mbox{$\sum_{a_1, ..., a_k} p^{t_1, ...,t_k}_{a_1,...,a_k} \phi = \phi$
\ \ for all \ \ $t_1, ..., t_k$} 
\right\},
\\ 
N  &  =  &  \left\{ \phi \in \H \, : \, \mbox{ for some $t_1,...,t_k$,  \ 
$p^{t_1, ...,t_k}_{a_1,...,a_k} \phi = 0$ \ 
for all 
$a_1, ..., a_k$}
\right\}.
\end{eqnarray*}

For examples of the setting above in a 
quantum mechanical context, see 
Subsections \ref{subsec:basic examples}
and \ref{subsec:particles}. 
For the relationship between $\Hp$ 
and the phenomenon of 
environmental decoherence, 
see Subsection~\ref{subsec:decoherence}.
For remarks 
on the meaning of the right-hand side of 
(\ref{Born}) in the following theorem, 
see Subsection~\ref{subsec:Born}.

\begin{thm}
\begin{itemize}
    \item[(a)] $N$ is a vector space and 
    \begin{equation} 
    \Hp = \Hp' = \Hp'' = N^{\perp}.
    \label{Hp=Hp}
    \end{equation}
    \item[(b)] For any $\phi \in \Hp \backslash \{0\}$, there exists a unique probability measure $\P_{\phi}$ on $(\Omega, \Sigma)$,
    such that 
    \begin{equation}
    \P_{\phi}(X_{t_i} = a_i, i = 1, ..., k) \ = \  
    || p^{t_1, ..., t_k}_{a_1,...,a_k} \hat{\phi} ||^2 \ = \  
    || p^{t_k}_{a_k} ... p^{t_1}_{a_1} \hat{\phi} ||^2,
    \label{Born}
    \end{equation}
    for every $t_1, ..., t_k$, and $a_1, ..., a_k$.
    \item[(c)] To each $A \in \Sigma$ there corresponds a subspace $\H_A \subset \Hp$, with the property that if $p_A$ 
    is the projection on $\H_A$ and $\t{p}_A$ is its restriction to $\Hp$, then $\{\t{p}_A : A \in \Sigma \}$ is the unique projection valued measure (p.v.m.) from $\Sigma$ to projections in $\Hp$ such that 
    \begin{equation}
    \t{p}_{\{X_{t_1} = a_1, ... ,X_{t_k} = a_k\}} \phi \ = \ p^{t_1, ..., t_k}_{a_1, ...a_k} \phi,
    \label{pvm}
    \end{equation}
    for all $t_1, ..., t_k$, $a_1, ..., a_k$
    and $\phi \in \Hp$. 
    In particular, $\H_{\Omega} = \Hp$,  
    and for any $A\in \Sigma$,
    \begin{equation}
    \H \ = \ \H_A \, \oplus \, \H_{A^c} \, \oplus \, \Hp^{\perp}
    \ = \ 
    \H_A \, \oplus \, \H_{A^c} \, \oplus  \, \overline{N}. 
    \label{Hsplit}    
    \end{equation}
    \item[(d)] For every $\phi \in \Hp \backslash \{0\}$ and every $A \in \Sigma$, 
    \begin{equation}
    \P_{\phi} (A) \ = \ || p_A \hat{\phi} ||^2.
    \label{thmd}
    \end{equation}
    And for every $A \in \Sigma$,
    \begin{equation}
    \H_A \ = \ 
    \left\{
    \phi \in \Hp \, : \, 
    \phi = 0 \ \ \mbox{or} \ \ 
    \P_\phi(A) = 1
    \right\}, 
    \label{thmd+}    
    \end{equation}
    and 
    \begin{equation}
\H_A^\perp  \ = \ 
    \left\{
    \phi \in \H \, : \, 
    \phi_\pi = 0 \ \ \mbox{or} \ \ 
    \P_{\phi_\pi}(A) = 0
    \right\}. 
    \label{lemmaG}  
    \end{equation}
    \item[(e)] For every $\phi \in \Hp \backslash \{0\}$ and every $A,B \in \Sigma$ such that $\P_{\phi}(A) \not= 0$ we have the following conditional probability formula
    \begin{equation}
    \P_{\phi}(B|A) \ = \ \P_{p_A \phi} (B).
    \label{thme}
    \end{equation}
    \item[(f)] To each measurable function $f$ 
    from $\Omega$ to $\R$ (endowed with the 
    Borel sigma-algebra $\cR$) there corresponds
    a self-adjoint operator 
    in $\Hp$, denoted by $Q_{f}$, 
    with domain $\mathcal{D}_f = 
    \{\phi \in \Hp : \phi = 0, \ \mbox{or} \ \int f^2 d\P_\phi < \infty
    \}$,
    such that
    \begin{equation}
    \int f d\P_\phi
    \ = \ 
    \langle \hat\phi, Q_f \hat\phi \rangle, 
    \label{Q_f}
    \end{equation}
    for every $\phi \in \mathcal{D}_f
    \backslash \{0\}$.
    The spectral decomposition of $Q_f$
    is given by the p.v.m. 
    \begin{equation}
    \left\{ 
    p_{\{f \in B\}} \, : \, B \in \cR
    \right\}.
    \label{Q_fspectrum}     
    \end{equation}
    All such operators $Q_f$ commute with each 
    other, in the sense 
    (\cite{RS}, Section VIII.5) 
    that the projections in their spectral decompositions  
    all commute with each other.  
\end{itemize}
\label{thm:basic}
\end{thm}

\noindent{\bf Remark on item (f):} 
If $f$ is bounded, then 
$\mathcal{D}_f = \Hp$, and $Q_f$ is a 
bounded operator. If also $g$ is a bounded
measurable function from $\Omega$ to $\R$, 
then the commutation stated in the theorem
takes the usual form $Q_f Q_g = Q_g Q_f$, 
thanks to the spectral theorem. 

Before proving Theorem~\ref{thm:basic}, we collect some technical results in two propositions.

\begin{prop}
Suppose that $p$ is the projection on the subspace $\S_p$ and $q$ is the projection 
on the subspace $\S_q$. Then the following are equivalent:
\begin{itemize}
    \item[(i)] $p \, \S_q \subset \S_q$.
    \item[(ii)] $p  \, \S_q^{\perp} \subset \S_q^{\perp}$.
    \item[(iii)] $p$ and $q$ commute. 
    \item[(iv)] $q \, \S_p \subset \S_p$.
    \item[(v)] $q \, \S_p^{\perp} \subset \S_p^{\perp}$.
\end{itemize}
\label{prop:reduction}
\end{prop}

\noindent{\bf Proof:}
Thm.H.23.2 implies that (i) and (ii) are equivalent, because $p$ is self-adjoint.
Thm.H.27.2 implies that (i) and (ii) together are equivalent to (iii). 
This establishes the equivalence of (i), (ii) and (iii). 
Interchanging the roles of $p$ and $q$, we obtain also the equivalence of 
(iv) and (v) with (iii), completing the proof. 
$\square$


\begin{prop} For every $t_1,...,t_k$ and $a_1, ..., a_k$,
\begin{itemize}
    \item[(a)]  
    $$
    p^{t_1,...,t_k}_{a_1, ...,a_k} \, \Hp \ \subset \Hp 
    \ \ \ \  \ \ \mbox{and}  \ \ \ \ \ \ 
    p^{t_1,...,t_k}_{a_1, ..., a_k} \, \Hp^{\perp} \ \subset \Hp^{\perp}. 
    $$
    \item[(b)] $p^{t_1,...,t_k}_{a_1, ..., a_k}$ commutes with $p_{\pi}$.
    \item[(c)] The restriction of $p^{t_1,...,t_k}_{a_1, ..., a_k}$ to $\Hp$,  
    $\t{p}^{t_1,...,t_k}_{a_1, ..., a_k}$, is the projection in $\Hp$ on the subspace  
    $\H^{t_1,...,t_k}_{a_1, ..., a_k} \, \cap \, \Hp$.  
    \item[(d)] For every $\phi \in \Hp, \ 
    p^{t_1,...,t_k}_{a_1, ..., a_k} \phi = p^{t_1}_{a_1} ... p^{t_k}_{a_k} \phi$.  
    \item[(e)] For every $\phi \in \Hp$,  
    \begin{equation}
    \sum_{a_i \in \I(t_i)} \, p^{t_1,...,t_{i-1},t_i,t_{i+1}, ..., t_k}_{a_1, ...,a_{i-1},a_i,a_{i+1}, ..., a_k} \, \phi
    \ = \ 
    p^{t_1,...,t_{i-1},t_{i+1}, ..., t_k}_{a_1, ...,a_{i-1},a_{i+1}, ..., a_k} \phi.
    \label{sump}
    \end{equation}
\end{itemize}
\label{prop:technical}
\end{prop}
\noindent{\bf Proof:}
\noindent {\bf (a) and (b):}
From (\ref{invariant}) and Proposition~\ref{prop:reduction}, we have 
$p_{\pi} \H^t_a \subset \H^t_a$, for each $t\in S$ and $a \in \I(t)$. 
Therefore
$$
p_\pi \, \H^{t_1,...,t_k}_{a_1,...,a_k} 
\ = \ 
p_\pi \left(    
\cap_{i=1}^k \, \H^{t_i}_{a_i} 
\right)
\ \subset \ 
\cap_{i=1}^k \, \H^{t_i}_{a_i}
\ = \ 
\H^{t_1,...,p_k}_{a_1, ...,a_k}.
$$
And invoking Proposition~\ref{prop:reduction} again we complete the proof of (a) and (b).

\noindent {\bf (c):}
The operator $\t{p}^{t_1,...,t_k}_{a_1, ..., a_k}$ in $\Hp$ inherits self-adjointness 
and idempotency from ${p}^{t_1,...,t_k}_{a_1, ..., a_k}$, so that it is indeed a projection in $\Hp$. And 
$$
\mbox{Range} \, \left( \t{p}^{t_1,...,t_k}_{a_1, ..., a_k} \right)
\ = \
\left\{ 
\phi \in \Hp \, : \, {p}^{t_1,...,t_k}_{a_1, ..., a_k} \, \phi \, = \, \phi 
\right\}
\ = \ 
\left\{ 
\phi \in \H \, : \, {p}^{t_1,...,t_k}_{a_1, ..., a_k} \, \phi \, = \, \phi 
\right\} \ \cap \ \Hp 
\ = \ 
\H^{t_1,...,t_k}_{a_1, ..., a_k} \, \cap \, \Hp.
$$
\noindent {\bf (d):} 
By the definition of $\Hp$, the projections $\t{p}^{t}_a$, $t \in S$, $a \in \I(t)$
commute with each other. Therefore Thm.H.29.1 implies that 
\begin{equation}
\t{p}^{t_1}_{a_1} ... \t{p}^{t_k}_{a_k} 
\ = \ 
\wedge_{i=1}^k \, \t{p}^{t_i}_{a_i}.
\label{label1}
\end{equation}
From (c) above, we have 
$$
\mbox{Range} \, \left( \wedge_{i=1}^k \t{p}^{t_i}_{a_i} \right)
\ = \
\bigcap_{i=1}^k \, 
\mbox{Range} \, \left( \t{p}^{t_i}_{a_i} \right)
 \ = \ 
\bigcap_{i=1}^k \,
\left(
\H^{t_i}_{a_i} \, \cap \, \Hp
\right)
\ = \ 
\H^{t_1,...,t_k}_{a_1, ..., a_k} \, \cap \, \Hp 
\ = \ 
\mbox{Range} \, \left(  
\t{p}^{t_1,...,t_k}_{a_1, ..., a_k}
\right),
$$
i.e., 
\begin{equation}
\wedge_{i=1}^k \, \t{p}^{t_i}_{a_i} \ = \ \t{p}^{t_1,...,t_k}_{a_1, ..., a_k}.
\label{label2}
\end{equation}
Combining 
(\ref{label1}) and (\ref{label2}), 
we have, for $\phi \in \Hp$,
$$
p^{t_1,...,t_k}_{a_1, ..., a_k} \phi 
\ = \ 
\t{p}^{t_1,...,t_k}_{a_1, ..., a_k} \phi
\ = \ 
\t{p}^{t_1}_{a_1} ... \t{p}^{t_k}_{a_k} \phi
\ = \ 
p^{t_1}_{a_1} ... p^{t_k}_{a_k} \phi.
$$
\noindent {\bf (e):} Using (d) above and (\ref{sumt}), we have,
for $\phi \in \Hp$,
\begin{eqnarray*}
    \sum_{a_i \in \I(t_i)} \, p^{t_1,...,t_{i-1},t_i,t_{i+1}, ..., t_k}_{a_1, ...,a_{i-1},a_i,a_{i+1}, ..., a_k} \, \phi
    & \ = \ & 
    \sum_{a_i \in \I(t_i)} \, p^{t_1}_{a_1}...p^{t_{i-1}}_{a_{i-1}} p^{t_i}_{a_i} p^{t_{i+1}}_{a_{i+1}} ... p^{t_k}_{a_k} \, \phi
    \\
    & \ = \ &
    p^{t_1}_{a_1}...p^{t_{i-1}}_{a_{i-1}} \, 
    \left( \sum_{a_i \in \I(t_i)} \, p^{t_i}_{a_i} \right) \, 
    p^{t_{i+1}}_{a_{i+1}} ... p^{t_k}_{a_k} \, \phi
    \\  
    & \ = \ &
    p^{t_1}_{a_1}...p^{t_{i-1}}_{a_{i-1}}  
    p^{t_{i+1}}_{a_{i+1}} ... p^{t_k}_{a_k} \, \phi
    \\  
    & \ = \ &
    p^{t_1,...,t_{i-1},t_{i+1}, ..., t_k}_{a_1, ...,a_{i-1},a_{i+1}, ..., a_k} \phi.
    \label{sump+'}
    \end{eqnarray*}
    The second equality is justified by Thm.H.28.1, 
    which states that since the sum inside the
    parenthesis is well defined, we can exchange 
    the order of the operations, as done there.
    $\square$

\noindent{\bf Proof of Theorem~\ref{thm:basic}:}

\noindent{\bf (a):} $N$ is clearly closed with respect to multiplication by scalars. That it is closed with respect to sums is a simple consequence of (\ref{monotone}). Indeed, if $\phi', \phi''\in N$, we can combine the corresponding sets $t'_1, ... , t'_{k'}$ and $t''_1, ... , t''_{k''}$, whose existence is implied by these assumptions, to produce a set $t_1, ... , t_k$ for which $p^{t_1, ...,t_k}_{a_1,...,a_k} \psi = 0$ for $\psi = \phi', \phi''$ and any $a_1, ..., a_k$. Hence also $p^{t_1, ...,t_k}_{a_1,...,a_k} (\phi' + \phi'') = 0$, completing the proof that $N$ is a vector space.

Next we will prove that 
\begin{equation}
\Hp \subset \Hp' \subset \Hp'' \subset \Hp.
\label{subsetchain}
\end{equation}

The first of these claims is a restatement of part (d) of Proposition~\ref{prop:technical}.

Suppose now that $\phi \in \Hp'$. Then, using (\ref{sumt}) and Thm.H.28.1 (as in the proof of 
part (e) of Proposition~\ref{prop:technical}),  
$$
\sum_{a_1, ..., a_k} p^{t_1, ...,t_k}_{a_1,...,a_k} \phi \ = \
\sum_{a_1, ..., a_k} p^{t_1}_{a_1} ... p^{t_k}_{a_k} \phi \ = \ \left(\sum_{a_1} p^{t_1}_{a_k} \right) ...  
\left(\sum_{a_k} p^{t_k}_{a_k} \right) \phi \ = \
I^k \phi \ = \
\phi,
$$
implying that $\Hp' \subset \Hp''$.

Finally, to prove the last claim in (\ref{subsetchain}),
suppose that $\phi \in \Hp''$. Suppose that 
$W = p^{s_1}_{b_1} ... p^{s_l}_{b_l}$. 
We need to show that $W \phi$ does not depend on the order 
of the factors defining $W$. 
The times $s_1, ... s_l$ may include repetitions, 
so let $t_1, ...,t_k$ be the same set of times, but without the repetitions. We will use the equation  
$\phi = \sum_{a_1, ..., a_k} p^{t_1, ...,t_k}_{a_1,...,a_k} \phi$. Apply $W$ to both sides of this equation,
and use the following two facts. First $\H^{t_i}_{c} \supset \H^{t_1, ...,t_k}_{a_1,...,a_k}$, if $c = a_i$,
so that in this case $p^{t_i}_c \, p^{t_1, ...,t_k}_{a_1,...,a_k} = p^{t_1, ...,t_k}_{a_1,...,a_k}$.
Second, $\H^{t_i}_{c} \perp \H^{t_1, ...,t_k}_{a_1,...,a_k}$, if $c \not= a_i$,
so that in this case $p^{t_i}_c \, p^{t_1, ...,t_k}_{a_1,...,a_k} = 0$. 
This gives us $W \phi = 0$, in case $W$ includes two factors $p^{t_i}_c$ with distinct $c$, and
$W \phi = p^{t_1, ...,t_k}_{a_1,...,a_k} \phi$ in case every factor $p^{t_i}_{c}$ in $W$ has $c=a_i$. In either case
$W\phi$ does not depend on the order of the factors that define $W$. Hence $\phi \in \Hp$.

This completes the proof of (\ref{subsetchain}) and hence of the first two equalities in (\ref{Hp=Hp}). 

The orthogonality in (\ref{perp}) implies that the statement $p^{t_1, ...,t_k}_{a_1,...,a_k} \phi = 0$ for all 
$a_1, ...a_k$, in the definition of $N$, is equivalent to the statement that 
$\sum_{a_1,...a_k} p^{t_1, ...,t_k}_{a_1,...,a_k} \phi = 0$. But this is equivalent to the statement that $\phi$
is orthogonal to the range of the projection $\sum_{a_1,...a_k} p^{t_1, ...,t_k}_{a_1,...,a_k}$, which is 
$\H^{t_1, ..., t_k}$. Therefore 
$$
N \ = \ \bigcup \left\{ \left( \H^{t_1, ...,t_k} \right)^{\perp} \ : \ 
t_1, ... , t_k \in S \right\}.
$$
And hence
$$
N^{\perp} \ = \ \bigcap \left\{ \H^{t_1, ..., t_k}  \ : \ 
t_1,  ... , t_k \in S \right\}
\ = \  \Hp'',
$$
finishing the proof of (\ref{Hp=Hp}).



\noindent{\bf (b):} We will use Kolmogorov's extension theorem, 
also called Kolmogorov's existence theorem.
(See, e.g., Section 36 of \cite{Billingsley},
or Section 4 of Chapter 9 of \cite{Folland}. 
To apply the theorem as usually 
stated, for real valued random variables,   
identify each $\I(t)$ with a subset of the set $\{0, 1, ...\} \subset \R$, 
so that the $X_t$ can be seen as real valued random variables.)
For this purpose, we first define probability 
measures on smaller spaces, corresponding to finitely many moments in time. For each 
$t_1, ..., t_k \in S$, define $\Omega^{t_1, ..., t_k} = \I(t_1) \times ... \times \I(t_k)$, and for each $G \subset \Omega^{t_1, ..., t_k}$ and $\phi \in \Hp \backslash \{0\}$ define
$$
\mu^{t_1, ..., t_k}_{\phi} (G) 
\ = \ 
\left|\left| \sum_{(a_1,...,a_k) \in G} p^{t_1, ...,t_k}_{a_1,...,a_k} \hat{\phi} \, \right|\right|^2.
$$
The sum in this expression is well defined since, by 
Thm.H.28.2, a sum of orthogonal projections is a 
projection, and orthogonality comes from 
(\ref{perp}). 
Since $\hat{\phi} \in \Hp = \Hp''$, we have 
$\mu^{t_1,...,t_k} (\Omega^{t_1,...,t_k})
= || \hat{\phi} ||^2 = 1$, so that these are indeed
probability measures. 
These probability measures satisfy the following two consistency 
conditions. 

First, suppose $G = G_1 \times ... \times G_k$, with $G_i \subset \I(t_i)$, $i=1, ..., k$. Let $\kappa$ be a permutation of the elements of the set $\{1, ..., k\}$, and 
set $\kappa(G) = G_{\kappa(1)} \times ... \times G_{\kappa(k)}$. 
Then $\mu^{t_{\kappa(1)}, ..., t_{\kappa(k)}}_{\phi} (\kappa(G)) = \mu_{\phi}^{t_1,...,t_k} (G)$, simply because $p^{t_{\kappa(1)}, ..., t_{\kappa(k)}}_{a_{\kappa(1)}, ..., a_{\kappa(k)}} = p^{t_1, ...,t_k}_{a_1,...,a_k}$.

Second, suppose that $G \subset \Omega^{t_1, ..., t_k}$,
$t_{k+1} \in S \backslash \{t_1,...,t_k\}$, 
and set $G' = G \times \I(t_{k+1}) \subset \Omega^{t_1, ..., t_{k+1}}$. 
Then (\ref{sump}) implies that 

$$
\mu^{t_1,...,t_{k+1}}_{\phi} (G')
\ = \ 
\left|\left| \sum_{(a_1,...,a_k) \in G} \, \sum_{a_{k+1} \in \I(t_{k+1})}
p^{t_1, ...,t_k, t_{k+1}}_{a_1,...,a_k, a_{k+1}} \hat{\phi} \, \right|\right|^2
\ = \ 
\left|\left| \sum_{(a_1,...,a_k) \in G} 
p^{t_1, ...,t_k}_{a_1,...,a_k} \, 
\hat{\phi} \, \right|\right|^2
\ = \
\mu^{t_1, ...,t_k}_{\phi} (G).
$$

These consistencies establish, thanks to Kolmogorov's extension theorem, the existence of a probability measure $\P_{\phi}$ on $(\Omega, \Sigma)$ that satisfies 
$\P_{\phi} ((X_{t_1}, ..., X_{t_k}) \in G) = \mu^{t_1, ..., t_k}_{\phi}(G)$ and in particular
the first equality in (\ref{Born}). The second equality there is
a consequence of $\hat{\phi} \in \Hp = \Hp'$.

Uniqueness of $\P_{\phi}$ is proved as follows. If (\ref{Born}) holds, then, 
for any $G \subset \I(t_1) \times ... \times \I(t_k)$ 
and $\phi \in \Hp \backslash \{0\}$,   
\begin{equation}
\P_{\phi} ((X_{t_1}, ..., X_{t_k}) \in G) 
\ = \
\sum_{(a_1,...,a_k) \in G} || p^{t_1, ...,t_k}_{a_1,...,a_k} \hat{\phi} ||^2
\ = \
\left|\left| \sum_{(a_1,...,a_k) \in G} p^{t_1, ...,t_k}_{a_1,...,a_k} \hat{\phi} \, \right|\right|^2
\ = \
\mu^{t_1, ..., t_k}_{\phi}(G),
\label{PA}
\end{equation}
where in the second equality we used 
the orthogonality (\ref{perp}).
This defines uniquely the probability measure $\P_{\phi}$ restricted to the algebra $\A$. And hence it can only have a unique
extension to the sigma-algebra $\Sigma$ generated by $\A$.

\noindent {\bf (c):} For each $A \in \A$ set   
\begin{equation}   
\H_A \ = \ \{ \phi \in \Hp \, : \, \mbox{$\phi = 0$ \, or \,
$\P_{\phi}(A) = 1$} \}.  
\label{HAP}
\end{equation}
We will show that 
each $\H_A$ is a subspace and that if $p_A$ is the projection on $\H_A$ and $\t{p}_A$ is its restriction 
to $\Hp$, then $\{\t{p}_A : A \in \A \}$ is the unique p.v.m. from the algebra $\A$ to projections in $\Hp$ that 
satisfies (\ref{pvm}).
The extension of these claims, from a p.v.m. on the algebra $\A$ to a p.v.m. 
on the sigma-algebra $\Sigma$, as stated in the theorem, then follows from Theorem~\ref{thm:extension}, in Section~\ref{sec:pvm}. 

If $A \in \A$, then  
$A = \{(X_{t_1}, ..., X_{t_k}) \in G \}$, 
for some $t_1, ..., t_k$ and $G$. Since (\ref{Born}) has already been proved, 
(\ref{PA}) is also true and it implies the following.
(Recall that 
$\t{p}^{t_1, ...,t_k}_{a_1,...,a_k}$ denotes the restriction of ${p}^{t_1, ...,t_k}_{a_1,...,a_k}$
to $\Hp$.)
\begin{equation}
\H_A \ = \ \left\{ \phi \in \Hp \, : \, 
\left|\left| \sum_{(a_1,...,a_k) \in G} \t{p}^{t_1, ...,t_k}_{a_1,...,a_k} \phi \, \right|\right|^2
\ = \ 
|| \phi ||^2 \right\} 
\ = \ 
\mbox{Range} \left( 
\sum_{(a_1,...,a_k) \in G} \t{p}^{t_1, ...,t_k}_{a_1,...,a_k}
\right),
\label{caprange}
\end{equation}
where, to obtain the second equality, we used the fact that a sum of orthogonal projections is a projection 
(Thm.H.28.2)
combined with (\ref{perp}), and the characterization of the range of a projection as the set of vectors whose norms are not affected by the projection 
(Thm.H.26.2).
As the range of a projection in $\Hp$, the right-hand side of this equation is a subspace of $\Hp$, and hence so is $\H_A$. And the equation means that 
\begin{equation}
\t{p}_A \ = \ \sum_{(a_1,...,a_k) \in G} \t{p}^{t_1, ...,t_k}_{a_1,...,a_k}. 
\label{tpA}    
\end{equation}
Moreover, combining (\ref{PA}) and (\ref{tpA}), 
we obtain that for any $\phi \in \Hp \backslash \{0\}$
and $A \in \A$,
\begin{equation}
\P_{\phi}(A) 
\ = \
\left|\left| \sum_{(a_1,...,a_k) \in G} \t{p}^{t_1, ...,t_k}_{a_1,...,a_k} \hat{\phi} \right|\right|^2
\ = \ ||\t{p}_A \hat{\phi} ||^2
\ = \ ||{p}_A \hat{\phi} ||^2.
\label{Pp}
\end{equation}
 
If $A,B \in \A$ there exists $t_1, ..., t_k$ such that 
$A = \{(X_{t_1}, ..., X_{t_k}) \in G_A \}$ and 
$B = \{(X_{t_1}, ..., X_{t_k}) \in G_B \}$, for 
some $G_A$ and $G_B$. If also 
$A \cap B = \emptyset$,
then $G_A \cap G_B = \emptyset$
and the ranges in the right-hand side 
of (\ref{caprange}) with $G=G_A$ or $G=G_B$ will be orthogonal to each other, due to (\ref{perp}), implying that $\H_A \perp \H_B$.
(In Section~\ref{sec:pvm} this property is called (PVM6).)

We will prove next the two conditions that define $\{ \t{p}_A : A \in \A \}$ as a p.v.m.. (See Section~\ref{sec:pvm} for this definition.) 

(PVM1) From (\ref{HAP}), it is immediate that $\H_\Omega = \Hp$, 
as required. 

(PVM2) If $A_1, A_2, ... $ are disjoint events in $\A$, and also $A = \cup_{i=1, ...} A_i \in \A$, then 
\begin{eqnarray}
\H_A & \ = \ & 
\{ \phi \in \Hp \, : \, \mbox{$\phi = 0$ \, or \, $\P_{\phi}(A) = 1$} \}
\ = \
\{ \phi \in \Hp \, : \, \mbox{$\phi = 0$ \, or \, $\sum_i\P_{\phi}(A_i) = 1$} \} 
\nonumber     \\   
& \ = \ &
\left\{ \phi \in \Hp \, : \, \sum_i ||\t{p}_{A_i} \phi ||^2 = ||\phi||^2 \right\}
\ = \ 
\left\{ \phi \in \Hp \, : \, \left| \left| \sum_i \t{p}_{A_i} \phi \right|\right| ^2 = \, ||\phi||^2 \right\}
\nonumber     \\
& \ = \ & 
\mbox{Range} \, \left(   
\sum_i \t{p}_{A_i}
\right),
\label{forunion}
\end{eqnarray}
where in the third equality we used (\ref{Pp}), 
in the fourth  the equality we used the orthogonality due to the disjointness of the events $A_i$, (PVM6), proved above, 
and in the fifth equality we used the same kind of arguments used to justify the second equality in (\ref{caprange}).
Equation (\ref{forunion}) implies that $\t{p}_A = \sum_i \t{p}_{A_i}$, as required. 

Uniqueness is immediate, since, by (PVM2) a p.v.m. that satisfies (\ref{pvm}) must satisfy (\ref{tpA}).

\noindent {\bf (d):} Since $\{\t{p}_A : A \in \Sigma \}$ is a p.v.m. in $\Hp$ and $\hat{\phi} \in \Hp$, the 
right-hand side of (\ref{thmd}) defines a measure on $(\Omega,\Sigma)$. Equation (\ref{Pp}) 
states that this measure coincides with $\P_{\phi}$ on the algebra $\A$. Hence it must coincide 
with $\P_{\phi}$ on the sigma-algebra $\Sigma$, generated by $\A$, proving (\ref{thmd}).

From Thm.H.26.3, 
that characterizes the range of a projection as the 
set of vectors whose norm is not 
affected by the projection, 
and (\ref{thmd}) we have 
\begin{eqnarray*}
\H_A 
& \ = \ & 
\left\{
\phi \in \Hp \, : \, || p_A {\phi} ||^2 
\, = \, ||\phi||^2
\right\}
\ = \ 
\left\{
\phi \in \Hp \, : \, 
\phi = 0 \ \ \mbox{or} \ \ 
|| p_A \hat{\phi} ||^2 
\, = \, 1
\right\}
\\
& \ = \ & 
\left\{
\phi \in \Hp \, : \, 
\phi = 0 \ \ \mbox{or} \ \ 
\P_\phi (A) = 1
\right\},
\end{eqnarray*}
proving (\ref{thmd+}).

To prove (\ref{lemmaG}), observe that  
the statement $\phi \in \H_A^\perp$ is
equivalent to the statement 
that $p_A \phi =0$.
And since $\H_A \subset \Hp$, this last
statement is equivalent to the statement 
that
$\phi_\pi = 0$, or 
$\P_{\phi_\pi} (A) = 
||p_A {\phi_\pi}||^2 / ||\phi_\pi||^2 =
||p_A {\phi}||^2 / ||\phi_\pi||^2 = 0$.

\noindent {\bf (e):} Set $\eta = p_A \phi$. Clearly $\eta \in \Hp$ and  
since $\P_{\phi}(A) \not= 0$, we have, from (\ref{thmd}) that $\eta \not= 0$.
Therefore $\P_{\eta}(B)$ is well defined.
Using now (\ref{thmd}) and the property $p_{AB} = p_B p_A$ of projection valued measures
(called (PVM8) in Section \ref{sec:pvm}), we have 
$$
\P_{\phi}(B | A) \ = \ 
\frac{\P_{\phi}(AB)}{\P_{\phi}(A)} \ = \
\frac{|| p_{AB} \hat{\phi} ||^2 }{ || p_{A} \hat{\phi} ||^2} \ = \ 
\frac{|| p_{AB}{\phi} ||^2 }{ || p_{A} {\phi} ||^2} \ = \
\frac{|| p_{B} p_A {\phi} ||^2 }{ || p_{A} {\phi} ||^2} \ = \ 
\frac{|| p_{B} {\eta} ||^2 }{ || {\eta} ||^2} \ = \ 
|| p_B \hat{\eta} ||^2 \ = \ 
\P_{\eta} (B). 
$$

\noindent {\bf (f):}
All the statements follow from the fact that 
$\{p_A :  A \in \Sigma \}$ is a p.v.m. in $\Hp$ that satisfies (\ref{thmd}), by applying these results to Theorems 13.24 and 13.28 of \cite{Rudin}.  

$\square$

\section{A characterization of $\H_A$ when $S$ is countable}
\label{sec:F}
For $A \subset \Omega$, we define 
$$
F_A \ = \ \left
\{\phi \in \H \, : \, 
\mbox{for any $\omega \in A$ 
there is $t_1, ..., t_k$ such that 
$p^{t_1,...,t_k}_{\omega_{t_1},...,\omega_{t_k}} \phi = 0$ 
}  \right\}. 
$$
If $S = \{s_1,s_2, ...\}$ is countable, we can use (\ref{monotone}) to see that 
\begin{equation}
F_A \ = \ \left
\{\phi \in \H \, : \, 
\mbox{for any $\omega \in A$ 
there is $k \in \{1,2, ...\}$ such that 
$p^{s_1,...,s_k}_{\omega_{s_1},...,
\omega_{s_k}} \phi = 0$ }  \right\}. 
\label{remark}
\end{equation}
Recall that $p_{\pi}$ is the projection on $\Hp$ and that $\phi_{\pi} = p_{\pi} \phi$.
Our goal in this section is to prove
\begin{thm} 
If $S$ is countable and $A \in \Sigma$, then
\begin{itemize}
\item[(a)] $\H_A = F_A^{\perp}$ \quad and \quad  
$\H_A^\perp = \overline{F_A}$.
\item[(b)] $\phi \in \overline{F_A} \ \ 
\Longleftrightarrow
\ \  \mbox{$\phi_{\pi} = 0$,\,  or \,  
$\P_{\phi_{\pi}}(A) = 0$}$.
\item[(c)] $\H_A \, = \, 
\overline{F_{A^c}} \cap \Hp \, = \, 
\overline{F_{A^c} \cap \Hp}$.
\end{itemize}
\label{thm:F}
\end{thm}

This theorem will be partially 
extended to arbitrary $S$ 
in Section~\ref{sec:Farbitrary}, building on the 
results in this section.

The proof of Theorem~\ref{thm:F} will rely on several lemmas. 
First we collect some elementary properties of $F_A$ in a proposition:
\begin{prop}
$F_A$ decreases as $A$ increases.  
For any family $\{A_\alpha\}$ of 
subsets of $\Omega$,
$F_{\cup_\alpha A_\alpha} = \cap_\alpha F_{A_\alpha}$.
And for any 
$A \subset \Omega$, $F_A$ is a vector space, and $N \subset F_{\Omega} \subset F_A$.
\label{prop:F}
\end{prop}
The proof that $F_A$ is a vector space is analogous to that used for $N$ 
(Theorem \ref{thm:basic}, item (a)). The other statements are immediate. Note that $N$ relates to $F_{\Omega}$ by an interchange in the order of quantifiers, amounting to uniformity in the choice of $t_1, ..., t_k$ in the definition of $N$. 
(Subsection~\ref{subsec:uniformity} 
explores this distinction.) 

\begin{lemma}
For any $A \subset \Omega$, if $\phi \in F_A$, then $\phi_{\pi} \in F_A$.
\label{lemmaA}
\end{lemma} 
\noindent{\bf Proof:}
For each $\omega \in A$ there exists $t_1,...,t_k$ such that
$p^{t_1,...,t_k}_{\omega_{t_1},...,\omega_{t_k}} \phi = 0$.
Using part (b) of Proposition~\ref{prop:technical}, we have 
$$
p^{t_1,...,t_k}_{\omega_{t_1},...,\omega_{t_k}} \, \phi_{\pi} 
\ = \ 
p^{t_1,...,t_k}_{\omega_{t_1},...,\omega_{t_k}} \, p_{\pi} \, \phi
\ = \ 
p_{\pi} \,  p^{t_1,...,t_k}_{\omega_{t_1},...,\omega_{t_k}} \, \phi
\ = \ 
0.
$$
This shows that $\phi_{\pi} \in F_A$.
$\square$



\begin{lemma}
For every $A \subset \Omega$,
\begin{equation}
\overline{F_A} 
\ = \ 
\left\{ \phi \in \H \, : \, \phi_{\pi} \in \overline{F_A \cap \Hp} \right\}
\ = \ 
\overline{F_A \cap \Hp} \, \oplus \, \Hp^{\perp},
\label{lemmaB1}
\end{equation}
\begin{equation}
F_A^{\perp} \ = \  (F_A \cap \Hp)^{\perp} \, \cap \, \Hp,
\label{lemmaB2}
\end{equation}
and 
\begin{equation}
\overline{F_A} \cap \Hp 
\ = \ 
\overline{F_A \cap \Hp}.
\label{lemmaB3}
\end{equation}
\label{lemmaB}
\end{lemma}

\noindent{\bf Proof:}
The second equality in (\ref{lemmaB1}) is 
immediate, (\ref{lemmaB2}) follows from 
(\ref{lemmaB1}), by taking the orthogonal complement,
and (\ref{lemmaB3}) follows from (\ref{lemmaB1}),
by taking the intersection with $\Hp$ on both sides.
So we only need to prove the first equality in (\ref{lemmaB1}). This will be done in two steps:
  
\noindent $\subset$) \ If $\phi \in \overline{F_A}$, then there is a sequence $(\phi_i)_{i = 1,2,...}$, such that $\phi_i \in F_A$, $\phi_i \to \phi$. 
If we set $\xi_i = p_{\pi} \phi_i$, then, by Lemma \ref{lemmaA}, $\xi_i \in F_A$, 
and hence $\xi_i \in F_A \cap \Hp$.
Since projections are continuous, $\xi_i \to p_{\pi} \phi = \phi_{\pi}$, and 
we conclude that $\phi_{\pi}
\in \overline{F_A \cap \Hp}$.

\noindent $\supset$) \ Suppose $\phi$ is such that 
$\phi_{\pi} \in \overline{F_A \cap \Hp}$. Then
$\phi_{\pi} \in \overline{F_A}$. But also $\phi-\phi_{\pi} \in \Hp^{\perp} =
\overline{N} \subset \overline{F_A}$, where we used part (a) 
of Theorem~\ref{thm:basic} and Proposition~\ref{prop:F}.
Therefore, $\phi = \phi_{\pi} + (\phi - \phi_{\pi})
\in \overline{F_A}$. 
$\square$

\begin{lemma}
If $S$ is countable and $A \in \A$, then $F_A \cap \Hp = \H_{A^c}$.
\label{lemmaC}
\end{lemma}

\noindent{\bf Proof}:
Since $A \in \A$, it can be represented as $A = \{(X_{t_1}, ..., X_{t_k}) \in G\}
= \{\omega \in \Omega : (\omega_{t_1}, ..., \omega_{t_k}) \in G \}$, 
for appropriate $t_1, ..., t_k \in S$ and 
$G \subset \I(t_1) \times ... \times \I(t_k)$.
And from (\ref{caprange}), 
\begin{eqnarray}
\H_{A^c}  & \ = \ &
\left\{
\phi \in \H_{\pi} \, : \, \sum_{(a_1, ..., a_k) \in G^c } 
\, p^{t_1,...,t_k}_{a_1,...,a_k} \phi = \phi
\right\}
\nonumber \\ 
& \ = \ & 
\left\{
\phi \in \H_{\pi} \, : \, \sum_{(a_1, ..., a_k) \in G } 
\, p^{t_1,...,t_k}_{a_1,...,a_k} \phi = 0
\right\}
\nonumber \\ 
& \ = \ & 
\left\{
\phi \in \H_{\pi} \, : \, p^{t_1,...,t_k}_{a_1,...,a_k} \phi = 0 \ \mbox {for every} \ 
(a_1, ..., a_k) \in G
\right\},
\label{HAc}
\end{eqnarray}
where in the second equality we used the fact
that $\Hp = \Hp''$, from part (a) of
Theorem~\ref{thm:basic}, and in the third 
equality we used (\ref{perp}) and the fact 
that a sum of orthogonal vectors can only
be 0 if all these vectors are 0. 

It is clear from (\ref{HAc}) that $\H_{A^c} \subset F_A \cap \Hp$. 

Suppose now that $\phi \not\in \H_{A^c}$.
We need to prove that then $\phi \not\in F_A \cap \Hp$.
If $\phi \not\in \Hp$ we are done, so we will also assume now that $\phi \in \Hp$.
Then (\ref{HAc}) implies that  
there must exist some $(a_1, ..., a_k) \in G$,
such that $p^{t_1,...,t_k}_{a_1,...,a_k} \phi \not= 0$.
Enumerate the elements of $S$ as $s_1,s_2, ...$,
starting with $s_1 = t_1, ..., s_k = t_k$, 
and continuing in an arbitrary fashion.
Using (\ref{sump})
we see that there exists $a_{k+1}$ such that 
$p^{s_1,...,s_k, s_{k+1}}_{a_1,...,a_k,a_{k+1} } \phi \not= 0$.
Proceeding inductively in this fashion, we conclude that there exists a 
sequence $a_1, a_2, ...$ such that 
$p^{s_1,...,s_l}_{a_1,...,a_l} \phi \not= 0$, for every $l \geq k$. 
This extends to all $l \geq 1$, thanks to (\ref{monotone}).
If we define $\omega$ by $\omega_{s_i} = a_{i}$, $i = 1,2, ...$, 
then $\omega \in A$, and, recalling (\ref{remark}), 
we have just proved that $\phi \not\in F_A$.
$\square$

\begin{lemma}
If $S$ is countable and $A \in \A_{\sigma}$, then $F_A \cap \Hp = \H_{A^c}
= \H_A^\perp \cap \Hp$,  
and  $F_A^{\perp} = \H_A$.
\label{lemmaD}
\end{lemma}

\noindent{\bf Proof}:
We have $A = \cup A_i$, for a countable 
disjoint collection of sets $A_i \in \A$. 
Now, 
\begin{eqnarray*}
F_A \cap \Hp 
& \ = \ &
\left( \cap_i F_{A_i} \right) \cap \Hp \ = \ 
\cap_i ( F_{A_i} \cap \Hp) \ = \
\cap_i \H_{A_i^c} 
\\  &  \ = \  & 
\cap_i (\H_{A_i}^{\perp} \cap \Hp) 
\ = \ 
(\cap_i \H_{A_i}^\perp) \cap \Hp
\ = \
(\left( \oplus_i \H_{A_i} \right)^{\perp})
\cap \Hp 
\ = \ 
\H_A^\perp \cap \Hp  
\ = \ 
\H_{A^c},
\end{eqnarray*}
where in the first equality we used Proposition~\ref{prop:F}, 
in the third equality we used Lemma~\ref{lemmaC}, in the fourth 
and eighth equalities we used (\ref{Hsplit}),
and in the sixth and seventh
equalities we used properties 
of projection valued measures
(called, respectively, (PVM6) and (PVM2) in Section \ref{sec:pvm}).

Combining this result with (\ref{lemmaB2}), 
we obtain 
$$
F_A^{\perp} \ = \ 
\left( \H_A^\perp \cap \Hp \right)^{\perp}
\, \cap \, \H_{\pi} \ = \ 
\left( \H_A  \, \oplus \, \H_{\pi}^{\perp} \right) \, \cap \, \H_{\pi} 
\ = \ 
\H_{A}.
$$
$\square$

For $\phi \in \H$, define 
$$
\Omega(\phi) \ = \ \left\{   
\omega \in \Omega \, : \, 
\mbox{ 
for all $t_1, ..., t_k$,  \
$p^{t_1,...,t_k}_{\omega_{t_1},...,\omega_{t_k}} \phi \not= 0$ 
}
\right\}.
$$
Then  
\begin{equation}
F_A \ = \ \left\{ \phi \in \H \, : \, 
A \, \subset \, \Omega^c(\phi)
\right\}.
\label{FOmega}
\end{equation}

\begin{lemma}
If $S$ is countable, then $\Omega(\phi) \in \A_{\delta}$, for every $\phi \in \H$.
\label{lemmaE}
\end{lemma}

\noindent{\bf Proof}:
Enumerate the elements of $S$ as $s_1, s_2, ...$. Then, (\ref{monotone}) implies 
(as in (\ref{remark})) that 
$$
\Omega(\phi) \ = \ 
\left\{ \omega \in \Omega \, : \, 
\mbox {for every $k \in \{1,2, ...\}, \ 
p^{s_1,...,s_k}_{\omega_{s_1},...,\omega_{s_k}} 
\phi \not= 0$}
\right\}
\ = \ 
\bigcap_{k=1}^{\infty} \, \Omega_k(\phi),
$$
where $\Omega_k(\phi) = \{\omega \in \Omega \, : \,  
p^{s_1,...,s_k}_{\omega_{s_1},...,\omega_{s_k}} 
\phi \not= 0\}
\in \A$. $\square$

\begin{lemma}
If $S$ is countable, then, for every $\phi \in \H$ and $A \subset \Omega$, 
\begin{itemize}
    \item[(a)] $\Omega(\phi) \subset A \ \ \Longleftrightarrow   \ \  \Omega(\phi) \subset B$, for some $B \in \A_{\delta}$, $B \subset A$.  
    \item[(b)] $A \subset \Omega^c(\phi) \ \ \Longleftrightarrow   \ \  
    C \subset \Omega^c(\phi)$, for some $C \in \A_{\sigma}$, $A \subset C$.
\end{itemize}
\label{lemmaF}
\end{lemma}

\noindent{\bf Proof}:

\noindent (a): The ($\Longleftarrow$) part is obvious, since $\Omega(\phi) \subset B \subset A$. 

For the ($\Longrightarrow$) part, set $B = \Omega(\phi)$, which belongs 
to $\A_{\delta}$ by Lemma~\ref{lemmaE}. By assumption $B \subset A$ and 
tautologically $\Omega(\phi) \subset B$.

\noindent (b): Apply (a) to $A^c$ in place of $A$ and set $C = B^c$. 
$\square$


\noindent{\bf Proof of Theorem~\ref{thm:F}}:

\noindent {\bf (a):} 
Since $F_A$ is a vector space, the two statements 
are equivalent. We will prove the first one. 

Combining (\ref{FOmega}) with part (b) of Lemma~\ref{lemmaF}, 
we have, for every $A \subset \Omega$,
$$
F_A 
\ = \ \bigcup \, 
\left\{
F_{C} \, : \,  C\in \A_{\sigma}, \, A \subset C
\right\}.
$$
Taking the orthogonal complement on both sides and using  
Lemma~\ref{lemmaD}, we obtain 
$$
F_{A}^{\perp} \ = \ 
\bigcap \, 
\left\{
F_{C}^{\perp} \, : \,  C\in \A_{\sigma}, \, A \subset C
\right\}
\ = \ 
\bigcap \, 
\left\{
\H_C \, : \,  C\in \A_{\sigma}, \, A \subset C
\right\}.
$$
In case $A \in \Sigma$, this amounts to $F_A^{\perp} = \H_A$, as claimed, thanks to (\ref{formula}) in Theorem~\ref{thm:extension}.

\noindent {\bf (b):}
Combine part (a) with (\ref{lemmaG}) in 
part (d) of Theorem~\ref{thm:basic}.

\noindent {\bf(c):}
From (\ref{Hsplit}) we have $\H_{A}^\perp = 
\H_{A^c} \oplus \H_\pi^\perp$. Hence 
$$
\H_{A} \ = \ 
\H_{A^c}^\perp \cap \Hp \ = \ 
\overline{F_{A^c}} \cap \Hp \ = \ 
\overline{F_{A^c} \cap \Hp},
$$
where in the second equality we used part (a) 
above, and the third equality is  
(\ref{lemmaB3}). 
$\square$

\section{Partial extensions of
Theorem~\ref{thm:F} to arbitrary $S$}
\label{sec:Farbitrary}

In this section we will partially extend 
Theorem~\ref{thm:F} to arbitrary $S$ in 
the following ways:
\begin{thm} 
If $ A \in \Sigma$, then
\begin{itemize}
\item [(a)] 
    $\H_A^\perp \ \subset \ \overline{F_A}.$
    \item [(b)] $\mbox{$\phi_{\pi} = 0$,\,  or \,  
$\P_{\phi_{\pi}}(A) = 0$}\ \ \ 
\Longrightarrow
\ \ \ \phi \in \overline{F_A}$.
 \end{itemize}
 If also $\I(t)$ is finite for all $t \in S$,
 then 
 \begin{itemize}
\item[(c)] $\H_A = F_A^{\perp}$ \quad and \quad  
$\H_A^\perp = \overline{F_A}$.
\item[(d)] $\phi \in \overline{F_A} \ \ 
\Longleftrightarrow
\ \  \mbox{$\phi_{\pi} = 0$,\,  or \,  
$\P_{\phi_{\pi}}(A) = 0$}$.
\item[(e)] $\H_A \, = \, 
\overline{F_{A^c}} \cap \Hp \, = \, 
\overline{F_{A^c} \cap \Hp}$.
\end{itemize}
\label{thm:Farbitrary}
\end{thm}

The proof of Theorem~\ref{thm:Farbitrary}
will 
build on the work done in Sections 
\ref{sec:basic results} and \ref{sec:F}.
The proof of the statements (c), (d) and (e) 
will use Zorn's Lemma and therefore depend 
on the acceptance of the Axiom of Choice.
Incidentally,
if one does not accept this axiom, the very nature
of the set $\Omega$ becomes unclear, when $S$ is not 
countable (see, e.g., \cite{Folland}, 
Section 2 in the Prologue).

It seems natural to conjecture that the 
limitation of parts (c), (d) and (e) of 
this theorem to $\pi$ with finite $\I(t)$,
for all $t \in S$, is purely technical, 
so that Theorem~\ref{thm:F} should fully
extend to arbitrary $\pi$. 

We start with a few observations and new definitions. 
Given $D \subset S$, let $\Sigma^D$ be the
sigma-algebra generated by $\{X_t : t \in D\}$. 
Theorem 36.3.(ii) of \cite{Billingsley}
states that 
\begin{equation}
\Sigma \ = \ \bigcup \left\{
\Sigma^D \ : \ D \subset S, \ 
\mbox{$D$ is countable}  
\right\}.
\label{allcountable}
\end{equation}
This important 
fact can easily be proved, by noting that 
the right-hand side is a sigma-algebra and that 
any sigma-algebra that contains all the sets 
$\{X_t = a\}$, $t \in S$, $a \in \I(t)$, must 
contain this right-hand side. 

Given $D \subset S$, we define also
$$
F_A^D \ = \  
\ \left
\{\phi \in \H \, : \, 
\mbox{for any $\omega \in A$ 
there is $t_1, ..., t_k \in D$ such that 
$p^{t_1,...,t_k}_{\omega_{t_1},...,\omega_{t_k}} 
\phi = 0$ 
}  \right\}, 
$$
so that $F^S_A = F_A$.
Proposition~\ref{prop:F} extends to $F^S_A$, with 
the exception that $N$ and $F^D_A$ are not comparable when $D$ is a proper subset of $S$:
\begin{prop} 
$F^D_A$ decreases as $A$ increases and increases
as $D$ increases.  
For any $D \subset S$ and any 
family $\{A_\alpha\}$ of 
subsets of $\Omega$,
$F^D_{\cup_\alpha A_\alpha} = \cap_\alpha F^D_{A_\alpha}$.
And for any 
$A \subset \Omega$ and $D \subset S$,
$F^D_A$ is a vector space.
\label{prop:Farbitrary}
\end{prop}

Again with $D \subset S$, we also set 
\begin{equation}
\pi(D) \ = \ \{p^t_a : t \in D, a \in \I(t) \}. 
\label{pi(D)}
\end{equation}
And, in a self-explanatory fashion, we denote by 
$\Omega(D)$, $\Sigma(D)$ and 
$\P^{(D)}_\phi$, $\phi \in \H$, the corresponding 
objects associated with $\pi(D)$. 
Clearly
\begin{equation}
\Hp \ \subset \ \H_{\pi(D)}.
\label{HpsubsetHpi(D)}    
\end{equation}

The next proposition collects some facts that are 
relatively easy consequences of the results in 
Sections \ref{sec:basic results} and \ref{sec:F}.
Note that, thanks to (\ref{allcountable}), for every 
$A \in \Sigma$ there is some $D$ in the conditions 
of this proposition. 
Note that some of the statements in this proposition 
do not involve $D$; those are 
identical to statements
(a) and (b) of Theorem~\ref{thm:Farbitrary}. 
Part (a) of this proposition 
will be used in the proof of
the other statements in 
Theorem~\ref{thm:Farbitrary}. 
\begin{prop} 
If $D \subset S$ is countable and $A \in \Sigma^D$, then 
\begin{itemize}
    \item [(a)] $\H_A \  = \ 
    \overline{F_{A^c}^D} \cap \Hp \ = \
    \overline{F_{A^c}^D \cap \Hp}$. 
    \item [(b)] $\overline{F_A^D} \ \subset  \
    \H_A^\perp \ \subset \ \overline{F_A}.$
    \item [(c)] $\mbox{$\phi_{\pi} = 0$,\,  or \,  
$\P_{\phi_{\pi}}(A) = 0$}\ \ \ 
\Longrightarrow
\ \ \ \phi \in \overline{F_A}$.
 \end{itemize}
If also $\H_{\pi(D)} = \Hp$, then  
 \begin{itemize} 
    \item[(d)] $\H_A^\perp \ = \ \overline{F_{A}^D}$.
    \item[(e)] $\phi \in \overline{F_{A}^D} \ \ \
\Longleftrightarrow
\ \  \ \mbox{$\phi_{\pi} = 0$,\,  or \,  
$\P_{\phi_{\pi}}(A) = 0$}$.
\end{itemize}
\label{prop:Farbitrary+}
\end{prop}

It is easy to produce examples 
in which the left-hand
side containment in part (b) is not 
tight. For
instance, making $\Hp = \{0\} \not= \H$ 
and taking $D$ with 
a single element. 
If $A = \Omega$, then 
$\overline{F_{A}^D} \ = \{0\} \ \not=  \
\H = \Hp^\perp = \H_A^\perp$.


For interesting examples in which the extra 
assumption needed in parts (d) and (e) holds, 
see Subsection~\ref{subsec:basic examples}, 
in which $S = \R$, and suppose that $D$ contains
all the rationals (natural assumptions in 
applications to quantum mechanics). 

To prove Proposition~\ref{prop:Farbitrary+}
we need one more concept. 
Given $D \subset S$, we define 
an equivalence relation in $\Omega$, 
by declaring as $D$-equivalent elements of $\Omega$ 
that have identical restrictions to $D$.
We say that a set $A \subset \Omega$ is $D$-determined
if any two $D$-equivalent
elements of $\Omega$ either both belong to $A$, 
or neither one does.
When $A \subset \Omega$ is $D$-determined 
we define 

\begin{eqnarray*}
A(D) &  \ = \  & 
\{\omega \in \Omega(D) \, : \, 
\mbox{$\omega_t = \omega'_t$ \, for all $t \in D$  
and some $\omega' \in A$} \},
\end{eqnarray*}
so that 
$$
A \ = \ 
\left\{ 
\omega \in \Omega \, : \, 
(\omega_t)_{t \in D} \in A(D)
\right\}.
$$
Note that if $A$ is $D$-determined, then also 
$A^c$ has this property and 
\begin{equation}
A^c(D) \ = \ (A(D))^c.
\label{Ac(D)}
\end{equation}
And if $\{A_\alpha\}$ is a family of disjoint 
$D$-determined subsets of $\Omega$, then 
also the sets $A_\alpha(D)$ are disjoint,  
$\cup_\alpha A_\alpha$ is $D$-determined 
and 

\begin{equation}
(\cup_\alpha A_\alpha)(D) = 
\cup_\alpha (A_\alpha(D)).
\label{cupA(D)}
\end{equation}

Also, if $A$ is $D$-determined, 
\begin{eqnarray}
F_{A(D)} & \ = \ & 
\ \left
\{\phi \in \H \, : \, 
\mbox{for any $\omega \in A(D)$ 
there is $t_1, ..., t_k \in D$ such that 
$p^{t_1,...,t_k}_{\omega_{t_1},...,\omega_{t_k}} 
\phi = 0$ 
}  \right\}
\nonumber  \\ & \ = \ &
\ \left
\{\phi \in \H \, : \, 
\mbox{for any $\omega \in A$ 
there is $t_1, ..., t_k \in D$ such that 
$p^{t_1,...,t_k}_{\omega_{t_1},...,\omega_{t_k}} 
\phi = 0$ 
}  \right\}
\ = \ F^D_A.
\label{FAD}
\end{eqnarray}
(Since $A(D) \subset \Omega(D)$, the notation 
$F_{A(D)}$ should be understood as identical 
to $F^D_{A(D)}$.)


\begin{lemma}
For any $D \subset S$, 
if $A \in \Sigma^D$, then 
$A$ is $D$-determined,  
$A(D) \in \Sigma(D)$,
and $\P_\phi^{(D)}(A(D)) = \P_\phi(A)$, for any
$\phi \in \H_\pi \backslash \{0\}$. 
\label{lemma:anotherlp}
\end{lemma}
\noindent {\bf Proof:}
The proof is a simple application of the 
$\pi$-$\lambda$ Theorem (see, e.g., Theorem 3.2 
of \cite{Billingsley}).

Consider the following two classes of subsets of
$\Omega$.
\begin{eqnarray*}
\mathcal{P} 
\ & = & \ 
\left\{ 
\{
X_{t_1} = a_1, ... , X_{t_k} = a_k
\}
\ : 
\ t_1, ..., t_k \in D, \, 
a_1 \in \I(t_1), ... , a_k \in \I(t_k)
\right\}
\ \cup \{ \emptyset \}. 
\\ 
\mathcal{L} \ & = & \ 
\{
A \subset \Omega \, : \, 
\mbox {$A$ is $D$-determined, 
$A(D) \in \Sigma(D)$ and
$\P_\phi^{(D)}(A(D)) = \P_\phi(A)$, for any
$\phi \in \H_\pi \backslash \{0\}$
}
\}. 
\end{eqnarray*}
It is clear that $\mathcal{P}$ is closed with 
respect to finite intersections, meaning that 
it is a $\pi$-system. And, 
using (\ref{Ac(D)}) and (\ref{cupA(D)}), 
it is also clear that 
$\mathcal{L}$ has the three properties that 
are required to be a $\lambda$-system: it contains
$\Omega$, is close with respect to taking the 
complement and with respect to taking 
countable disjoint unions. 

It is also not difficult to see
that $\mathcal{P} \subset 
\mathcal{L}$. Each $A \in \mathcal{P}$ 
is clearly $D$-determined, 
$\emptyset(D) = \emptyset \in \Sigma(D)$, 
for \\ $A = \{X_{t_1} = a_1, ..., X_{t_k} = a_k\}$, 
$$
A(D) \ = \ 
\{ 
\omega \in \Omega(D) \, : \, 
\omega_{t_1} = a_1, ... , \omega_{t_k} = a_k
\} 
\ \in \ \Sigma(D),
$$
and, if $\phi \in \H_\pi \backslash \{0\}$,
then also $\phi \in \H_{\pi(D)} \backslash \{0\}$, 
by (\ref{HpsubsetHpi(D)}), and 
$$
\P_\phi^{(D)}(A(D)) \ = \ 
|| p^{t_k}_{a_k} ... p^{t_1}_{a_1} \hat{\phi} ||^2 
\ = \ 
\P_\phi(A), 
$$
where we used (\ref{Born}) in part (b) of Theorem~\ref{thm:basic} (for $\pi$ and for 
$\pi(D)$).

And since the sigma-algebra generated by 
$\mathcal{P}$ is $\Sigma^D$, the $\pi$-$\lambda$
Theorem implies that  $\Sigma^D \subset \mathcal{L}$.
$\square$

\begin{lemma}
For any $D \subset S$, 
if $A \in \Sigma^D$, then 
$$
\H_A \ = \ \H_{A(D)} \, \cap \, \Hp.
$$
\label{insteadofthm6}
\end{lemma}
\noindent{\bf Proof:} 
Using (\ref{thmd+}) in part (d) of Theorem~\ref{thm:basic} (for $\pi$ and for 
$\pi(D)$), (\ref{HpsubsetHpi(D)}), 
and Lemma~\ref{lemma:anotherlp}, we have
\begin{eqnarray*}
\H_{A(D)} \, \cap \, \Hp & \ = \ &
\left\{
\phi \in \H_{\pi(D)} \, : \, 
\mbox{ $\phi = 0$ \, or \, 
$\P_\phi^{(D)}(A(D)) = 1$} 
\right\}
\, \cap \, \Hp
\\  & \ = \ &
\left\{
\phi \in \Hp \, : \, 
\mbox{ $\phi = 0$ \, or \, 
$\P_\phi^{(D)}(A(D)) = 1$} 
\right\}
\\ & \ = \ & 
\left\{
\phi \in \Hp \, : \, 
\mbox{ $\phi = 0$ \, or \, $\P_\phi(A) = 1$} 
\right\}
\ = \ \H_A.
\end{eqnarray*}
$\square$

The next lemma is a counterpart for $F_A^D$ of
what Lemmas \ref{lemmaA} and \ref{lemmaB} are 
for $F_A$. It is weaker than the latter one, because 
$N$ and $F^D_A$ are not comparable sets, 
when $D$ is a proper subset of $S$. 
\begin{lemma}
For any $D \subset S$ and $A \subset \Omega$, 
\begin{itemize}
    \item [(a)] If \, $\phi \in F_{A}^D$, \,
    then \, $\phi_\pi \in F_A^D$.
    \item [(b)] $\overline{F_A^D}\cap \Hp \ = \
    \overline{F_A^D \cap \Hp}$.
\end{itemize}
\label{lemma:FarbitraryA}
\end{lemma}
\noindent {\bf Proof:}
The proof of part (a) is analogous to the proof of 
Lemma~\ref{lemmaA}. And part (b) is a simple 
consequence of part (a), 
since if $\phi \in \Hp$ can be approximated 
by $\phi_i \in F_A^D$, then it can also 
be approximated by $p_\pi \phi_i 
\in F_A^D \cap \Hp$. $\square$

\noindent {\bf Proof of Proposition~\ref{prop:Farbitrary+}:}

\noindent {\bf (a):}
$$
\H_A \ = \ \H_{A(D)} \cap \Hp \ = \ 
\left( \overline{F_{A^c(D)}} \cap \H_{\pi(D)} \right)
\cap \Hp 
\ = \ 
\overline{F_{A^c(D)}} \cap \Hp \ = \
\overline{F_{A^c}^D} \cap \Hp \ = \
\overline{F_{A^c}^D \cap \Hp},
$$
where the first equality is from  Lemma~\ref{insteadofthm6},
the second one is from part (c) of 
Theorem~\ref{thm:F} (applied to $\pi(D)$),
combined with (\ref{Ac(D)}),
the third one is from 
(\ref{HpsubsetHpi(D)}),
the fourth one is from 
(\ref{FAD}),  
which can be used thanks to
Lemma~\ref{lemma:anotherlp},
the fifth one is from  
and Lemma~\ref{lemma:FarbitraryA}.

\noindent{\bf (b) - (e):}
Using (\ref{Hsplit}) and part (a) above,
$$
\H_A^\perp \ = \ \H_{A^c} \, \oplus \, \overline{N}
\ = \  \left( \overline{F_A^D} \cap \Hp \right)
\, \oplus \, \overline{N}.
$$
But $F_A^D \subset F_A$ and also 
$N \subset F_{\Omega}
\subset F_A$. Therefore we obtain $\H_A^\perp 
\subset \overline{F_A}$. 

Combining this with 
(\ref{lemmaG}) in 
part (d) of Theorem~\ref{thm:basic},
we obtain 
the statement in part (c) of the proposition.

By Lemma~\ref{insteadofthm6}, 
parts (c) and (a) of Theorem~\ref{thm:F}
(applied to $\pi(D)$), 
and (\ref{FAD})
again, 
$$
\H_A \ = \ \H_{A(D)} \cap \Hp \ \subset \ \H_{A(D)} 
\ = \ F_{A(D)}^\perp \ = \ (F_A^D)^\perp, 
$$
with equality in case $\Hp = \H_{\pi(D)}$. Taking 
the orthogonal complement, we complete the proof of 
(b) and (d). Part (e) follows from part (d) and 
(\ref{lemmaG}). 
$\square$

The main technical work in the proof of  
parts (c), (d) and (e) of
Theorem~\ref{thm:Farbitrary} is contained in 
the proof of the following theorem, that is 
interesting also in its own right. 

\begin{thm}
Suppose that $\I(t)$ is finite, for every 
$t \in S$. 
For any $D \subset S$, if $A \subset \Omega$ 
is $D$-determined, then
\begin{equation}
F_A \cap \Hp \ = \ F^D_A \cap \Hp.
\label{Zorngoal}
\end{equation}
\label{thm:useZorn}
\end{thm}

\noindent{\bf Proof:}
Clearly we only have to prove that the left-hand 
side is contained in the right-hand side. So we  
suppose that $\phi \not\in F^D_A \cap \Hp$ and will 
prove that $\phi \not\in F_A \cap \Hp$. If 
$\phi \not\in \Hp$ we are done, so we also 
assume $\phi \in \Hp$, which implies that we are 
assuming that $\phi \not\in F^D_A$. 

This assumption states that there is 
$\omega = (\omega_t)_{t \in D } \in A(D)$ such that 
$p^{t_1,...t_k}_{\omega_{t_1}, ... , 
\omega_{t_k}}\phi \not= 0$,
for any $t_1, ..., t_k \in D$.
Since $A$ is $D$-determined, any extension 
of this $\omega$ to $(\omega_t)_{t \in S}$ 
will be an element of $A$.

We will be considering extensions of $\omega$ to
$(\omega_t)_{t \in T}$, $D \subset T \subset S$.  
Such an extension will be said to be ``good'' if 
$p^{t_1,...t_k}_{\omega_{t_1}, ... , 
\omega_{t_k}}\phi \not= 0$,
for any $t_1, ..., t_k \in T$.
Our goal is to show that there is a good extension 
of $\omega$ on $T=S$. 

We proceed now in typical 
Zorn-Lemma-application  fashion. 
Partially order the good extensions of $\omega$, by 
declaring $(\omega^1_t)_{t \in T^1} \leq 
(\omega^2_t)_{t \in T^2}$, when $T^1 \subset T^2$ 
and $\omega^1_t = \omega^2_t$, for all $t \in T^1$.
Given a linearly ordered family of good extensions, 
$\{ \omega^\lambda : \lambda \in \Lambda \}$, where 
$\Lambda$ is some index set, we can present an 
upper bound for it as follows. 
Let $T^\lambda$ be the
domain of $\omega^\lambda$. Set 
$T^{\Lambda} = \cup_{\lambda \in \Lambda} 
T^\lambda$, and define 
$\omega^{\Lambda}$ as the extension of $\omega$ on
$T^\Lambda$ given by $\omega^\Lambda_t = \omega^\lambda_t$, where $\lambda$ is such that
$t \in T^\lambda$. The fact that 
$\{ \omega^\lambda : \lambda \in \Lambda \}$ is 
a linearly ordered family 
assures the consistency of this definition. 
It is clear that $\omega^\Lambda$ is a good 
extension of $\omega$ and that it is an 
upper bound for the family 
$\{ \omega^\lambda : \lambda \in \Lambda \}$.

Zorn's Lemma therefore implies the existence of 
a maximal good extension of $\omega$, that we 
denote by $\omega^M$, and whose domain we 
denote by $T^M$. 

If $T^M = S$, then the existence of the good 
$\omega^M = (\omega^M_t)_{t \in S} \in A$ means
that $\phi \not\in F_A$, and we are done. 

So suppose instead that $T^M \not= S$. Then 
there exist $s \in S \backslash T^M$. And 
for any such $s$, 
any extension of $\omega^M$ to $T^M \cup \{s\}$
must not be good, by the maximality of $\omega^M$
in the class of good extensions of $\omega$. 
For $a \in \I(s)$, let $\omega^a$ be the extension of
$\omega^M$ to $T^M \cup \{s\}$ defined by 
$\omega^a_s = a$. 
As $\omega^M$ is good and $\omega^a$ is 
not good, there exists
$k(a)$ and 
$t^a_1,..., t^a_{k(a)} \in T^M$ such that 
\begin{equation}
p^{t^a_1, ..., t^a_{k(a)},s}_{\omega^M_{t^a_1}, ..., 
\omega^M_{t^a_{k(a)}},a} \phi \ = \ 0.
\label{indices}
\end{equation}
Let $\{t_1, ..., t_k\} = \cup_{a \in \I(s)} 
\{t^a_1, ..., t^a_{k(a)}\}$.
Since $\I(s)$ is finite, this set is also 
finite, and using (\ref{monotone}) and 
(\ref{indices}), we obtain 
$$
p^{t_1, ..., t_k,s}_{\omega^M_{t_1}, ...,
\omega^M_{t_k}, a} \phi \ = \ 0, 
$$
for each $a \in \I(s)$. Summing over $a \in \I(s)$, 
using (\ref{sump}), applicable since 
$\phi \in \Hp$, we obtain 
$$
p^{t_1, ..., t_k}_{\omega^M_{t_1}, ...,
\omega^M_{t_k}} \phi \ = \ 0.  
$$
Since $t_1, ..., t_k \in T^M$, this is 
in contradiction with the fact that $\omega^M$ 
is good. This contradiction shows that the 
maximality of $\omega^M$ implies $T^M = S$, 
and concludes the proof of (\ref{Zorngoal}).
$\square$


\noindent {\bf Proof of Theorem~\ref{thm:Farbitrary}:}
Parts (a) and (b) are contained in 
Proposition~\ref{prop:Farbitrary+}. 

Since $A \in \Sigma$, (\ref{allcountable}) implies 
that there exists $D \subset S$ countable such that 
$A \in \Sigma^D$. 
We can therefore combine part (a) of
Proposition~\ref{prop:Farbitrary+}, with
Theorem~\ref{thm:useZorn} and part (b) of 
Lemma~\ref{lemma:FarbitraryA}, to prove 
part (e): 
$$
\H_A \ = \ 
\overline{F^D_{A^c} \cap \Hp}
\ = \ 
\overline{F_{A^c} \cap \Hp}
\ = \ 
\overline{F_{A^c}} \cap \Hp.
$$
We can now use (\ref{lemmaB1}) to prove 
one of the equivalent statements in part (c):
$$
\overline{F_A} 
\ = \ 
\overline{F_A \cap \Hp} \, \oplus \, \Hp^{\perp}
\ = \ 
\H_{A^c} \, \oplus \, \Hp^{\perp}
\ = \ 
\H_A^\perp,
$$
where the last step is from (\ref{Hsplit}).

Finally, part (d) follows from part (a) and 
(\ref{lemmaG}) in 
part (d) of Theorem~\ref{thm:basic}.
$\square$

\section{Extension of projection valued measures}
\label{sec:pvm}

In this section we will prove an analogue of Carath\'eodory's extension theorem for projection valued measures.
(For the classical Carath\'eodory's extension theorem for measures,
see, e.g., Section 4 of Chapter 1 of \cite{Folland}, or 
Section 2 of Chapter 12 of \cite{Royden}, or Section 3 of Chapter 1 of \cite{Billingsley}.)
Our setting includes a Hilbert space $\H$, an arbitrary set $\Omega$, and a family $\A$ of subsets of $\Omega$ that form an algebra. 
Those do not have to be the ones that appeared in other sections of this paper. This section of the paper is independent of the other
sections, except for terminology and notation introduced in the first paragraph of Section~\ref{sec:basic results}.

Let $\{p_A : A \in \A\}$ be a set of projections in $\H$, and, for each $A \in \A$, denote by $\H_A$ the range of $p_A$.  
We say that $\{p_A : A \in \A\}$ is a projection valued measure (p.v.m.; called a ``spectral measure'' in \cite{Halmos},
and a ``resolution of the identity'' in 
\cite{Royden}), 
if it satisfies the following two axioms:
\begin{itemize}
    \item [(PVM1)] $p_{\Omega} = I$, the identity operator.
    \item [(PVM2)] If $A_i \in \A$, $i=1,2, ...$ are disjoint sets in $\A$ and also $A = \cup_{i=1}^{\infty} A_i \in \A$, then $p_A = \sum_{i=1}^{\infty} p_{A_i}$.
\end{itemize}
(If $\A$ is a sigma-algebra, the condition $A \in \A$ in (PVM2) is redundant. Often one reserves the name p.v.m. only for 
this case. But for the purpose in this paper it is more natural to also define a p.v.m. indexed by an algebra $\A$, as done
above.)

From the axioms (PVM1) and (PVM2) a number of other properties can be deduced, including
the following, where all sets are assumed to be in $\A$:
\begin{itemize}
    \item [(PVM3)] $p_{\emptyset} = 0$, the operator that maps every vector to the 0 vector.
    \item[(PVM4)] If $A_1, ..., A_n$ are disjoint sets,
    then $p_A = \sum_{i=1}^{n} p_{A_i}$.
    \item[(PVM5)] $p_A + p_{A^c} = I$, or equivalently, $\H_{A^c} = \H_A^{\perp}$.
    \item[(PVM6)] If $A \cap B = \emptyset$, then $\H_A \perp \H_B$, or equivalently, 
    $p_A p_B = p_B p_A = 0$.
    \item[(PVM7)] If $A \subset B$, then $\H_A \subset \H_B$.
    \item[(PVM8)] $p_{A \cap B} = p_A p_B = p_B p_A$, so that in particular all the $p_A$, $A \in \A$, commute with each other.
\end{itemize}
(PVM3) follows from (PVM2) by taking $A_i = \emptyset$, for all $i$. 
(PVM4) follows from (PVM2) and (PVM3) by taking $A_i = \emptyset$ for $i > n$.
(PVM5) follows from (PVM1) and (PVM4).
(PVM6) follows from (PVM4) and Thm.H.28.2, according
to which a sum of projections 
is a projection if and only if 
the added projections are orthogonal to each other. 
(PVM7) follows from 
(PVM4) and (PVM6), 
as they imply 
$\H_B = \H_{A} \oplus \H_{B \cap A^c}$.
(PVM8) follows from using (PVM4) for writing
$p_{A} = p_{A \cap B} + p_{A \cap B^c}$,
then multiplying both sides by $p_B$, 
once on the left, once on the right, 
and then using (PVM6) for $B$ and $A \cap B^c$,
and (PVM7) for $B$ and $A \cap B$.

Later, when dealing with more than one algebra, we will use the notation $\A$-(PVM$x$) to indicate the statement (PVM$x$) for sets assumed to be in $\A$. 

For each $\phi \in \H$ and $A \in \A$ define 
\begin{equation} 
M_{\phi}(A) \ = \ ||p_A \phi ||^2.
\label{defofMA}
\end{equation}
Then we have, from (PVM1), (PVM2), (PVM3) and (PVM6), that  
\begin{itemize}
    \item[(M1)] $M_{\phi} (\emptyset) = 0$ and $M_{\phi} (\Omega) = ||\phi||^2$.
    \item[(M2)] If $A_i \in \A$, $i=1,2, ...$ are disjoint sets and also $A = \cup_{i=1}^{\infty} A_i \in \A$, then 
    $M_{\phi} (A) = \sum_{i=1}^{\infty} M_{\phi} (A_i)$.
\end{itemize}
Therefore $\{M_{\phi} : A \in \A \}$ is a finite measure on $\A$.
(Sometimes the name ``premeasure'' is used and the name ``measure'' reserved for the case in which $\A$ is a sigma-algebra.)

Given $\phi \in \H$, 
for each $A \subset \Omega$, we define its 
outer measure relative to $M_\phi$ by
\begin{equation}
M_\phi^*(A) \ = \ 
\inf \, 
\left\{
\sum_{i=1}^{\infty}
\, M_\phi(A_i) \, : \, 
\ A \subset \cup_{i=1}^\infty A_i,  \ 
A_i \in \A, \, i=1,2, ...
\right\}.
\label{outermeasure}    
\end{equation}

The set of $M_\phi^*$-measurable sets is
defined as 
$$
\M_\phi \ =\ 
\left\{
A \subset \Omega \ : \ 
M^*_\phi(B) = M_\phi^*(B \cap A) +
M_\phi^*(B \cap A^c), 
\ \mbox{for all $B \subset \Omega$}
\right\},
$$
and turns out to be a sigma-algebra that 
contains $\A$. Therefore, if we denote by
$\Sigma$ the sigma-algebra generated by 
$\A$, we have $\A \subset \Sigma \subset \M_\phi$, for each $\phi \in \H$.

For each $A \in \A$,
we have 
\begin{equation}
M_\phi(A) = M_\phi^*(A),
\label{MAM*A}
\end{equation}
so that this equality can be extended
consistently as a definition of 
$M_\phi(A)$, for $A \in \M_\phi$.

Carath\'eodory's extension
theorem states that, for each
$\phi \in \H$, 
$\{M_\phi(A) : A \in \M_\phi \}$ is a 
measure, which 
extends the measure   
$\{M_\phi(A) : A \in \A \}$. 
Furthermore, uniqueness holds on $\Sigma$, in that 
$\{M_\phi(A) : A \in \Sigma \}$
is the only extension of 
$\{M_\phi(A) : A \in \A \}$ 
to a measure on $\Sigma$.

Denote by $\A_\sigma$ the family of 
subsets of $\Omega$
that can be expressed as countable 
unions of sets in $\A$. 
It is clear from Carath\'eodory's 
extension theorem 
and the definitions above 
that, for $A \in \M_\phi$,
\begin{equation}
M_\phi(A) 
\ = \ 
\inf \, \{
M_\phi (B) \, : \, 
B \in \A_\sigma, \, A \subset B
\}.
\label{fromRoyden-}    
\end{equation}

Our goal in this section is to prove a 
counterpart to Carath\'eodory's extension
theorem and the identity 
(\ref{fromRoyden-}) for p.v.m. 
Define $\M = \cap_{\phi \in \H} \M_\phi$.
Then $\M$ is also a sigma-algebra and 
$\A \subset \Sigma \subset \M$.
Our main result in this section is:
\begin{thm}
Suppose that $\{p_A : A \in \A \}$ is a p.v.m.. Then there exists a 
p.v.m. $\{p_A : A \in \M \}$ that extends it to $\M$.
For any $\phi \in \H$ and $A \in \M$ we have 
\begin{equation}
||p_A \phi||^2 \ = \ M_\phi(A).
\label{Mp-}    
\end{equation}
For $A \in \M$ the range of $p_A$ is 
\begin{equation}
\H_A \ = \ \{\phi \in \H \, : \, M_{\phi} (A) = ||\phi ||^2 \},
\label{HAMA}
\end{equation}
and the following relation holds: 
\begin{equation}
\H_A \ = \ \bigcap \ \, \{ \H_B \, : \, B \in \A_\sigma \, , \, A \subset B \}.
\label{formula}
\end{equation}
Furthermore, 
$\{p_A : A \in \Sigma \}$ 
is the unique p.v.m. that extends 
$\{p_A : A \in \A \}$ to $\Sigma$.
\label{thm:extension}
\end{thm}

Before we can prove this theorem, we need to 
prove some properties of projections, the first 
of which is well known, but for which we could not
find a reference.

Given a sequence of subspaces $(\S_i)_{1=1,2,...}$, we will indicate with $\S_i \nearrow \S$ the statement that $\S_{i} \subset \S_{i+1}$, $i=1,2,...$, and $\S = \overline{\cup_{i=1}^{\infty} \S_i}$.
And we will indicate with $\S_i \searrow \S$ the statement that $\S_{i+1} \subset \S_{i}$, $i=1,2,...$, and $\S = {\cap_{i=1}^{\infty} \S_i}$.

\begin{prop}
Suppose that $\S_i$, $i=1,2, ...$ and $\S$ are subspaces and denote by $p_i$ the projection on $\S_i$ and by $p$ the projection on $\S$. 
Suppose that $\phi \in \H$.
\begin{itemize}
    \item[(a)] If $\S_i \nearrow \S$, 
    then 
    \begin{equation}
        \lim_{i \to \infty} \, p_i \phi \ = \ p \phi.
    \label{propmonotone1}
    \end{equation}
    \item[(b)] If $\S_i \searrow \S$, 
    then (\ref{propmonotone1}) holds as well.
\end{itemize}
\label{prop:monotonep}
\end{prop}

\noindent {\bf Proof:}
{\bf (a):}
Since $\S_i \subset \S$ 
\begin{equation}
|| p \phi - p_i \phi ||
\ = \ 
|| p \phi - p_i (p \phi) ||
\ = \ 
\mbox{dist} (p \phi, \S_i),
\label{monotonep1}    
\end{equation}
where the right-hand side is the distance between the point $p \phi$
and the subspace $\S_i$.

Since 
$p \phi \in \S = \overline{\cup_{i=1}^{\infty} \S_i}$,
for any $\epsilon > 0$ there is $\zeta \in \cup_{i=1}^{\infty} \S_i$
such that $||p \phi - \zeta|| \leq \epsilon$.  
But this implies that, there is $j$ such that $\zeta \in \S_j$ and 
hence $\mbox{dist} (p \phi, \S_j) \leq \epsilon$.
Since $\S_i$ increases with $i$, we conclude that for $i \geq j$,
\begin{equation}
\mbox{dist} (p \phi, \S_i) 
\ \leq \ \epsilon. 
\label{monotonep2}    
\end{equation}
Combining (\ref{monotonep1}) with (\ref{monotonep2}) proves 
(\ref{propmonotone1}).

\noindent {\bf (b):}
Apply part (a) to $\S_i^{\perp}$ and $\S^{\perp}$.
$\square$

\begin{thm}
Suppose that $ \{ \S_{\alpha} \}_{\alpha \in \Lambda}$ is a family of subspaces of $\H$, where $\Lambda$ is an arbitrary index set. 
Assume that it satisfies the following condition: For any $\alpha, \beta \in \Lambda$, there exists $\gamma \in \Lambda$ such that 
$\S_{\gamma} \subset \S_{\alpha} \cap \S_{\beta}$. 
For each $\alpha \in \Lambda$, denote by $p_{\alpha}$ the projection on $\S_{\alpha}$, and let $p_{\Lambda}$ be the projection on $\cap_{\alpha \in \Lambda} \S_{\alpha}$.
Then, for any $\phi \in \H$, 
\begin{equation} 
|| p_{\Lambda} \phi || \ = \ 
\inf_{\alpha \in \Lambda} \, || p_{\alpha} \phi ||.
\label{infnormthesis}
\end{equation}
\label{thm:infnorm}
\end{thm}

\noindent {\bf Remark:} The need for some condition on the family 
$\{\S_{\alpha}\}_{\alpha \in \Lambda}$ in this theorem is made 
clear by a simple 
counter-example in which the family contains only two
orthogonal subspaces $\S$ and $\S^{\perp}$, and $\phi$ is not 
contained in either one of these. 
In this case the left-hand side of 
(\ref{infnormthesis}) is 0, while the 
right-hand side is positive.

\noindent {\bf Proof:}
There exists a sequence of indices $(\alpha_i)_{i=1,2, ...}$ such 
that $||p_{\alpha_i} \phi|| \,  \to \, \inf_{\alpha \in \Lambda} \, 
|| p_{\alpha} \phi ||$, as $i \to \infty$.
Set $\beta_1 = \alpha_1$, and for $i=2,3, ...$, recursively choose $\beta_i$
such that $\S_{\beta_i} \subset \S_{\alpha_i} \cap \S_{\beta_{i-1}}$.
Since $\S_{\beta_i} \subset \S_{\alpha_i}$, we have 
$\inf_{\alpha \in \Lambda} \, || p_{\alpha} \phi ||
\, \leq \, ||p_{\beta_i} \phi|| \, \leq \, 
||p_{\alpha_i} \phi||$, 
and hence  
\begin{equation}
\lim_{i \to \infty}
||p_{\beta_i} \phi|| \  = \  \inf_{\alpha \in \Lambda} \, 
|| p_{\alpha} \phi ||.
\label{infnorm1}    
\end{equation}
Let $q$ be the projection on $\cap_{i=1}^{\infty} \S_{\beta_i}$, and $\eta = q \phi$.
Since $\S_{\beta_i} \subset \S_{\beta_{i-1}}$, $i=2,3, ...$, 
we have from part (b) of Proposition~\ref{prop:monotonep} that
$\eta \, = \, \lim_{i \to \infty} \, p_{\beta_i} \phi$, 
and therefore, using (\ref{infnorm1}), 
\begin{equation}
|| \eta ||
\ = \ 
\inf_{\alpha \in \Lambda} \, 
|| p_{\alpha} \phi ||.
\label{infnorm2}    
\end{equation}
Since 
$ 
\cap_{\alpha \in \Lambda} \S_{\alpha}
\subset  \cap_{i=1}^{\infty} \S_{\beta_i},
$
we have  
\begin{equation}
|| p_{\Lambda} \phi ||
\ = \ 
|| p_{\Lambda} q \phi ||
\ = \ 
|| p_{\Lambda} \eta ||.
\label{infnorm3}
\end{equation}
If we had $\eta \in \cap_{\alpha \in \Lambda} \S_{\alpha}$, 
we would have $p_{\Lambda} \eta = \eta$, and then 
from (\ref{infnorm3}) and (\ref{infnorm2}), 
$$
|| p_{\Lambda} \phi ||
\ = \ 
|| \eta || 
\ = \ 
\inf_{\alpha \in \Lambda} \, 
|| p_{\alpha} \phi ||.
$$
Therefore, for 
(\ref{infnormthesis}) 
to be false, there must 
exist $\gamma \in \Lambda$ such that $\eta \not\in \S_{\gamma}$. 
Assuming this to be the case, choose $\delta_1$ such that 
$\S_{\delta_1} \subset \S_{\beta_1} \cap \S_{\gamma}$, and 
for $i =2, 3, ...$, recursively choose $\delta_i$ such that  
$\S_{\delta_i} \subset \S_{\beta_i} \cap \S_{\delta_{i-1}}$. 
Let $r$ be the projection on $\cap_{i=1}^{\infty} \, \S_{\delta_i}$.
We would then have
\begin{equation}
\inf_{\alpha \in \Lambda} \, || p_{\alpha} \phi ||
\ \leq \
\lim_{i \to \infty} || p_{\delta_i} \, \phi ||
\ = \ 
|| r \phi ||
\ \leq \ 
|| r q \phi ||
\  =  \ 
|| r \eta ||
\ \leq \ 
|| p_{\gamma} \eta ||
\  <  \ 
|| \eta ||,
\label{infnorm4}    
\end{equation}
where 
in the second step we used part (b) of Proposition~\ref{prop:monotonep} and 
the fact that $\S_{\delta_i} \subset \S_{\delta_{i-1}}$, 
$i=2,3,...$, 
in the third step we used the fact that 
$\S_{\delta_i} \subset \S_{\beta_i}$, $i=1,2, ...$, 
and therefore 
$\cap_{i=1}^{\infty} \S_{\delta_i} \subset 
\cap_{i=1}^{\infty} 
\S_{\beta_i}$,
in the fifth step we used the fact that 
$\S_{\delta_i} \subset \S_{\gamma}$, $i=1,2, ...$,
and therefore 
$\cap_{i=1}^{\infty}
S_{\delta_i} \subset \S_{\gamma}$, 
and in the sixth step we used the assumption that 
$\eta \not\in \S_{\gamma}$.

The contradiction between (\ref{infnorm2}) and (\ref{infnorm4}) shows that the assumption that led to (\ref{infnorm4}) must be false,
and therefore (\ref{infnormthesis}) must be true. 
$\square$

\noindent {\bf Proof of Theorem~\ref{thm:extension}:}

Thm.H.26.3 implies that for any projection $p$,
\begin{equation}
\mbox{Range} \, (p) \ = \ 
\{ 
\phi \in \H \, : \, ||p \phi|| = ||\phi|| 
\}. 
\label{Thm.H.26.3}    
\end{equation}
This and the definition of $M_\phi (A)$ when
$A \in \A$, (\ref{defofMA}), show that 
we can define 
\begin{equation}
\H_A \ = \ \{   
\phi \in \H \, : \, M_{\phi}(A) = ||\phi||^2
\},
\label{defHA}
\end{equation}
for all $A \in \M$, consistently with the previous 
definition in case $A \in \A$
(in the second paragraph of this section). 
Once we show that for each $A \in \M$, $\H_A$ is a subspace, we can, also consistently,  define $p_A$ as the projection on $\H_A$.
For later use, note that $\M$-(PVM7) is satisfied, 
since for any $\A \in \M$, $M_\phi(A) \leq M_\phi(\Omega) = ||\phi||^2$.
 
We will show next that for each $A \in \M$, $\H_A$
is indeed a subspace and
\begin{equation}
M_{\phi} (A) \ = \ ||p_A \phi ||^2.      
\label{Mp}    
\end{equation}
This will be done in two steps. First we consider $A \in \A_{\sigma}$.
In this case we can write $A = \cup_{i=1}^{\infty} A_i$, where 
$A_i \in \A$, $i=1,2, ...$ are disjoint sets. Hence 
\begin{equation}
M_\phi (A) 
\ = \ 
\sum_{i=1}^{\infty} \, M_\phi (A_i)
\ = \ 
\sum_{i=1}^{\infty} \, ||p_{A_i} \phi||^2
\ = \  
\left| \left|   
\sum_{i=1}^{\infty} \, p_{A_i} \phi
\right| \right|^2,
\label{forMpAsigma}
\end{equation}
where in the second equality we used the definition of $M_\phi$, 
and in the third equality we used the orthogonality stated in $\A$-(PVM6). 
Thm.H.28.2 states that a sum of orthogonal
projections is a projection. Therefore 
$\sum_{i=1}^{\infty} p_{A_i}$ is a projection 
and from (\ref{defHA}), (\ref{forMpAsigma}) and 
(\ref{Thm.H.26.3}), 
\begin{equation} 
\H_A 
\ = \ 
\left\{
\phi \in \H \ : \ 
\left| \left|   
\sum_{i=1}^{\infty} \, p_{A_i} \phi
\right| \right|^2 
\, = \, 
||\phi||^2
\right\}
\ = \ 
\mbox{Range} 
\left( \sum_{i=1}^{\infty} \, p_{A_i}
\right),
\label{forPVM2}
\end{equation}
implying that $\H_A$ is indeed a subspace and that 
\begin{equation}
p_A = \sum_{i=1}^{\infty} {p_{A_i}}.
\label{psum}
\end{equation} 
Feeding (\ref{psum}) back into 
(\ref{forMpAsigma}), we obtain (\ref{Mp}) in case $A \in \A_{\sigma}$.

We turn now to general $A \in \M$. 
Define $\A_\sigma(A) = \{B \in \A_\sigma : A \subset B \}$.
Since $M_\phi(B) \leq 
\M_\phi(\Omega) =
||\phi||^2$ for any $B \in \M$, 
identity (\ref{fromRoyden-}), 
in conjunction with (\ref{defHA}), implies that 
\begin{equation}
\H_A 
\ = \ 
\{
\phi \in \H \, : \, 
M_\phi(B) = ||\phi||^2 \ \ \mbox{for all} \ \ B \in \A_\sigma(A)
\}
\ = \ 
\bigcap \, \{
\H_B \, : \, B \in \A_\sigma(A)
\}.
\label{HAfromRoyden}    
\end{equation}
Since intersections of subspaces are subspaces, 
this implies that $\H_A$ is a subspace. 

We will now apply Theorem~\ref{thm:infnorm} with $\Lambda = \A_\sigma(A)$, and 
for $B \in \Lambda$, $\S_B = \H_B$. To verify the condition in that theorem, 
given $B',B'' \in \Lambda$, take $B = B' \cap B''$, which does belongs to
$\Lambda$, since $A \subset B$ and  
intersections of finitely many elements of $\A_\sigma$ are 
also in $\A_\sigma$. And since we already know that $\M$-(PVM7) holds (as observed after 
(\ref{defHA})), we have $\H_B \subset \H_{B'} \cap \H_{B''}$, 
as required.
Theorem~\ref{thm:infnorm} 
and (\ref{HAfromRoyden}) 
give us then 
$$
|| p_A \phi ||^2 
\ = \ 
\inf \, \{ ||p_B \phi ||^2 \, : \, 
B \in \A_\sigma(A)
\}
\ = \ 
\inf \, \{ M_\phi(B) \, : \, 
B \in \A_\sigma(A) 
\}
\ = \ 
M_\phi(A),
$$
where in the second step we used the fact that (\ref{Mp})
has already been proved for sets in $\A_\sigma$,
and in the last step we used (\ref{fromRoyden-}).
This concludes the proof that (\ref{Mp}) holds for 
every $A \in \M$.

Our next task is to show that (\ref{Mp}) implies $\M$-(PVM2).  
Computations and arguments identical to the ones 
involving 
(\ref{forMpAsigma}), (\ref{forPVM2})
and (\ref{psum}) 
show that 
this task will be fulfilled if we prove that 
$\M$-(PVM6) holds. 
To do it, first we observe that, using (\ref{Mp}), 
we obtain, for each $A \in \M$, 
\begin{equation}
\H_{A^c} 
\ = \ 
\{ \phi \in \H \, : \, M_\phi(A^c) = ||\phi||^2 \}
\ = \ 
\{ \phi \in \H \, : \, M_\phi(A) = 0 \}
\ = \ 
\{ \phi \in \H \, : \, ||p_A\phi||^2 = 0 \}
\ = \ 
\H_A^{\perp}.
\label{HAc2}    
\end{equation}

Since we already know that $\M$-(PVM7) holds 
(as observed after (\ref{defHA})), we have 
that if $A \cap B = \emptyset$, then $B \subset A^c$, and hence 
$\H_B \subset \H_{A^c}$. 
Therefore (\ref{HAc2}) implies 
$\H_A \perp \H_B$. This establishes $\M$-(PVM6) and completes the proof of $\M$-(PVM2). 

Since $\M$-(PVM1) is 
the same as $\A$-(PVM1), it is 
already assumed to be true, and we have completed the proof that 
$\{p_A : A \in \M \}$ is a p.v.m..  

This proof also provided us with the claims 
(\ref{Mp-}),
(\ref{HAMA}) 
and (\ref{formula}), which appeared above as (\ref{Mp}), (\ref{defHA}) and (\ref{HAfromRoyden}),
respectively.


To show the uniqueness of the extension to 
$\Sigma$, suppose that 
$\{p'_A : A \in \Sigma \}$ is a p.v.m. 
that extends 
$\{p_A : A \in \A \}$.
For $\phi \in \H$ and $A \in \Sigma$, define
$M'_\phi(A) = ||p'_A \phi||^2$.
Then 
$\{M'_\phi(A) : A \in \Sigma \}$ 
is a measure on $\Sigma$ that agrees 
with $M_\phi(A)$ when $A \in \A$. 
By uniqueness of extension of finite measures 
from the algebra 
$\A$ to the sigma-algebra $\Sigma$ that 
it generates, we must also have 
$M'_\phi(A) = M_\phi(A)$, 
for all $A \in \Sigma$.

Using now (\ref{Thm.H.26.3}), we have, 
when $A \in \Sigma$, 
$$
\mbox{Range} \, (p'_A) 
\ = \ 
\left\{
\phi \in \H \, : \, 
M'_\phi(A) = ||\phi||^2
\right\}
\ = \ 
\left\{
\phi \in \H \, : \, 
M_\phi(A) = ||\phi||^2
\right\}
\ = \ 
\mbox{Range} \, (p_A),
$$
showing that $p'_A = p_A$.
$\square$.

\vspace{2mm} 

\noindent {\bf Remark on alternative proof:}
The proof of the existence 
part given above 
goes in steps, from 
$\A$ to $\A_\sigma$ to $\M$.
There is a more direct approach, at the 
cost of more abstraction, that is 
worth pointing out. We will only indicate
the ideas, leaving the details to the 
interested reader. 

The key tool is again 
Theorem~\ref{thm:infnorm}. 
But now, given $A \in \M$, we take
$\Lambda = \Lambda(A)$ given by 
$$
\Lambda \ = \ 
\{
\{A_i\}_{i=1,...} \, : \ 
 A_i \in \A, \ A_i \cap A_j = \emptyset 
 \ \mbox{if} \ i \not= j, \ 
 A \subset \cup_i A_i
\}.
$$
And for $\alpha = \{A_i\}_{i=1, ...}
\in \Lambda$, 
we define 
$$
\S_\alpha \ = \ 
\bigoplus_{i=1}^\infty \, \H_{A_i}. 
$$
Given two elements of $\Lambda$: 
$\alpha = \{A_i\}_{i=1, ...}$ 
and $\beta = \{B_j\}_{j=1, ...}$,
one can check that if 
$\gamma = \{A_i \cap B_j
\}_{i,j = 1, ...}$, then 
$\gamma \in \Lambda$ and 
$\S_\gamma \subset \S_\alpha \cap \S_\beta$. 
Therefore we can apply 
Theorem~\ref{thm:infnorm}. This gives us,
for any $\phi \in \H$,
\begin{eqnarray}
||p_\Lambda \phi ||^2 \ =  \ 
\inf_{\alpha \in \Lambda} 
|| p_\alpha \phi ||^2
\ & = & \ 
\inf \, \left\{\sum_i ||p_{A_i} \phi||^2 \, : \, 
\{A_i\}_{i=1, ...} \in \Lambda \right\}
\nonumber  \\  \ & = & \ 
\inf \, \left\{\sum_i M_\phi(A_i) \, : \, 
\{A_i\}_{i=1, ...} \in \Lambda \right\}
\ = \  M_\phi(A),
\label{extraproof1}
\end{eqnarray}
where in the third equality we used 
(\ref{defofMA}) and in the fourth 
equality we used (\ref{outermeasure}),
the observation that the infimum is 
not altered by taking only disjoint 
sets $A_i$, and the definition of $M_\phi(A)$,
for $A \in \M_\phi$, given by (\ref{MAM*A}).
By its definition, $p_\Lambda$ is 
the projection on 
$\cap_{\alpha \in \Lambda} \S_\alpha$. 
From this and the properties of a p.v.m.,
one can readily verify that 
in case $A \in \A$, 
$$
p_A \ = \ p_\Lambda.
$$
Therefore, we can consistently extend this 
equation as the definition of $p_A$,  for 
$A \in \M$. Equation 
(\ref{Mp-}) follows then from
(\ref{extraproof1}). Equation (\ref{HAMA}) 
follows from (\ref{Mp-}) and 
(\ref{Thm.H.26.3}). The proofs of
(\ref{formula}) and of the claim that 
$\{p_A : A \in \M \}$ is a p.v.m. follow
from (\ref{HAMA}) and 
(\ref{Mp-}) by the arguments in 
the proof above of the Theorem.

\section{A partial converse to part (b) of
Theorem~\ref{thm:basic}}
\label{sec:converse}

In this section we return to the setting
introduced in 
Section~\ref{sec:basic results}, but 
we will suppose that the set $S$ is
totally ordered. In case $S \subset \R$, which is the 
case in applications to quantum mechanics, in which 
elements of $S$ are moments in time, we can think that 
$S$ inherits the order from $\R$. Using $\leq$ for the 
order relation, we will, as usual, write $s < t$ in case
$s \leq t$ and $s \not= t$. 

In the statement of the theorem below we 
refer to a subset $V$ of $\H$ that is
dense in $\H$ (i.e., $\overline{V} = \H$).
Important examples 
of such sets are the domains of 
self-adjoint operators, including the 
Hamiltonian (see 
Subsection~\ref{subsec:basic examples}).  
The relevance of stating the theorem in 
terms of such a subset of $\H$ rather than 
$\H$ itself relates to its applicability to 
typical 
pilot-wave theories, including Bohmian 
mechanics, in 
Subsection~\ref{subsec:pilot-wave}. 

\begin{thm}
Suppose $S$ is a totally ordered set and $V \subset \H$ is 
dense in $\H$.
If for every $\phi \in V$
there is a probability measure $\P_\phi$ 
on $(\Omega,\Sigma)$ such that 
\begin{equation}
\P_\phi(X_{t_i} = a_i, i=1,...,k) 
\ = \ 
|| p^{t_k}_{a_k} ... p^{t_1}_{a_1} \hat{\phi} ||^2,
\label{converse}    
\end{equation}
for every $t_1 < ...< t_k$ and $a_1, ..., a_k$, 
then $\Hp = \H$, i.e., $p^s_a$ and $p^t_b$ commute,
for every $s,t \in S$, $a \in \I(s)$, $b \in \I(t)$.
\label{thm:converse}
\end{thm}

\noindent {\bf Proof:}
If $s = t$, then $p^s_a p^t_b = p^t_b p^s_a = 0$, if 
$a \not= b$ and $p^s_a p^t_b = p^t_b p^s_a = p^s_a$, 
if $a =b$. So, with no loss, we only need to consider
the case in which $s < t$, to which we turn now. 

For $\phi \in V \backslash \{0\}$
and $b \in \I(t)$,
we have from (\ref{converse}), 
$$
\sum_a \, || p^t_b p^s_a \hat{\phi} ||^2 
\ = \ 
\sum_a \, \P_\phi (X_s = a, X_t = b) 
\ = \ 
\P_\phi (X_t=b) 
\ = \
||p^t_b \hat{\phi}||^2.
$$

This implies that for every $\phi \in V$, 
\begin{equation}
\sum_a \, || p^t_b p^s_a {\phi} ||^2 
\ = \
||p^t_b {\phi}||^2.
\label{converse1}
\end{equation}


On the other hand, for any $\phi \in \H$, 
we have
\begin{eqnarray}
||p^t_b {\phi}||^2
\ & = &\ 
\left\langle 
p^t_b \phi \, , \, p^t_b \phi 
\right\rangle 
\ = \ 
\left\langle 
\sum_a \, p^t_b p^s_a \phi \, , \, p^t_b \phi
\right\rangle 
\ = \ 
\sum_a \,
\left\langle 
p^t_b p^s_a \phi \, , \, p^t_b \phi
\right\rangle 
\nonumber  \\ 
\ & = & \ 
\sum_a \,
\left\langle 
p^t_b p^s_a \phi \, , \,  p^t_b 
(I - p^s_a) \phi 
\ + \ 
p^t_b p^s_a \phi
\right\rangle
\ = \ 
\sum_a \,
\left(
\left\langle 
p^t_b p^s_a \phi \, , \,  p^t_b 
(I - p^s_a) \phi 
\right\rangle
\ + \ 
\langle p^t_b p^s_a {\phi} \, , \, 
p^t_b p^s_a {\phi} \rangle 
\right)
\nonumber \\
\ & = & \ 
\sum_a \,
\left(
\left\langle 
\phi \, , \, p^s_a p^t_b 
(I - p^s_a) \phi 
\right\rangle
\ + \ 
|| p^t_b p^s_a {\phi} ||^2
\right),
\label{converse2}    
\end{eqnarray}
where in the second equality we used 
(\ref{sumt}) and Thm.H.28.1 
(as in the proof of part (e) 
of Proposition~\ref{prop:technical}), 
and the third equality is 
justified by Thm.H.7.3, 
since $\sum_a p^t_b p^s_a \phi = p^t_b \phi$
is well defined.

Combining (\ref{converse1}) with (\ref{converse2}), 
we now have, for every $\phi \in V$, 
\begin{equation}
\sum_a \, 
\left\langle 
\phi \, , \, 
Q_a \phi
\right\rangle 
\ = \ 
0, 
\quad \quad 
\mbox{where}
\quad \quad 
Q_a = p^s_a p^t_b (I-p^s_a).
\label{VthenH}
\end{equation}

Next we will show that $\sum_a Q_a \phi$
converges for every $\phi \in \H$, 
defining a bounded operator
$\sum_a Q_a$. 
Since the projections $p^s_a$, 
$a \in \I(s)$ are orthogonal to each other,
so are also the vectors $Q_a \phi$.
Hence the claimed convergence is 
equivalent to the statement that 
$\sum_a ||Q_a \phi||^2 < \infty$, which 
we easily verify:
\begin{eqnarray*}
\sum_a ||Q_a \phi||^2  
\ & \leq & \ 
2 \sum_a \left(    
||p^s_a p^t_b \phi||^2 \, + \, 
||p^s_a p^t_b p^s_a \phi||^2
\right)
\ \leq \ 
2 \sum_a \left(    
\left| \left|p^s_a p^t_b \phi 
\right| \right|^2 \, + \, 
||p^s_a \phi||^2 
\right)
\\ \ & = & \  
2 \left(
\left| \left |\sum_a p^s_a p^t_b \phi
\right | \right |^2 \, + \, 
\left| \left|\sum_a p^s_a \phi
\right| \right |^2
\right)
\ = \ 
2 \left(
\left| \left |p^t_b \phi
\right | \right |^2 \, + \, 
\left| \left| \phi
\right| \right |^2
\right)
\ \leq \ 
4 \, || \phi ||^2,
\end{eqnarray*}
where in the second and in the last 
steps we used the fact that projections 
cannot increase the norm of a vector, 
and in the third and fourth steps
we used again the orthogonality  
of the projections $p^s_a$, $a \in \I(s)$, 
and (\ref{sumt}), respectively. 
The norm of $\sum_a Q_a $ can be 
estimated from $||\sum_a Q_a \phi ||^2 
= \sum_a ||Q_a \phi||^2 \leq 4 ||\phi||^2$, 
as being at most 2. 

The convergence of $\sum_a Q_a \phi$ allows
the application of Thm.H.7.3 to  
(\ref{VthenH}), to obtain 
\begin{equation}
\left\langle
\phi \, , \,  \sum_a Q_a \phi
\right\rangle
\ = \ 0,
\label{converse10}    
\end{equation}
for any $\phi \in V$. But since $\sum_a Q_a$
is a bounded (and hence continuous)
operator and inner products 
are jointly continuous in their
two arguments, 
(\ref{converse10}) extends by continuity,
from the dense $V$, 
to all $\phi \in \H$. 
And Theorem 12.7 of \cite{Rudin}
tells us that (since our Hilbert 
space is over the Complex field)
this implies 
\begin{equation}
    \sum_a Q_a \phi \ = \ 0,
    \label{converse11}
\end{equation}
for all $\phi \in \H$. 
Recall that since the projections $p^s_a$, 
$a \in \I(s)$ are orthogonal to each other,
so are also the vectors $Q_a \phi$.
And since a sum of orthogonal vectors can 
only be 0 if each one of them is, 
(\ref{converse11}) yields, 
for each $a \in \I(s)$ and $\phi \in \H$, 
$$
Q_a \phi \ = \ 0.
$$

So we have proved that 
$$
p^s_a p^t_b
\ = \ 
p^s_a p^t_b p^s_a, 
$$
for each $a \in \I(s)$ and $b \in \I(t)$.
Since the right-hand side of this equation is a 
self-adjoint operator, so has to be the left-hand side.
But the adjoint of $p^s_a p^t_b$ is 
$p^t_b p^s_a$. So we have learned that 
$p^s_a p^t_b = p^t_b p^s_a$, completing the proof.
$\square$


\vspace{3mm}

It is natural to ask if when $\Hp \not = \H$, there could
still be some exceptional $\phi \not\in \Hp$ for which 
(\ref{converse}) holds. An example with $S = \{s,t\}$, 
$s < t$, shows that this is possible. 
By (\ref{sumt}) we have 
$\oplus_a \H^s_a = \H$. Therefore, if $\Hp \not = \H$,
there must exist some
$c \in \I(s)$ for which there is some $\phi \in \H^s_c$,
with $\phi \not\in \Hp$. 
Obviously $\phi \not= 0$, so that we can compute
\begin{equation} 
|| p^t_b p^s_a \hat{\phi} ||^2
\ = \ 
|| p^t_b \hat{\phi} ||^2 \, \delta_{a,c},
\label{singlephi}
\end{equation}
where $\delta_{a,c} = 1$ if $a = c$ and 
$\delta_{a,c} = 0$ if $a \not= c$.
The numbers in the right-hand side of (\ref{singlephi})
are non-negative and satisfy 
$$
\sum_{a,b} \, || p^t_b \hat{\phi} ||^2 \, \delta_{a,c}
\ = \ 
\sum_{b} \, || p^t_b \hat{\phi} ||^2
\ = \ 
\left| \left|
\sum_b \, 
p^t_b \hat{\phi} 
\right| \right|^2
\ = \ 
|| \hat{\phi} ||^2
\ = \ 
1,
$$
where we used (\ref{perp}) and (\ref{sumt}).
Therefore (\ref{singlephi}) defines a probability 
measure $\P_{\phi}$ on $(\Omega, \Sigma)$, that 
satisfies 
$$
\P_\phi(X_s = a, X_t = b) \ = \ 
|| p^t_b p^s_a \hat{\phi} ||^2
\ = \ 
|| p^t_b \hat{\phi} ||^2 \, \delta_{a,c}.
$$
This probability measure satisfies also
$$
\P_\phi (X_s = a) \ = \  
\sum_b \, \P_\phi(X_s = a, X_t = b) \ = \
\sum_b \, || p^t_b \hat{\phi} ||^2 \, \delta_{a,c}
\ = \ 
\delta_{a,c}
\ = \ 
|| p^s_a \hat{\phi} ||^2.
$$
And 
$$
\P_\phi (X_t = b) \ = \  
\sum_a \, \P_\phi(X_s = a, X_t = b) \ = \
\sum_a \, || p^t_b \hat{\phi} ||^2 \, \delta_{a,c}
\ = \ 
|| p^t_b \hat{\phi} ||^2. 
$$
The last three displays show that (\ref{converse}) 
is satisfied by $\P_\phi$.

\section{Refinements and coarsenings}
\label{sec:refinements}
The concepts of refinement and
coarsening discussed in this section 
are the same as those in the 
consistent, or decoherent,
approach to quantum mechanics 
(see, e.g., \cite{Omnes}, 
\cite{GMH}, \cite{Griffiths}). 

Suppose that $\pi$ is as defined in 
Section~\ref{sec:basic results}.
A refinement of $\pi$ is another set of projections in $\H$, 
$$
\pi' \ = \ 
\{
p^{t}_{b} \, : \, t \in S', \, b \in  \I'(t) 
\},
$$
where $S \subset S'$ and for each
$t \in S$, $\I'(t)$ is the disjoint union of 
some sets $\I'_a(t)$, $a \in \I(t)$, with the 
property that 
\begin{equation}
p^t_a \ = \ \sum_{b \in \I'_a(t)} \, p^{t}_{b}.
\label{sumrefinement}    
\end{equation}
This implies that the condition 
$$
\sum_{b \in \I'(t)}  \, p^t_b \ = \  I
$$
is satisfied for every $t \in S$, and we 
assume that it is satisfied for
every $t \in S'$.
Informally, we are increasing the set of times from 
$S$ to $S'$ and, for each $t \in S$, 
breaking each $p^t_a$ into a sum of smaller 
orthogonal projections,
according to (\ref{sumrefinement}).
(By smaller projections we mean as usual that 
their ranges are smaller subspaces. And the 
orthogonality of the ranges of the $p^t_b$ in 
(\ref{sumrefinement}) is a consequence of 
Thm.H.28.2  according to which a sum of projections 
can only be a projection if their ranges are 
orthogonal to each other.)

From the definition above, it is clear that 
\begin{equation}
\H_{\pi'} \ \subset \ \Hp.    
\label{Hp'subHp}    
\end{equation}
 
We will use primes to denote,
in a self-explanatory fashion,
the following objects associated to
$\pi'$:  
$\Omega'$, $\Sigma'$, $\A'$, $X'_{s}$, $s \in S'$. 

Define now, for each $A \subset \Omega$,
\begin{equation}
A' \ = \ 
\left\{
\omega' \in \Omega' \, : \, 
\mbox{
for some $\omega \in A, \ 
\omega'(t) \in \I'_{\omega(t)}(t)$, \
for all $t \in S$
}
\right\}.
\label{A'}    
\end{equation}
The following properties of $A'$ are immediate:
\begin{itemize}
    \item The notation $\Omega'$ was defined twice above, 
    but consistently. 
    \item If $A$ and $B$ are disjoint subsets of $\Omega$,
    then $A'$ and $B'$ are disjoint subsets of $\Omega'$.
    \item If $\{ A_\alpha \}$ is an arbitrary family of 
    subsets of $\Omega$ and $A = \cup_\alpha A_\alpha$, 
    then $A' = \cup_\alpha A'_\alpha$. 
    \item For any $A \subset \Omega$, \ 
    $(A^c)' = (A')^c$.
\end{itemize}

When $\pi'$ is a refinement of $\pi$, we say that 
$\pi$ is a coarsening of of $\pi'$. A simple example,
from Section~\ref{sec:Farbitrary}, is $\pi(D)$,
defined by (\ref{pi(D)}), as a coarsening of $\pi$.
Lemma~\ref{insteadofthm6} in that section
is an instance of 
one of the statements in Theorem~\ref{thm:refinement}, below.
A number of interesting examples will 
appear in 
Subsection~\ref{subsec:ontology+}.

\begin{thm}
If $A \in \Sigma$, then $A' \in \Sigma'$ and 
\begin{equation}
p_{A'} \  = \  p_A p_{\pi'} 
\ = \ 
p_{\pi'} p_A 
\ = \ 
p_A \wedge p_{\pi'}. 
\label{refinementformula}    
\end{equation}
So that in particular $p_A$ and $p_{\pi'}$ commute 
and $\H_{A'} = \H_A \cap \H_{\pi'}$.
\label{thm:refinement}
\end{thm}

\noindent {\bf Proof:}
We will use twice the $\pi$-$\lambda$ Theorem (see, e.g., 
Theorem 3.2 of \cite{Billingsley}).

Consider the following class of subsets of $\Omega$
$$
\mathcal{P} 
\ = \ 
\left\{ 
\{
X_{t_1} = a_1, ... , X_{t_k} = a_k
\}
\ : 
\ t_1, ..., t_k \in S, \, 
a_1 \in \I(t_1), ... , a_k \in \I(t_k)
\right\}
\ \cup \{ \emptyset \}. 
$$
This class is clearly closed with respect to finite
intersections, which means that it is a $\pi$-system.   

Consider now the class 
$$
\mathcal{L}_1
\ = \ 
\left\{
A \subset \Omega \ : \ A' \in \Sigma'
\right\}.
$$
The properties of the mapping from $A$ to $A'$ listed 
above imply that $\mathcal{L}_1$ has the three properties
that define a $\lambda$-system:
\begin{itemize}
    \item $\Omega \in \mathcal{L}_1$.
    \item $\mathcal{L}_1$ is closed with respect to 
    complements.
    \item $\mathcal{L}_1$ is closed with respect to 
    countable disjoint unions.  
\end{itemize}

We claim that for any $A \in \mathcal{P}$, we have 
$A' \in \A'$. Indeed, if $A = \emptyset$, then 
$A' = \emptyset$, and for $A 
= \{X_{t_1} = a_1, ..., X_{t_k} = a_k \}
\in \mathcal{P}$, we have 
\begin{equation}
A' \  =  \ 
\left\{
 X'_{t_1} \in  \I'_{a_1}(t_1), ... , 
X'_{t_k} \in \I'_{a_k}(t_k)
\right\}
\ \in \ \A'.
\label{A'forAinP}
\end{equation}
This means that $\mathcal{P} \subset \mathcal{L}_1$, 
and, since the smallest sigma-algebra that contains  
$\mathcal{P}$ is $\Sigma$, we learn 
from the $\pi$-$\lambda$ Theorem that
$\Sigma \subset \mathcal{L}_1$, completing 
the proof that $A' \in \Sigma'$ whenever 
$A \in \Sigma$.

Thm.H.29.1 states that a product of projections is a 
projection if and only if they commute, and in this case 
their product in any order is equal to their meet.
Therefore we only need to prove the first equality 
in (\ref{refinementformula}), and the others follow. 

If $\phi \perp \H_{\pi'}$, then
$p_{A'}\phi = 0 = p_A p_{\pi'} \phi$, so it is 
sufficient to prove that if $\phi \in \H_{\pi'}$, then
$p_{A'}\phi = p_A p_{\pi'} \phi$. and this is the same
as the statement that 
\begin{equation}
\mbox{
if \ $\phi \in \H_{\pi'}$, \ then
$p_{A'}\phi = p_A \phi$.
}
\label{goal}    
\end{equation}
Set 
$$
\mathcal{L}_2 
\ = \ 
\{
A \in \Sigma \, : \, 
\mbox{$p_{A'}\phi = p_A \phi$ \ for every \ 
$\phi \in \H_{\pi'}$  
}
\}.
$$
The class $\mathcal{L}_2$ is a $\lambda$-system, since for 
every $\phi \in \H_{\pi'}$ we have:
\begin{itemize}
    \item For $A = \Omega$, \ $p_{A'} \phi = 
     p_{\Omega'} \phi = p_{\pi'} \phi = 
     p_{\pi} \phi = p_{\Omega} \phi = 
     p_A \phi$, \  where we used (\ref{Hp'subHp}), in the 
     third equality. Therefore $\Omega \in \mathcal{L}_2$.
     \item If $A \in \mathcal{L}_2$, then, \ 
     $p_{(A^c)'} \phi = (p_{\Omega'} - p_{A'}) \phi
     = (p_{\Omega} - p_{A}) \phi = p_{A^c} \phi$,
     where we used the already proved facts that 
     $A'$ and $(A^c)'$ are in $\Sigma'$, part (c) of 
     Theorem~\ref{thm:basic} for $\pi$ and for $\pi'$ and
     specifically property (PVM5) of a p.v.m.
     (see Section~\ref{sec:pvm}), as well as the fact 
     from the previous item. 
     Therefore $A^c \in \mathcal{L}_2$.
     \item If $A_1,A_2, ...$ are disjoint sets in 
     $\mathcal{L}_2$
     and $A = \cup_{i=1}^{\infty} A_i$, then, 
     from the properties of the mapping from $A$ to
     $A'$ listed before this theorem, we have that 
     also $A'_1,A'_2, ...$ are disjoint sets 
     and $A' = \cup_{i=1}^{\infty} A'_i$. And 
     since we already know that $A'_i \in \Sigma'$, 
     we can use again part (c) of 
     Theorem~\ref{thm:basic} for $\pi$ and for $\pi'$ and
     property (PVM2) of a p.v.m.
     (see Section~\ref{sec:pvm}), to write: 
     \ 
     $p_{A'} \phi = \sum_{i=1}^{\infty} p_{A_i'} \phi 
     = \sum_{i=1}^{\infty} p_{A_i} \phi =
     p_A \phi$. 
     Therefore $A \in \mathcal{L}_2$.
\end{itemize}

Our next task is to show that 
\begin{equation}
    \mathcal{P} \subset \mathcal{L}_2.
    \label{PinL2}
\end{equation}
Clearly $\emptyset \in \mathcal{L}_2$,
and for $A = \{X_{t_1} = a_1, ..., X_{t_k} = a_k \}
\in \mathcal{P}$, we have (\ref{A'forAinP}), so that, 
for $\phi \in \H_{\pi'}$, 
\begin{eqnarray*}
p_{A'} \phi 
& \ = \ &
\sum_{b_1 \in \I'_{a_1}(t_1), ..., 
b_k \in \I'_{a_k}(t_k)}
\, p^{t_1,...,t_k}_{b_1,...,b_k} \, \phi 
\ = \ 
\sum_{b_1 \in \I'_{a_1}(t_1), ..., 
b_k \in \I'_{a_k}(t_k)}
\, p^{t_1}_{b_1}...p^{t_k}_{b_k} \, \phi 
\\
& \ = \ &
\left(
\sum_{b_1 \in \I'_{a_1}(t_1)} \, p^{t_1}_{b_1}
\right)
\, \cdots \, 
\left(
\sum_{b_k \in \I'_{a_k}(t_k)} \, p^{t_k}_{b_k}
\right) \, \phi 
\ = \ 
p^{t_1}_{a_1} ... p^{t_k}_{a_k} \, \phi
\ = \ 
p^{t_1,...,t_k}_{a_1,...,a_k} \, \phi 
\ = \ 
p_A \phi.
\end{eqnarray*}
In the first and in the last 
equalities, we used part (c)
of Theorem~\ref{thm:basic}, first 
for $\pi'$, then for 
$\pi$, and in the former we also 
used property (PVM2)
of a p.v.m (see Section~\ref{sec:pvm}). 
In the second and in the 
next-to-last equalities, we used 
part (d) of Proposition~\ref{prop:technical},
first for $\pi'$ (fine since 
$\phi \in \H_{\pi'}$) 
then for $\pi$ (fine since, thanks 
to (\ref{Hp'subHp}),
also $\phi \in \Hp$).
And in the third and fourth equalities, we used 
Thm.H.28.1 (as in the proof of part (e) of Proposition~\ref{prop:technical})
and (\ref{sumrefinement}).

Since $\mathcal{P}$ is a $\pi$-system that generates
$\Sigma$ and $\mathcal{L}_2$
is a $\lambda$-system, the $\pi$-$\lambda$ Theorem tells
us that (\ref{PinL2}) implies the stronger statement 
$$
\Sigma \ \subset \ \mathcal{L}_2,
$$
which means that (\ref{goal}) holds for 
every $A \in \Sigma$, 
completing the proof of the theorem.
$\square$

If $\pi'$ is a refinement of $\pi$, we 
write 
$\pi' \leq \pi$, as this is a 
partial order in
the set of possible $\pi$. 

In the set of possible $\pi$ with a given fixed 
$S$, the minimal element is the one in which, 
for each $t \in S$, $\I(t)$ has 
a single element $a_t$ and $p^{t}_{a_t} = I$.
(This is unique modulo the  
choice of the labels $a_t$.)
This set of $\pi$ also 
has maximal elements, those being characterized 
by the sets $\H^t_a$ having dimension 1, for 
all $t \in S$, $a \in \I(t)$.

\section{Examples, Remarks and Applications}
\label{sec:examples and remarks}

This section combines mathematical issues with 
issues of interpretation. It includes some 
applications that illustrate the use of 
the theorems proved in the 
previous sections to evaluate proposed 
interpretations, or to propose different ones. 
The first application described in the abstract of 
the paper appears in
Subsection~\ref{subsec:pilot-wave} and 
is further elaborated in 
Subsection~\ref{subsec:comparison}. 
The second one appears in
Subsection~\ref{subsec:ontology} 
and is further elaborated in the following 
three subsections.
The third one appears in 
Subsection~\ref{subsec:Born-superposition}.

\subsection{Basic examples}
\label{subsec:basic examples}
In the standard quantum mechanics setting, in 
addition to the Hilbert space $\H$, there is a 
strongly continuous group of unitary operators 
$(U_t)_{t \in \R}$, that provide the evolution of 
the state in the Schr\"odinger picture, or of the 
operators in the Heisenberg picture. In the 
Schr\"odinger picture
the state at time $t$ is given by 
$\Psi_t = U_t \Psi$, 
if at time $0$ it is $\Psi \in \H$.
In the Heisenberg picture $\Psi$ does not change
with time, but each operator $Q$ evolves to 
$Q_t = U_{-t} Q U_t$ at time $t$.

The assumptions on $(U_t)_{t \in \R}$ are 
expressed as 
\begin{equation}
U_{0} = I, \quad U_{t+s} = U_t U_s, \quad
U^*_t = U^{-1}_t,
\label{Ugroup} 
\end{equation}
for each $t, s \in \R$, 
where the star denotes the adjoint. And 
\begin{equation}
\lim_{s \to t} \, U_s \phi \ = \  U_t \phi,
\label{strongly continuous}    
\end{equation}
for each $\phi \in \H$, $t \in \R$.

Note that (\ref{Ugroup}) implies that 
\begin{equation}
U_{-t} \, = \,  U^{-1}_t \, = \, U^*_t.    
\label{Ugroup+}    
\end{equation}

Stone's Theorem and its converse
(see Theorems VIII.7 and VIII.8 of \cite{RS})
state that the conditions above on 
$(U_t)_{t \in \R}$ 
are equivalent to the existence of a 
self-adjoint operator $H$, in this context 
called the Hamiltonian, such that 
$$
U_t \ = \ \exp (- i t H),
$$
for all $t \in \R$. 

Suppose that $\I$ is a countable set and 
$\{p_a : a \in \I\}$ is a set of projections 
in $\H$ that satisfy 
\begin{equation}
\sum_a p_a = I.    
\label{suma}    
\end{equation}
In other words, this set of projections is  
a partition of the identity. 

Define now, for each $t \in \R$, 
$$
p^t_a  \ = \  U_{-t} \, p_a \, U_t.   
$$
Then, using (\ref{Ugroup+}), 
$$
\sum_a \, p^t_a \ = \ 
\sum_a \, U_{-t} \,  p_a \,  U_t \ = \ 
U_{-t} \left(\sum_a  p_a \right) U_t \ = \ 
U_{-t} \, I \, U_t \ = \ 
I,
$$
so that (\ref{sumt}) is satisfied.
If $S \subset \R$, 
then 
$$
\pi = \{p^t_a : t \in S, a \in \I \}
$$
is an example of the sort of 
family of projections studied 
in this paper, with the particular 
feature that $\I(t) = \I$, for all $t \in S$.

Every such $\pi$ has a natural refinement 
$$
\pi' = \{p^t_a : t \in \R, a \in \I \}.
$$
And in case $S$ is dense in $\R$, 
i.e., $\overline{S} = \R$, 
(\ref{strongly continuous}) implies that 
\begin{equation}
\Hp = \H_{\pi'}.
\label{Hp=Hp'}
\end{equation} 

The group property in (\ref{Ugroup}) yields,
for each $t \in \R$, 
\begin{eqnarray}
p^{t_1}_{a_1} ... p^{t_k}_{a_k} U_t \phi 
& \ = \ & 
U_{-t_1} p_{a_1} U_{t_1} ... 
U_{-t_k} p_{a_k} U_{t_k} \, U_t \phi 
\nonumber \\  & \ = \ & 
U_t \left(
U_{-(t_1+t)} p_{a_1} U_{t_1 + t} ... 
U_{-(t_k+t)} p_{a_k} U_{t_k + t}
\right) \, \phi
\nonumber \\  & \ = \ & 
U_t \, p^{t_1 + t}_{a_1} ... p^{t_k + t}_{a_k}
\phi.
    \label{pU}
\end{eqnarray}
And this implies that $\Hp$ is invariant under
$U_t$, i.e., 
\begin{equation}
U_t \, \Hp \ \subset \ \Hp.
    \label{Uinvariant}
\end{equation}
This conclusion applied to $-t$,  
in conjunction with (\ref{Ugroup+}) implies 
that $\Hp$ is also invariant under
$U^*_t$. And Thm.H.23.2 implies then that 
$\Hp^{\perp}$ is invariant under
$U_t$:
\begin{equation}
U_t \, \Hp^\perp \ \subset \ \Hp^\perp.
    \label{Uinvariantperp}
\end{equation}
Together, (\ref{Uinvariant}) and 
(\ref{Uinvariantperp}) are expressed by saying 
that $\Hp$ reduces $U_t$. And Thm.H.26.2 then 
tells us that $U_t$ commutes with $p_{\pi}$, 
for every $t \in \R$. 

\subsection{Particle models}
\label{subsec:particles}

We turn now to ``particle models'', that are 
important examples of the setting in 
Subsection~\ref{subsec:basic examples}.  
For simplicity, we consider first 
a universe with a single type of particle, 
and no creation or annihilation of 
particles. Suppose that the dimension of
the physical  
space is 3 and that there are n particles.
In this case $\H = L^2(\R^{3n})$, and 
$\mathcal{C} = \R^{3n}$ 
is called the configuration space, 
and is endowed with its Borel sigma-algebra
and Lebesgue measure.
For each measurable $R \subset \R^{3n}$, let 
$I_R$ denote its indicator function, i.e., 
$I_R (x) = 1$, if $x \in R$, and  
$I_R (x) = 0$, if $x \not\in R$.

Let $\I$ be a countable set
and let $\{R_a : a \in \I\}$ be a partition of 
$\R^{3n}$ into measurable disjoint sets $R_a$
that have boundaries of Lebesgue 
measure 0.
Define now the projections $p_a$ by 
$$
(p_a \phi)(x) \ = \ I_{R_a}(x) \,  \phi(x),  
$$
$x \in \R^{3n}$. It is clear that (\ref{suma}) 
is satisfied, and hence we have an example of
the setting discussed in 
Subsection~\ref{subsec:basic examples}. 
Clearly also, the range of $p_a$ is 
$$
\H_a  \ = \ 
\{
\phi \in \H \, : \, 
\mbox{supp} \, \phi \, \subset \, 
\overline{R}_a
\},
$$
where $\mbox{supp}  \, \phi$ 
denotes the essential support of $\phi$
i.e., the smallest closed subset of $\R^{3n}$ such that $\phi = 0$ almost everywhere 
on the complement of this set.
If we use the notation, $\phi_t = U_t \phi$, then 
it follows that, 
for each $t\in S$ and $a \in \I(t)$,
\begin{eqnarray*}
\H^t_a 
\ = \ 
\left\{
\phi \in \H \, : \, p^t_a \phi = \phi 
\right\}
\ = \ 
\left\{
\phi \in \H \, : \, U_{-t} p_a U_t \phi = \phi 
\right\}
\ = \ 
\left\{
\phi \in \H \, : \, p_a \phi_{t} = \phi_t 
\right\}
\\
\ = \ 
\left\{
\phi \in \H \, : \, \phi_t \in \H_a 
\right\}
\ = \ 
\left\{
\phi \in \H \, : \, 
\mbox{supp} \, \phi_t \, \subset \, 
\overline{R}_a
\right\},
\end{eqnarray*}
where in the third equality we used 
(\ref{Ugroup+}). It follows that 
for each $t_1,...,t_k$ and $a_1,...a_k$, 
$$
\H^{t_1, ..., t_k}_{a_1, ...,a_k} 
\ = \ 
\left\{ \phi \in \H \, : \, 
\mbox{supp} \, \phi_{t_i} \, \subset \, \overline{R}_{a_i}, 
\, i = 1,...,k
\right\}.
$$
In words, this subspace is the set of wave 
functions 
at time 0 that 
evolve with time in the Schr\"odinger 
picture in such a way that at 
each time $t_i$,
$i=1, ...,k$ their 
essential support is contained in $\overline{R}_{a_i}$,
and hence they are  
almost everywhere identically 
0 outside of $R_{a_i}$
(recall that the boundary of each $R_{a_i}$
has Lebesgue measure 0). 
In applications, the sets $R_a$ may correspond to
physically meaningful macroscopic descriptions. For 
instance, in one of these sets some of the particles 
may form a healthy 
cat, or a measuring device with 
a pointer indicating some outcome to an experiment,
or a computer in a certain computational state, 
or human beings with brains in 
configurations that correspond to certain 
mental states.  
The subspace 
$\H^{t_1, ..., t_k}_{a_1, ...,a_k}$
should then be understood as 
the set of time-0 wave functions 
with the property that at the 
times $t_1, ..., t_k$ the respective physical 
descriptions indexed by $a_1, ...,a_k$
correspond to our unique macroscopic reality.

There is no difficulty in modifying the 
example above to allow for creation and 
annihilation of particles. In this case 
the configuration space should be taken as
the disjoint union 
$\mathcal{C} = \cup_{n=o}^\infty \R^{3n}$,
where $\R^0$ is a set with a single 
element, 
called the ``vacuum configuration'', and 
the Hilbert space will be the Fock space 
$$
\H \  = \ \bigoplus_{n=0}^\infty \, L^2(\R^{3n})
\ = \ 
\left\{
(\phi_0, \phi_1, ...) \, : \, \phi_n \in 
L^2(\R^{3n}), n = 0, 1, ..., \, 
\sum_{n=0}^\infty ||\phi_n||^2 < \infty 
\right\},
$$
where $L^2(\R^0) = \mathbb{C}$, the set of
complex numbers.

Let $\I$ be a countable set
and, 
for each 
$n = 0,1, ...$, let 
$\{R_{a,n} : a \in \I\}$ be 
a partition of the corresponding $\R^{3n}$ 
into measurable disjoint sets $R_{a,n}$
that have boundaries of Lebesgue 
measure 0.
Define now the projections $p_a$ by
$$
p_a \phi \ = \ p_a (\phi_0, \phi_1, ...) 
\ = \ 
(p_{a,0} \phi_0, p_{a,1} \phi_1, ... ), 
$$
where 
$$
p_{a,n} \, \phi_n (x) \ = 
\ I_{R_{a,n}}(x) \, \phi_n (x), 
$$
$x \in \R^{3n}$.
Then, similarly to the previous example, 
$$
\H_a  \ = \ 
\{
(\phi_0,\phi_1,...) \in \H \, : \, 
\mbox{supp} \, \phi_n \, \subset \, 
\overline{R}_{a,n}, \
n=0,1, ...
\}.
$$
And, if we use the notation 
$U_t \phi = U_t (\phi_0, \phi_1, ...) = 
(\phi_{t,0}, \phi_{t,1}, ...)$, then 
for any $t_1, ..., t_k$ and $a_1, ..., a_k$, 
we have 
$$
\H^{t_1, ..., t_k}_{a_1, ...,a_k} 
\ = \ 
\left\{ (\phi_0, \phi_1, ...) \in \H \, : \, 
\mbox{supp} \, \phi_{t_i,n} \, \subset \,
\overline{R}_{a_i,n}, 
\, i = 1,...,k, \ n=0,1, ...
\right\}.
$$

And as in the previous example, this subspace 
admits the same sort of interpretation that 
that one has. 
Suppose that we take 
the partitions $\{R_{a,n} : a \in \I \}$ 
in such a way that 
configurations in any of the sets 
$R_{a,n}$, $n=0,1, ...$ correspond to the 
same macroscopic 
description indexed by $a \in \I$. 
Then the subspace 
$\H^{t_1, ..., t_k}_{a_1, ...,a_k}$
should be understood as 
the set of time-0 wave functions 
with the property that at the 
times $t_1, ..., t_k$ the respective physical 
descriptions indexed by $a_1, ...,a_k$
correspond to our unique macroscopic
reality.

We can also include 
different kinds of particles, possibly 
with different spins, 
without any further difficulty, 
by replacing in 
the Fock space
$L^2(\R^{3n})$ with the appropriate
tensor products (see, e.g., Section II.4 
of \cite{RS}). 
The configuration space becomes then a disjoint union
$\mathcal{C} = \cup_{n_1=0}^\infty ... 
\cup_{n_l=0}^\infty \, \R^{3n_1} \times ...
\times \R^{3n_l}$, where the indices $1,...,l$ 
correspond to the different types of particles. 

\subsection{Remarks on the expression 
$||p^{t_k}_{a_k} ... p^{t_1}_{a_1} \hat{\phi} ||^2$}
\label{subsec:Born}
This expression, that appears in the right-hand side 
of (\ref{Born}), in part (b) of 
Theorem~\ref{thm:basic}, can be rewritten in ways 
that more explicitly show its relation with 
Born's rule and (apparent) 
collapse of the wave function at ``obsevation'' 
times. Here we are assuming that $S \subset \R$ and 
$t_1 < ... < t_k$.
We are also supposing that at each time $t_i$, 
$i=1, ..., k$ an ``observation''
is being made which has 
possible outcomes in $\I(t_i)$ and that if $a_i$ is
``observed'', standard quantum mechanics with collapse
postulates collapse of the  
(Heisenberg-picture) wave function into its 
projection on the subspace $\H^{t_i}_{a_i}$. 
In the usual jargon, 
and assuming that the indexes $a_i$ are 
identified with real numbers, 
at time $t_i$ the observable
corresponding to the self-adjoint operator
$\sum_{a_i \in \I(t_i)} a_i \, p^{t_i}_{a_i}$ is
being measured. 

In the case $k=1$, 
we have 
$||p^{t_k}_{a_k} ... p^{t_1}_{a_1} \hat{\phi} ||^2
\, = \,  
||p^{t_1}_{a_1} \hat{\phi}||^2$, which indeed 
is the probability given by Born's rule, for
an ``observation'' of $a_1$ at time $t_1$, when
the state (in the Heisenberg picture) is $\phi$.

Set $\phi_0 = \phi$, and, for $i = 1, ..., k$, 
recursively define 
$\phi_i = p^{t_i}_{a_i} \hat{\phi}_{i-1}$, if 
$\phi_{i-1} \not= 0$, and 
$\phi_i = 0$, if $\phi_{i-1} = 0$. 
If $\phi_{k} = 0$,
let 
$$
i_0 = \min \{i \in \{1, ..., k\} : 
\phi_i = 0\}.
$$  
Then, when $\phi_{k} \not= 0$, we have  
$$
||p^{t_k}_{a_k} ... p^{t_1}_{a_1} \hat{\phi} ||^2
\ = \ 
||p^{t_k}_{a_k} ... p^{t_2}_{a_2} \hat{\phi_1} ||^2
\, ||\phi_1||^2
\ = \ 
... \ = \  
||\phi_k||^2 \cdots ||\phi_1||^2
\ = \ 
||p^{t_k}_{a_k} \hat{\phi}_{k-1}||^2 
\cdots ||p^{t_1}_{a_1}\hat{\phi}_0||^2.
$$

This is precisely the probability that standard
quantum mechanics with collapse gives to the 
successive ``observations'' of $a_1$ at $t_1$,
..., $a_k$ at $t_k$, with collapse of the wave 
function at each ``observation''. 

When $\phi_{k} = 0$, 
$$
||p^{t_k}_{a_k} ... p^{t_1}_{a_1} \hat{\phi} ||^2
\ = \ 
\cdots 
\ = \ 
||p^{t_k}_{a_k} ... p^{t_{i_0}}_{a_{i_0}} \hat{\phi}_{{i_0}-1} ||^2
\, ||\phi_{{i_0}-1}||^2 \cdots ||\phi_1||^2
\ = \ 0.
$$
This also agrees with standard quantum mechanics
with collapse, since then the 
$({i_0}-1)$-th ``observation''
would have collapsed the wave function to 
$\hat{\phi}_{{i_0}-1}$ which is incompatible with the
observation of $a_{i_0}$ at time $t_{i_0}$, because
$p^{t_{i_0}}_{a_{i_0}} \hat{\phi}_{{i_0}-1} = \phi_{i_0} = 0$. 

\subsection{Pilot-wave theories in configuration space
and physical space that are fully 
equivalent to standard 
quantum mechanics in a 
path-wise sense}
\label{subsec:pilot-wave}
Part (b) of Theorem~\ref{thm:basic} can be seen as 
stating the existence of a pilot-wave theory in
``$\I$-space'' that
is in full agreement with 
standard quantum mechanics. 
This can be used to build pilot-wave theories 
in configuration space, and hence also in 
physical space, for the particle models discussed
in Subsection~\ref{subsec:particles}. 

In these models the configuration space 
$\mathcal{C}$ is partitioned into sets 
$R_{a,i}$, where $a \in \I$ and $i$ specifies 
the number of particles of each kind present. 
The interpretation being that all points in 
each $R_a = \cup_i R_{a,i}$
correspond to the same physically meaningful
macroscopic description, labeled by $a \in \I$. 

Suppose now that $x$ is a function from 
$\I$ to $\mathcal{C}$, with the property that,
$x(a) \in R_a$, for each $a \in \I$. 
For $\omega \in \Omega$ and $t \in \R$, 
set $x_t (\omega) = x(\omega_t)$.
Then, for each $\phi \in \Hp$, 
$(x_t)_{t \in \R}$ is a stochastic process on 
the probability space $(\Omega, \Sigma, \P_\phi)$
(measurability issues are automatically satisfied 
because $\I$ is a discrete space). And from 
(\ref{Born}) we obtain 
\begin{equation} 
\P_\phi (x_{t_1} \in R_{a_1}, ... , 
x_{t_k} \in R_{a_k}) \ = \ 
 || p^{t_k}_{a_k} ... p^{t_1}_{a_1} \hat{\phi} ||^2,
 \label{Bornx}
\end{equation}
for every $t_1 <  ...< t_k$, and $a_1, ..., a_k$. (Actually we obtain (\ref{Bornx}), 
under these assumptions, for every 
$t_1, ..., t_k$. But for our purposes 
in this subsection and subsequent ones, 
when we quote (\ref{Bornx}) we mean it 
with the times in the stated order.) 

Since a point in $\mathcal{C}$ specifies how many
particles of each kind are present, and where they 
are located, one can see $(x_t)_{t \in \R}$ as 
describing particles in physical space moving and 
being created and annihilated. This all 
happening in fashions that, 
through $\P_\phi$, are guided by the wave
function $\phi$ and the unitary group 
$(U_t)_{t \in \R}$. 

The argument above is one of existence of 
processes $(x_t)_{t \in \R}$ with the 
described properties. From that
construction, it is clear that
uniqueness 
is not at all true. And unfortunately, 
it is not clear what 
properties, including Markovianity, 
smoothness properties of the paths, etc, 
a process $(x_t)_{t \in \R}$ that
satisfies (\ref{Bornx}) may,
or may not have. 

Compare the construction above with the more 
standard pilot-wave theories, including the 
paradigmatic Bohmian mechanics. Those are 
usually Markovian and have continuous paths, 
except when particles are created or annihilated.
But in those, one usually is satisfied with 
a weaker  
condition than (\ref{Bornx}), namely:
\begin{equation} 
\P_\phi (x_{t} \in R_{a}) \ = \ 
|| p^t_a \hat{\phi} ||^2,
 \label{equivariance}
\end{equation}
for every $t$ and $a$.
To prove (\ref{equivariance}), one usually 
shows a property called ``equivariance'', 
which states
that if (\ref{equivariance})
holds at one time, then it holds at any 
other time.
One then assumes that it holds at one given
time,
sometimes with the support of 
some plausibility argument.
There is also a competing idea, that 
(\ref{equivariance}) was not always true 
in our universe, but that it is 
a sort of equilibrium condition, that 
resulted from good mixing properties 
of the underlying pilot-wave process.
The first of these two approaches
to (\ref{equivariance}) appears 
in most of the papers on pilot-wave 
theories listed in the introduction. 
The second view is defended in 
\cite{Valentini}. For  
appraisals of both approaches, see
\cite{Callender} and \cite{Norsen}.

One should point out that in 
typical pilot-wave theories, the partition 
of the configuration space into the sets
$R_a$ so that (\ref{equivariance}) holds
can be fairly arbitrary, with only 
measurability requirements being necessary.
And $\phi$ can then typically be chosen 
arbitrarily from a dense, linearly 
closed, subset of the full 
Hilbert space $\H$, not just $\Hp$.

The expression ``fully equivalent to 
quantum mechanics'' in the title of this
section refers to pilot-wave theories 
that satisfy (\ref{Bornx}), rather than 
simply satisfying (\ref{equivariance}). 

There are arguments,
related to the idea of an ``effective 
collapse of the wave function'', 
that suggest that Bohmian 
mechanics may satisfy the full (\ref{Bornx}),
at least approximately (see, e.g., Section 9.2 
of \cite{DT}, 
Section 5.1.6 of \cite{Bricmont}, and 
Section 8 
of \cite{Goldstein}). But it seems that whether
exact agreement with this equation holds for 
Bohmian mechanics is an open question. 

Lack of full agreement with quantum mechanics 
in the sense discussed here was one of the
criticisms of stochastic mechanics (another 
well known pilot-wave theory) by its own 
first developer, in Section 10.2 of \cite{Nelson1},
and Section 5 of \cite{Nelson2}.

As pointed out in \cite{Bell}, Section 5, 
it is easy to produce stochastic processes 
that satisfy (\ref{equivariance}),
but do not satisfy 
(\ref{Bornx}). 
For instance one can take a point
$x_t \in \mathcal{C}$ at each time $t \in \R$ 
independently of anything else, with probability
$\P_\phi(x_t \in R_a) = ||p^t_a \hat\phi||^2$. 

An important philosophical question that arises 
is if (\ref{equivariance})
should be considered sufficient 
to make a pilot-wave theory plausible. 
The point, 
made in \cite{Bell}, \cite{Nelson1} and 
\cite{Nelson2}, is that 
if we had (\ref{equivariance}) 
we would not be able to perceive that we do not 
have the full (\ref{Bornx}), based on 
experiments. We could nevertheless have incorrect 
records (including those in our brains)
of our true history. 
Think of the example in the last paragraph,
for a dramatic case of complete 
lack of correlation across time, 
and in particular between memories and 
true pasts.
Similarly, the models introduced in 
\cite{Davidson} are diffusions with 
arbitrarily large diffusion coefficients, 
and will show very low correlation 
between memories and true pasts when this 
coefficient is large, despite the paths 
being continuous.
If one is not bothered by this, then one can simply 
propose the independent choices of $x_t$ at different
times as a satisfactory interpretation of quantum 
mechanics. But if one finds this possibility 
unacceptable, as emphasized in 
\cite{Bell}, \cite{Nelson1} and 
\cite{Nelson2} then 
one should ask which pilot-wave theories 
satisfy (\ref{Bornx}). 
(We should observe that, due to tunneling, 
quantum mechanics may produce false records 
of the past. What (\ref{Bornx}) entails 
is that the correlations between the records 
and the true past are given correctly by the 
quantum dynamics, and not modified by 
additional phenomena pertaining to the 
pilot-wave theory, as in the examples given 
in this paragraph.) 

A second question is whether approximately 
satisfying (\ref{Bornx}), 
which may turn out to be the case for 
Bohmian mechanics, should be considered 
philosophically satisfactory. 
And what one then
means by a satisfactory approximate
fulfilment of this condition. 

A most interesting mathematical 
question stressed and left open here is 
whether Bohmian mechanics satisfies (\ref{Bornx}) exactly, for the kind of 
$\pi$ discussed in
Subsection~\ref{subsec:particles}, 
with the corresponding sets $R_a$ 
corresponding to certain macroscopic  
descriptions labeled by $a \in \I$. 
Note that since Bohmian mechanics 
can be defined for all $\phi$ 
in the domain of the Hamiltonian, 
which is a dense
subset of $L^2(\R^{3n})$, 
Theorem~\ref{thm:converse} 
would imply, if the answer is positive,
that $\Hp=\H = L^2(\R^{3n})$ in this case.
And from part (b) of 
Theorem~\ref{thm:basic}, we would then learn
that there is a pilot-wave theory that 
satisfies (\ref{Bornx}) for all 
$\phi \in  \H = L^2(\R^{3n})$.

A related important open problem is how regular 
the paths of pilot-wave theories that satisfy
(\ref{Bornx}) can be. Can they be continuous in the 
case in which particles are not created or 
annihilated? Can they be continuous from one 
side, with limits from the other when
particles 
can be created and annihilated? 

Especially in view of
Theorem~\ref{thm:converse}, 
one can ask what is the value of 
having (\ref{Bornx}), that applies 
to $\phi \in \Hp$, if it turns out
that we live in a universe that is in 
a Heisenberg-picture 
state $\Psi \not\in \Hp$, 
for the relevant $\pi$. We will 
answer this question in 
Subsection~\ref{subsec:comparison}, 
where, building on previous subsections,
we will propose that a pilot-wave 
theory that satisfies (\ref{Bornx}) 
with $\phi = \Psi_{\pi}$ should be a 
good candidate for an interpretation of 
quantum mechanics. 

\subsection{When $S$ is finite}
\label{subsec:finiteS}
When $S = \{t_1,...,t_K\}$ is a finite set, there are 
major simplifications to many of the proofs in this
paper. 

In this case $\Omega = \I(t_1) \times ... \times \I(t_K)$ is countable, and $\Sigma$ = $\A$ contain 
all the subsets of $\Omega$. Also 
$$
\Hp'' \ = \ \H^{t_1, ..., t_K} 
\ = \ \bigoplus_{a_1, ..., a_K} \, \H^{t_1, ..., t_K}_{a_1, ..., a_K}
\ = \ 
\mbox{Range} \, \left\{ 
\sum_{a_1,...,a_K} \, 
p^{t_1, ..., t_K}_{a_1, ..., a_K}
\right\},
$$
and, using (\ref{monotone}), (\ref{perp}) and the 
fact that a sum of orthogonal vectors can only be 0
if all these vectors are 0, 
\begin{eqnarray*}
N  & \ = \ &  
\{ \phi \in \H \, : \, 
p^{t_1, ..., t_K}_{a_1, ..., a_K} \phi = 0 \ 
\mbox{for all} \ a_1, ..., a_K \}
\\ & \ = \ &  
\left\{ \phi \in \H \, : \, 
\sum_{a_1,...,a_K} \,
p^{t_1, ..., t_K}_{a_1, ..., a_K} \phi = 0 \right\}
\ = \ 
\mbox{Kernel} \, \left\{ 
\sum_{a_1,...,a_K} \, 
p^{t_1, ..., t_K}_{a_1, ..., a_K}
\right\}.
\end{eqnarray*}
Therefore it is immediate that $N$ is a vector
space that, in this case,
is topologically closed, i.e., $N = \overline{N}$, 
and that we have $\Hp'' = N^{\perp}$. 

I am not aware of any other simplification in the
proof of part (a) of Theorem~\ref{thm:basic}. But 
it is worth pointing out that once one has 
proved these statements in case $S$ is finite, 
the general case follows simply by taking 
intersections over $\{t_1,...,t_k\}$.

Parts (b), (c), (d) and (f) of 
Theorem~\ref{thm:basic}, as well as parts (a)
and (c) of 
Theorem~\ref{thm:F} are greatly simplified.
And, as with $N$, also the sets $F_A$ are 
topologically closed. 

One can start by defining, 
for $A \in \Sigma$, 

$$
\H_A \ = \ 
\bigoplus_{(a_1, ..., a_K) \in A}
\, \H^{t_1, ..., t_K}_{a_1, ..., a_K}
\ = \ 
\mbox{Range} \, \left\{ 
\sum_{(a_1,...,a_K) \in A} \, 
p^{t_1, ..., t_K}_{a_1, ..., a_K}
\right\}.
$$
And noting, again 
using (\ref{monotone}), (\ref{perp}) and the 
fact that a sum of orthogonal vectors can only be 0
if all these vectors are 0, that for all 
$A \in \Sigma$,  
\begin{eqnarray*}
F_A  & \ = \ &  
\{ \phi \in \H \, : \, 
p^{t_1, ..., t_K}_{a_1, ..., a_K} \phi = 0 \ 
\mbox{for all} \ (a_1, ..., a_K) \in A \}
\\ & \ = \ &  
\left\{ \phi \in \H \, : \, 
\sum_{(a_1,...,a_K) \in A} \,
p^{t_1, ..., t_K}_{a_1, ..., a_K} \phi = 0 \right\}
\ = \ 
\mbox{Kernel} \, \left\{ 
\sum_{(a_1,...,a_K) \in A} \, 
p^{t_1, ..., t_K}_{a_1, ..., a_K}
\right\}.
\end{eqnarray*}
It is then easy to check that 
$\{\H_A : A \in \Sigma\}$ has the 
properties claimed in part (c) of 
Theorem~\ref{thm:basic}.
And it is immediate that $F_A = \overline{F_A}$
and that $F_A^{\perp} = \H_A$, as stated in 
part (a) of Theorem~\ref{thm:F}. 
Because $F_A$ is closed, part (c) of 
Theorem~\ref{thm:F} now reads 
$\H_A = F_{A^c} \cap \Hp$, and it follows 
immediately from part (b) of that theorem 
and (\ref{Hsplit}) in part (c) of 
Theorem~\ref{thm:basic}. 

If one now recalls that $p_A$ is the projection 
on $\H_A$ and defines, for 
$\phi \in \Hp \backslash \{0\}$ and $A \in \Sigma$,  
$$
\P_\phi (A) \ = \ ||p_A \hat{\phi}||^2, 
$$
then it is easy to check that $\P_\phi$ has 
the properties claimed in 
part (b) of Theorem~\ref{thm:basic}, 
and that the second claim in part (d) 
of that theorem also holds. 

Finally, for part (f) of 
Theorem~\ref{thm:basic}, given $f: \Omega \to \R$, 
we can simply set 
$$
Q_f \ = \ \sum_{\omega \in \Omega} \, 
f(\omega) p_{\{\omega\}},  
$$
with domain $\mathcal{D}_f$ as defined there, 
and check easily the required properties. 

\subsection{Do we need to consider infinite $S$? 
Uncountable $S$? Infinite $\I(t)$? }
\label{subsec:needof}
In light of the remarks in 
Subsection~\ref{subsec:finiteS}
it is natural to ask what is gained, as far as
applications to foundations of quantum mechanics are 
at stake, from considering infinite $S$. One 
important reason
for considering countably infinite sets $S$ is to 
be able to use the limit theorems of probability
theory, like the strong law of large numbers,  
that apply to idealized settings with infinitely 
many random variables. Those would correspond, for
instance, to idealized sequences of experiments.

Less clear is if, for the sake of physics, there is 
a need for considering uncountably large sets $S$. An 
argument in favor is in the fact that we usually 
consider physical time to be a real number, so that 
we should consider $S = \R$ as our fundamental 
setting. But is there really a reason for thinking 
that physical time is not limited to rational values?
And that the real line comes in simply as a
mathematical tool, providing completeness in the 
mathematical sense as a convenience, but not an 
additional physical reality? This is an
interesting philosophical issue that 
will not affect the applicability of the results 
in this paper in situations in which the sets 
$\I(t)$, $t \in S$, of interest are all finite,
thanks to the results in
Section~\ref{sec:Farbitrary}.

And this raises the question whether there is 
any reason for considering infinite $\I(t)$ in 
applications to foundations of quantum 
mechanics. In applications of the kind proposed
in Subsection~\ref{subsec:particles}, when 
the number of particles in the universe 
is fixed (so that the configuration 
space is $\mathcal{C} = \R^{3n}$), there should 
be only a finite number of sets $R_a$ that are 
macroscopically distinguishable from each other
and meaningful to us. After all, in such a 
universe, there can only be a finite number of 
computational devices (including human brains), 
each one capable of holding some finite number of 
distinct computational states. Even if particles 
can be created, energy considerations may limit 
the number of particles and hence the number 
of bits that all the computers (including 
our brains) can hold. 

In any case, we will see in
Subsection~\ref{subsec:Born-superposition}
that the partial results obtained in case 
of infinite $\I(t)$ and uncountable $S$ in 
Section~\ref{sec:Farbitrary} are 
sufficient to 
draw the conclusion that, if we accept
certain intuitive assumptions, 
then events for which we compute
Born-probability 0
should not happen, even if 
$S$ is uncountable and $\I$ is infinite. 
What is 
currently missing in this case is 
the converse.
So we have not ruled out that, 
in this case, there could be 
events of positive probability that will not 
happen. 

\subsection{Can $F_A$ in Section~\ref{sec:F}
be replaced with a set $N_A$
that provides more uniformity in time?}
\label{subsec:uniformity}

For each $A \subset \Omega$, define 
\begin{equation}
N_A \ = \ \left\{\phi \in \H \, : \, 
\mbox{for some $t_1, ...t_k$, \, 
$p^{t_1, ..., t_k}_{\omega_{t_1}, ..., \omega_{t_k}} \phi = 0$ \, for all $\omega \in A$ } \right\}.
\label{N_A}
\end{equation}
 
Note that $N_\Omega = N$, and $N_A \subset F_A$. 
The difference between $N_A$ and $F_A$ is the 
extra uniformity, with respect to $\omega \in A$, in the choice of $t_1, ..., t_k$ in $N_A$. It is 
natural to ask if in Section~\ref{sec:F} we could
have used $N_A$ instead of $F_A$, and in particular
whether in part (b) of Theorem~\ref{thm:F}, which is directly related to 
interpretation, we could 
replace $F_A$ with $N_A$. 

The answer is that in some parts of 
Section~\ref{sec:F} we can make this replacement,
but not in others that include 
Theorem~\ref{thm:F}. This discussion  
highlights some of the technical details of 
the proofs in that section. 

In Lemmas \ref{lemmaA}, \ref{lemmaB} and \ref{lemmaC} we can indeed replace $F_A$ with $N_A$, keeping the 
same proofs, as the reader can check. 
In the case of Lemma~\ref{lemmaC}, this can 
also be understood more quickly by observing that 
from (\ref{HAc}), we have 
$$
\H_{A^c} \ \subset  \ N_A  \cap \Hp \
\subset
\ F_A \cap \Hp = \H_{A^c},
$$
where the last step is the statement of
Lemma~\ref{lemmaC}.

But the proof of Lemma~\ref{lemmaD}  
fails if we replace $F_A$ with $N_A$. 
In this proof we are using the only statement of
Proposition~\ref{prop:F} in which $F_A$ cannot
be replaced with $N_A$. For an infinite family of 
subsets of $\Omega$, $\{A_\alpha\}$, 
$N_{\cup_\alpha A_\alpha}$ is, in general, 
not equal to $\cap N_{A_\alpha}$, because of 
loss of uniformity. 

The following example shows that the problem is 
not only with the proof, but with the conclusion
in this lemma, which is a special case of part (a)
of Theorem~\ref{thm:F}. 

Suppose that $S = \{t_1, t_2, ...  \}$. 
For $i=1, 2, ...$, choose $G_i \subset
\I(t_i)$, \, $G_i \not= \emptyset$, \, 
and set $A_i = \{X_{t_i} \in G_i\}$, \,
$A = \cup_{i=1}^\infty A_i$. It is clear that 
for each $i$, $A_i \in \A$, and hence 
$A \in \A_\sigma$. Now, using (\ref{monotone})
(as in (\ref{remark})), 
\begin{eqnarray*}
N_A \ &  = & \
\bigcup_{k=1}^\infty \, \left\{
\phi \in \H \, : \, 
p^{t_1, ..., t_k}_{a_1, ..., a_k} \phi = 0, 
\ \mbox{for all} \ (a_1,a_2, ...) \in A
\right\}
\\ & = & \ 
\bigcup_{k=1}^\infty \, \left\{
\phi \in \H \, : \, 
p^{t_1, ..., t_k}_{a_1, ..., a_k} \phi = 0, 
\ \mbox{for all} \ (a_1, ..., a_k)
\right\}
\ = \ N.
\end{eqnarray*}
In the second equality, we used the fact that for
any $(a_1, ..., a_k)$ there is
some $a_{k+1} \in G_{k+1} \subset \I(t_{k+1})$, 
such that 
$(a_1, ..., a_k, a_{k+1}, ...) 
\in A_{k+1} \subset A$.

If we could replace $F_A$ with $N_A$ in the 
statement of Lemma~\ref{lemmaD}, or part (a) 
of Theorem~\ref{thm:F}, we would then have 
$$
\H_A \ = \ N_A^{\perp} \ = \ N^\perp \ = \ \Hp, 
$$
where the last equality is from part (a) of
Theorem~\ref{thm:basic}.
In particular, for every $\phi \in \Hp \backslash 
\{0\}$, we would have 
$\P_\phi(A) = ||p_A \hat{\phi}||^2 = 1$. This  
is certainly absurd in many applications, 
since the sets $G_i$ can be very small subsets 
of the corresponding $\I(t_i)$, only assumed to 
be non-empty above. 

For a counter-example, 
let $\H = L^2([0,1])$, \,  $S = \{ 1, 2, ...\}$, 
\, $\I(t) = \{1,2\}$ and 
$p^t_1$ be defined by 
$(p^t_1 \phi)(x) = I_{[0,3^{-t}]}(x) \phi(x)$,
where (as before in this paper) $I_R(x) = 1$, 
if $x \in R$ and 0 otherwise. 
By necessity, $p^t_2 = I - p^t_1$.
In this setting, $\H^t_a \subset \H^s_a$,  
whenever $s < t$, and this implies 
$p^t_1 p^s_1 = p^s_1 p^t_1 = p^t_1$. 
It follows that $\Hp = \H$. 

Take $G_i = \{1\}$ for each $i$. Then, for 
$\phi$ defined by $\phi(x) = 1$, we have 
$$
\P_\phi(A) \ \leq \ \sum_{i=1}^\infty \P_\phi(A_i) 
\ = \ 
\sum_{i=1}^\infty \P_\phi(X_i = 1) 
\ = \ 
 \sum_{i=1}^\infty ||p^i_1 \hat{\phi} ||^2
\ = \
\sum_{i=1}^\infty \, \left( \frac{1}{3} \right)^i
\ < \ 1. 
$$

It is worth pointing out that the
counter-example above has $\I(t)$
finite for all $t \in S$, so that 
this extra assumption (as made in 
parts of Section~\ref{sec:Farbitrary})
would not change the conclusion here.

\subsection{Can we eliminate the topological
closure of $F_A$ in part (b) of Theorem~\ref{thm:F}, or part (d) of 
Theorem~\ref{thm:Farbitrary}?}
\label{subsec:noclosure}
It is natural to ask if we can replace
$\overline{F_A}$ with $F_A$ in part (b) of 
Theorem~\ref{thm:F},
or part (d) of Theorem~\ref{thm:Farbitrary}.
This is important for 
interpretations of quantum mechanics, as 
the condition $\Psi \in F_A$ can naturally 
be proposed to imply 
that if the state of the 
universe is $\Psi$, then the event $A$ 
should not be part of our experiences.
But that the condition $\Psi \in \overline{F_A}$
should also have this implication is a  
more delicate philosophical issue. In \cite{Sch1} and 
\cite{Sch2} this lead to the consideration 
of a version of the superposition principle 
to reach this conclusion.  
(See Subsection~\ref{subsec:superposition} below.)

A simple argument, though, shows that in 
relevant situations the closure of $F_A$ 
is needed to make part (b) of 
Theorem~\ref{thm:F} 
and part (d) of 
Theorem~\ref{thm:Farbitrary} true.
As observed in 
Proposition~\ref{prop:F}, for 
any family of events $A_\alpha \in \Sigma$, 
if $\phi \in F_{A_\alpha}$, for each $\alpha$,
then $\phi \in F_{\cup_\alpha A_\alpha}$.
But $\P_\phi (A_\alpha) = 0$, for all $\alpha$
does not imply that 
$\P_\phi (\cup_\alpha A_\alpha) = 0$, 
unless this family of events is countable.

\subsection{$\overline{F_A}$, \, 
$\overline{N_A}$ \, and 
superposition of states}
\label{subsec:superposition}

We recall now how in \cite{Sch1} and 
\cite{Sch2} the statement $\Psi \in \overline{F_A}$
was translated into the 
statement that $\Psi$ is a 
superposition of states in $F_A$.
 $\Psi \in \overline{F_A}$ means that there are 
 $\zeta_1, \zeta_2, ...$ such that $\zeta_i \in F_A$
 and $\zeta_i \to \Psi$, as $i \to \infty$.
 Equivalently, $\zeta_1 + (\zeta_2-\zeta_1) +
 (\zeta_3 - \zeta_2) + ...$ converges to $\Psi$.
 We can apply the Gram-Schmidt orthonormalization 
 procedure (see p.46 of \cite{RS}, or p.167 of
 \cite{Folland}) to the vectors $\zeta_1,
 \zeta_2-\zeta_1, \zeta_3 - \zeta_2, ...$ to produce
 a sequence of orthonormalized  
 vectors $\eta_1, \eta_2, ...$ that 
 have the same closed span,
 to which $\Psi$ belongs. 
 Since $F_A$ is a vector 
 space and $\zeta_i \in F_A$, this procedure
 (which only involves linear operations)
 gives us that also $\eta_i \in F_A$. 
 Set $\Psi_i 
 = \langle \Psi, \eta_i \rangle \, \eta_i$, 
 $i=1,2, ...$.
 Then the vectors $\Psi_i$ are orthogonal to
 each other, $\Psi_i \in F_A$ for each $i=1,2, ...$
 and $\sum_{i=1}^\infty \Psi_i = \Psi$. 
 A converse statement is trivial, any convergent 
 series of vectors in $F_A$ converges to a vector 
 in $\overline{F_A}$. 
 
 Referring to $N_A$, as defined by 
 (\ref{N_A}), since those are also 
 vector spaces, a similar derivation applies 
 to $\overline{N_A}$. 
 
 In words, using common quantum-mechanics
 jargon: Belonging 
 to $\overline{F_A}$ ({\it resp.} 
 $\overline{N_A}$) is the 
 same as being a superposition of 
 orthogonal states in $F_A$
 ({\it resp.} $N_A$). 
 
 The version of the superposition  
 principle proposed in \cite{Sch1} and 
 \cite{Sch2} can be rephrased,
 replacing {\it prediction} with {\it ontology},
 in the following 
 fashion. Here all the mentioned universes
 are supposed to be described by the same 
 Hilbert space $\H$ and group of unitary 
 evolution operators $\{U_t\}$, and their
 state is given in the Heisenberg picture 
 by an element of $\H$. 
 
 {\bf One-Sided Superposition 
 Principle:} {\it If an event $A$ is 
 not realized  
 in universes that are
 in states $\Psi_i$, $i=1,2,...$, then 
 it is also not 
 realized in a  
 universe in state $\Psi = \sum_{i=1}^\infty
 \Psi_i$.}
 
 The reason for the title of ``one-sided 
 superposition principle'' is that if we remove
 the word ``not'' in the two places that
 it appears, we obtain a statement that is 
 certainly false, due to interference. 
 Superpositions cannot create new realities, 
 but they can eliminate realities by interference.
 
 If we accept the idea that in a universe in 
 a state $\Phi \in F_A$ 
 ({\it resp.} $\Phi \in N_A$), 
 the event $A$ is not 
realized and accept also the one-sided 
 superposition principle, then we conclude that 
 the same is the case in a universe 
 in a state 
 $\Psi \in \overline{F_A}$
 ({\it resp.} $\Psi \in \overline{N_A}$).
 
 In particular, if we accept
 the one-sided superposition principle 
 and the idea that in a universe in 
 a state $\Phi \in N$
 no event in $\Sigma$ 
 is realized,  
 then  
 we conclude that no event in $\Sigma$ 
is 
realized in a universe in a state
$\Psi \in \overline{N} = \Hp^\perp$.

\subsection{Should we believe that 
in our universe $\Hp \not= \{0$\}, and 
$\Psi_\pi \not= 0$? The role of decoherence, the 
``we-are-here'' argument, and the 
ordinary nature 
of the present time on a cosmological scale}
\label{subsec:decoherence}

In this subsection we are considering 
one of the 
particle models of 
Subsection~\ref{subsec:particles} as a 
model for our universe. And we are considering the 
sort of $\pi$ discussed there, associated to
a partition of the configuration space according to
sets with macroscopically meaningful descriptions.
But we should make one modification in 
how $\pi$ is chosen, because we are 
only interested in times that, on a 
cosmological scale, are not too early nor 
too late. For this reason we will assume
$S = (t_-,t_+)$, where $-\infty < t_-
< s < t_+ < \infty$, with $s$ being the 
present moment, and the differences 
$s-t_-$ and $t_+ - s$ being of the 
order of cosmological times. 

It is natural to ask whether, for 
some values of $t_-$ and $t_+$ as 
above, we should believe that 
$\Hp \not= \{0\}$, and more specifically 
$\Psi_\pi \not= 0$, where $\Psi$ is the
Heisenberg-picture 
state of our model universe. 

There are three complementary ideas to discuss. 

The first one 
is the role of environmental decoherence,
\cite{JZKGKS}, 
\cite{Zurek}, \cite{Omnes}, \cite{GMH}, 
\cite{Schlosshauer},
\cite{Bacciagaluppi}. The subsets $R_{a,i}$ into 
which $\mathcal{C}$ is partitioned (where $a \in \I$ 
and $i$ specifies how many particles of each kind 
are present) correspond to macroscopically meaningful
descriptions (labeled by $a$), 
and therefore involve large numbers of 
particles, that should interact with the environment
producing records of the history. 
Now, these environmental particles that produce 
records should also be described by the vector 
$\Psi$. And it may happen that $\Psi$ includes 
components on which such environmental memories 
do not form. Let us therefore 
leave $\Psi$ aside for the moment, but assume that 
there is 
$\Phi \in \H \backslash \{0\}$ 
that supports a rich enough
environment such that all events pertaining to $\pi$ 
are recorded in this environment. 
This means that 
if $t_- < t_1 < t_2 ...< t_k \leq t < t_+$, then $\Phi_t =
U_t \Phi$ should decompose as 
$$
\Phi_t \ = \ \sum_{a_1,...,a_k} \, 
U_t \, \Phi^{t_1, ..., t_k}_{a_1, ..., a_k},
$$
where $\Phi^{t_1, ..., t_k}_{a_1, ..., a_k}
\in \H^{t_1, ..., t_k}_{a_1, ..., a_k}$, 
for each $a_1, ..., a_k$.
Therefore $p^{t_1, ..., t_k}_{a_1, ..., a_k} \Phi
= p^{t_1, ..., t_k}_{a_1, ..., a_k}
U_{-t} \Phi_t 
= \Phi^{t_1, ..., t_k}_{a_1, ..., a_k}$.
And
$$
\sum_{a_1,...,a_k} \, 
p^{t_1, ..., t_k}_{a_1, ..., a_k} \,
\Phi
\ = \ 
\sum_{a_1,...,a_k} \, 
\Phi^{t_1, ..., t_k}_{a_1, ..., a_k}
\ = \ U_{-t} \Phi_t \ = \ \Phi. 
$$ 
This means that $\Phi \in \Hp''= \Hp$, where 
we used part (a) of Theorem~\ref{thm:basic}. 

In conclusion: the 
assumption that we have enough 
decoherence in our universe
that such a $\Phi$ as above exists
implies $\Hp \not= \{0\}$. And $\Hp$ 
should be the set of elements in $\H$ that support
lasting environmental memories of all 
events associated to $\pi$. 

The second idea, to which we turn now, 
will explain why we may believe in a 
statement that is related to, but weaker than
$\Psi_\pi \not= 0$.
To explain what this weaker 
statement is, we define, for 
$s \in (t_-,t_+)$, 
the following coarsening of $\pi$:
$$
\pi(s) \ = \ \{p^t_a \, : \, 
t_- < t \leq s, \, a \in
\I(t) \}.
$$
Clearly $\H_{\pi(s)}$ decreases to $\Hp$, as $s$
increases to $t_+$. 
We will also use the self-explanatory
notation $\Omega(s)$, $\Sigma(s)$ and  
$$
N(s) \ = \ 
\left\{ \phi \in \H \, : \, \mbox{ for some $t_1,...,t_k \in (t_-, s]$,  \ 
$p^{t_1, ...,t_k}_{a_1,...,a_k} \phi = 0$ \ 
for all 
$a_1, ..., a_k$}
\right\}.
$$
Suppose that $s$ is the present moment.
It seems very reasonable to believe that 
if $\Psi \in N(s)$, then no $A \in \Sigma(s)$
would be realized.
After all, in this case there are 
$t_1, ..., t_k \in (t_-,s]$ such that 
$\Psi \perp \H^{t_1, ...,t_k}_{a_1,...,a_k}$,
for each $a_1, ..., a_k$. 
And this should mean that each event 
$\{X_{t_1} = a_1, ..., X_{t_k} = a_k\}$
should not be realized. But the union of 
these events over $a_1, ..., a_k$ 
is $\Omega(s)$ and so contains any 
$A \in \Sigma(s)$.

Now, if $\Psi_{\pi(s)} = 0$, then 
$\Psi \in 
\H_{\pi(s)}^\perp = \overline{N(s)}$, 
where we used part (a) 
of Theorem~\ref{thm:basic}, applied to $\pi(s)$. 
If we accept the 
one-sided superposition principle stated in 
Subsection~\ref{subsec:superposition},
we should conclude, as explained at the 
end of that subsection, that all 
our experiences up to the present time
would not be realized. 
And since we are here,
an have one experience or another, 
we should believe that 
$\Psi_{\pi(s)} \not= 0$.
This is the weaker statement alluded to above.

Should we upgrade this belief to the 
belief that $\Psi_\pi \not=0$? 
For this we apply a third idea.
One argument in this direction 
evokes the absence of anything special about
the present moment, as compared to other times 
that are of the same order of magnitude 
in a cosmological 
scale. If $\Psi$ includes a component with a 
rich enough environment to allow for 
$\Psi_{\pi(s)} \not=0$, when $s$ is the present 
moment, we should expect 
$\Psi_{\pi(s')} \not= 0$ for at least $s'$ 
larger than $s$ by a cosmological extension of time. 
And this perhaps is the most that we can argue for
and believe.  
And it is certainly good enough for practical purposes. 

What would it mean if $\Psi_{\pi(s')} = 0$ 
at some future time? The one-sided superposition 
principle would then imply that we (in the 
way we understand ourselves, with the kind of 
possible experiences labeled by $\I$) would not 
be part of this universe after time $s'$. 
And perhaps this is the way things will be
in a cosmological time in which the universe 
will look very different from its present state.

In this scenario, 
whatever event in $\Sigma$
we would have predicted not to happen after 
time $s'$ will indeed not happen. But if we assumed 
$\Psi_\pi \not= 0$ when making predictions, 
we would have erred in the opposite
direction, incorrectly predicting events to happen
that will actually not happen. 
In this scenario, as our own existence would not go 
beyond time $s'$, we would not be there to 
realize that we were wrong. 

\subsection{A minimalistic ontology for non-collapse
quantum mechanics}
\label{subsec:ontology}
The mathematical results in this paper support a 
minimalistic ontology for non-collapse 
quantum mechanics, that 
conforms with our experiences, including our 
perceptions of apparent collapses of the wave 
function according to Born's rule. This ontology 
is also compatible with the one-sided superposition
principle of Subsection~\ref{subsec:superposition}.

In this subsection we will build the 
theory based 
only on the mathematical results in 
Section~\ref{sec:basic results}. 
In the next two, we will further elaborate
on this theory, using also notions from 
Section~\ref{sec:refinements}. 
And in 
Subsection~\ref{subsec:P1P2P3} 
we will see how this ontology 
relates to the results in Sections
\ref{sec:F} and \ref{sec:Farbitrary}. 

In this ontology, the primary 
physical reality is limited to
a vector $\Psi$ which belongs to a Hilbert space $\H$
and a strongly continuous group of unitary operators 
on $\H$, $(U_t)_{t \in T}$, where  
$T = \R$, or $T = \Q$. 
In the latter case we are 
assuming that only rational times have physical 
meaning, as discussed in 
Subsection~\ref{subsec:needof}. 
We could also entertain the idea of 
assuming $T = \epsilon \N$, 
where $\epsilon$ is a time interval shorter than 
anything that we can (presently) measure. 

At this point it is important to explain what we 
mean by ``primary reality'' and how it differs 
from the broader use of ``reality'' below.
For a good illustration of the distinction 
consider the concept of cellular 
automata, as the well known Game of 
Life, \cite{GL}.
The primary reality is limited to a grid,
a deterministic 
updating rule in discrete time and an initial
configuration of alive and dead cells of
the grid. (Those are analogous, respectively, to
our $\H$, $(U_t)_{t \in T}$ and $\Psi$.)
But in addition to this primary reality, there are
patterns of alive and dead cells that develop 
and propagate in time. And this is actually the 
reason for the interest in the model.
In particular because these propagating
patterns can produce the same  
computations as a Turing machine. 
In our terminology, such patterns are elements of 
the ``derived reality'', or simply 
``reality'' of the system. If the grid and the 
deterministic 
updating rule are fixed, we may regard the 
patterns that develop and propagate 
in time as features of the
initial configuration. 

We need to propose a theory about the nature of our 
experiences, compatible with the primary 
quantum ontology 
proposed above, and with the fact that these 
experiences are well described by textbook quantum 
mechanics (with collapse according to Born's rule).
Theorem~\ref{thm:basic} and the analogy above 
suggest an answer: Our 
experiences are 
in one-to-one correspondence 
with a class of patterns in $\Psi$.

We should think of each possible $\pi$, with 
$S \subset T$, as a 
tool for analysing the features of $\Psi$. 
For this reason, we will call each such $\pi$ an
``analyser''. Given such an analyser, we have its 
associated sets $\Omega$ and $\Sigma$. And 
Theorem~\ref{thm:basic} provides us with 
$\{ p_A : A \in \Sigma \} $, which is a projection
valued measure on 
$\Hp$. Given $A \in \Sigma$, we say that $A$ is
a $\pi$-pattern in $\Psi$ if $p_A \Psi \not= 0$.

The proposal is to regard any $\pi$-pattern $A$ 
for any analyser $\pi$
as part of the reality defined
by (or derived from) $\Psi$. 
The corresponding postulate is:

\noindent {\bf Ontological Postulate:}
{\it For any analyser $\pi$ and any $A \in \Sigma$,  
}
\begin{equation}
\mbox{$A$ is part of reality} \ \ \ 
\Longleftrightarrow \ \ \ 
p_A \Psi \not= 0,
\label{axiom}
\end{equation}
\label{postulate:1}

And the idea is that our experiences are 
in one-to-one correspondence with the  
$\pi$-patterns of $\Psi$ for an appropriate $\pi$. 
In short: that our experiences are $\pi$-patterns 
of $\Psi$, for a certain $\pi$. 

There are several interesting 
aspects of such a theory. 

First, it satisfies the one-sided superposition
principle of Subsection~\ref{subsec:superposition}:
$$
\mbox{If \, $p_A \Psi_i = 0$, \, for $i=1,2, ...$, 
\, then \, $p_A \left( \sum_i \Psi_i\right) = 0$. }
$$

Second, 
\begin{equation}   
p_A \Psi \not= 0  \ \ \ 
\Longleftrightarrow
\ \ \  p_A \Psi_\pi \not= 0, 
\label{axiomPno}
\end{equation}
so that, for the relevant $\pi$, 
our experiences are only affected by 
$\Psi_\pi$, not by what $\Psi - \Psi_\pi$ may be.
As a consequence, we have no information, through 
our experiences, of what $\Psi - \Psi_\pi$ is. 
For us, it is as if the 
Heisenberg-picture 
state of the universe were 
$\Psi_\pi$, rather than $\Psi$.
And since we have experiences, it must be the case
that $\Psi_\pi \not= 0$.

Third, Theorem~\ref{thm:basic} implies that, 
if $\Psi_\pi \not= 0$, then (\ref{axiom}) is 
equivalent to 
\begin{equation}
\mbox{$A$ is part of reality}
\ \ \ 
\Longleftrightarrow \ \ \ 
\P_{\Psi_\pi}(A) \not= 0.
\label{axiomP}
\end{equation}

Now, which $\pi$ is relevant in describing 
our human experiences? We will start with a 
broad proposal, then scrutinize it and 
settle for   
a very precise instance of that proposal as our $\pi$. 

The natural starting point is to assume 
that the setting is one of the particle 
models of Subsection~\ref{subsec:particles}. 
And that $\pi$ is as defined there, with each 
$p_a$ associated to a subset 
$R_a = \cup_i R_{a,i}$ of the 
configuration space $\mathcal{C}$,
which admits a macroscopically meaningful description
to us, labeled by $a \in \I$, and where $i$
gives the number of particles of each type present in
each component $R_{a,i}$ of $R_a$. 
To assure that $\Psi_\pi \not= 0$, 
the set $S$ may need to be limited 
to an interval 
$(t_-, t_+) \cap T$, 
for some $t_-$ that is finite and
significantly smaller 
than the present time on a cosmological
scale, and some $t_+$ that is finite 
and significantly larger 
than the present time on a cosmological
scale,
as explained in
Subsection~\ref{subsec:decoherence}.
And, as emphasized in that subsection, the
physical 
phenomenon responsible for $\Psi_\pi \not= 0$ 
is decoherence. And the meaning of ``macroscopic'' 
in the definition of $\pi$ relates to our sets 
$R_a$ being defined by 
the positions of large numbers of particles that 
allow for environmental decoherence to happen. 

We can now theorize that our experiences are 
the events $A$ that are $\pi$-patterns of $\Psi$,
for a $\pi$ as defined in the last paragraph.

This works well in that 
(\ref{axiomP})
implies that human experiences are precisely the
ones that have positive probability according 
to standard Born-collapse quantum mechanics.
Here it is important to stress that we
perceive the Heisenberg-picture
state of the universe as $\Psi_\pi$, rather 
than $\Psi$, as $\Psi - \Psi_\pi$ does not 
affect us. And that the ``effective state of 
the universe for us'', $\Psi_\pi$, is what we use 
in computing the Born probabilities. 
It is also important to understand that if 
$\P_{\Psi_\pi}(A) > 0$
and also $\P_{\Psi_\pi}(A^c) > 0$, 
then both are 
human experiences, but in the Everettian
sense 
that humans branch, and along each 
branch only 
perceive one of these two events. What is 
accomplished here is to eliminate the naive,
but important, 
criticism of non-collapse quantum mechanics, 
according to which if no collapse happens, then
every event $A \in \Sigma$ would happen. 
Events with $\P_{\Psi_\pi}(A) = 0$ have 
$p_A \Psi = 0$ and, according to the postulate 
above, do not happen in any branch!

There is a way to explain the last statement 
above that may be helpful in convincing  
skeptical readers, who would insist that without 
collapses every $A \in \Sigma$ would happen. 
Consider a fictitious model universe, in which 
nature picks an infinite collection of
independent
realizations of one of the stochastic processes
$(x_t)_{t \in \R}$, defined in 
Subsection~\ref{subsec:pilot-wave}, that satisfy (\ref{Bornx}) with 
$\phi = \Psi_{\pi}$.
In this model universe, we have infinitely many 
trajectories $x_t$ in configuration 
space, and we can think of this ensemble of 
independent trajectories as an 
ensemble of different worlds 
that do not interact with each other. In each 
one there are humans whose experiences are 
identical to those predicted by 
Born-collapse 
quantum mechanics, in a universe in 
state $\Psi_{\pi}$. And since there 
are infinitely 
many of these worlds, each event $A \in \Sigma$ that 
has $\P_{\Psi_\pi}(A) > 0$ is experienced in some
of them (actually in infinitely many of them).
But no event $A \in \Sigma$ that has 
$\P_{\Psi_\pi}(A) = 0$
is experiences in any of them. 
The assumption that 
our experiences correspond to events 
that are $\pi$-patterns of $\Psi$ is equivalent,
thanks to Theorem~\ref{thm:basic}, 
to the statement that our experiences are 
identical to those of the humans in this 
model-infinite-ensemble universe.
Now, imagine that in this 
model-infinite-ensemble universe
a group of scientists is performing 
a sequence of identical independent 
quantum experiments that may each time 
result in outcome 1
with Born-probability 
0.9, or outcome 2 with Born-probability
0.1. In each experiment there are worlds 
in which outcome 1 happens and worlds in
which outcome 2 happens. But in no 
world does the frequency of outcomes 1 
converge to a number different from 
0.9. Having every possible outcome 
in each single experiment
in this universe does not 
imply having every possible outcome in 
infinite series of experiments.  

Another issue that may be raised is whether 
(\ref{axiomP}) captures all the ways in which
Born's-rule-collapse probabilities are used 
in standard collapse quantum mechanics. 
Argumentation answering this question in 
the affirmative is presented in 
Sections 3 and 6 of \cite{Sch1}. 

Now to some essential 
criticism of the kind of choice above of $\pi$.
There is a substantial
amount of subjectivity involved. 
What is ``meaningful''?
Would we all agree that a certain family of
sets $R_{a}$ correspond to each label $a \in \I$
that we describe in a certain way? 
How fine can the partition of the configuration 
space be to still allow decoherence to happen?
Fortunately,
there is a good way to solve these issues, 
if we accept the (currently standard)
view that our perceptions are encoded in the 
physical state of our brains, and that our 
mental processes are in one-to-one 
correspondence with computational processes 
produced as our brains behave according to the
same physical laws that apply to everything else. 

Before returning to humans, it is helpful to 
consider a computer of the kind that we build 
with silicon. In this context, we can
introduce the relevant 
analyser $\pi$ by partitioning the configuration
space into the sets $R_{a}$ labeled by the
computational states of the computer, including 
a set in which the computer is not present in the 
universe and a set in which the computational 
state of the computer may not be well defined. 
Instead of ``experiences that it has'', 
we should talk of ``computations that the computer performs''. If occasionally the computer receives 
bits of input that correspond to outcomes of
quantum experiments, the computer will branch
in an Everettian sense, with each branch continuing 
to compute separately from the others. 
If we accept (\ref{axiom}),
and suppose that, due to decoherence 
(expected, since each bit of information is encoded 
by the state of a very large number of particles)
$\Psi_\pi \not= 0$, 
then, by (\ref{axiomP}), 
events $A \in \Sigma$ (which now pertain to the 
computational processes of this computer) will 
be part of the collective reality of the branching 
versions of the computer when and only when 
$\P_{\Psi_\pi}(A) > 0$. 
The important point that we want to 
emphasize is that here 
the partition of the configuration
space into the parts $R_a$ is objective! 

If the story above involving a computer is 
understood and we accept the hypothesis 
that our mental processes
are manifestations of computational processes 
in our brains, then there is no 
relevant difficulty in replacing
the computer by the family of humans. We 
should partition 
the configuration space $\mathcal{C}$ according to 
the computational
states of our brains, including a set 
in which there are no humans present and (possibly) sets in which some humans have 
undefined states of mind.
More precisely, we should define the sets $R_a$ 
so that two points of the configuration 
space, $x$ and $y$, belong to the same $R_a$
if and only if in these two configurations we 
have the same humans present and each one has 
the same state of mind in $x$ as in $y$. 
The rest of the story is the same as 
that of the computer, with one complication. 
We should account for any possible number of 
humans at any time. 
And, if our particle-model universe admits
creation of particles, in principle there is no
limit for this number, so that $\I$ will be 
infinite. 
We will use the notation 
$\pi_{H}$ for this analyser, 
and use the notation 
$\Omega_H$ and 
$\Sigma_H$ accordingly. 
If we accept the (standard) 
premises made above, about 
how our mental processes relate to the physical 
universe, then $\pi_H$ has been defined in an
objective way. 
And therefore also $\Phi = \Psi_{\pi_H}$
has an objective definition. 
And the fact that $\Phi \not= 0$
is once more due to 
decoherence, which is expected 
to happen if each computational bit
in our brains is encoded by the state
of a large number of particles.
And all our perceptions are the same that would
happen if the state of our universe were $\Phi$ 
instead of being $\Psi$.
And in making predictions about our 
future perceptions,
we should compute Born probabilities
$\P_{\Phi}(A)$, $A \in \Sigma_H$ 
and deem as impossible to happen 
events that have probability 0.
And this is what
we indeed do, using for $\Phi$ 
approximate inferences  
based on our previous and current perceptions.

One can make the case that in interpreting 
quantum mechanics all we have to accomplish 
is to produce a coherent, logically consistent, 
theory and predict our human perceptions 
correctly from it.
(See, e.g., Chapter 9 of 
\cite{Schlosshauer} and references 
therein.)
The theory presented 
here, with $\pi_H$ as the relevant 
analyser from the human perspective, 
fulfils these requirements!

One should not misunderstand the statement 
that $\pi_H$ is the relevant analyser for the 
purpose of predicting human perceptions, with 
the incorrect idea that in the theory only 
$\pi_H$-patterns are part of reality. The 
postulate above applies to every analyser 
$\pi$, as stated. And there are good 
philosophical reasons for thinking about what 
we can learn from considering other related
analysers. This is the subject of the 
next subsection. 


\subsection{Refinements of $\pi_H$ and coarsenings
of refinements of $\pi_H$. 
The realm of classical physics $\pi_C$}
\label{subsec:ontology+}
Among the partitions of the
configuration space into sets with ``macroscopically
meaningful descriptions'', an important class 
is that of refinements of $\pi_{H}$.
It is true that we may sometimes disagree on the 
precise borders of the sets $R_a$, 
but often we all 
agree with several of the relevant macroscopic
descriptions, as for instance with the description 
of a measuring device pointing to a certain result,
or with the letters that are printed on a piece of 
paper.
And this allows us to consider various interesting 
refinements of $\pi_{H}$. 
It is worth looking at this in some more detail.
The analyser $\pi_H$ corresponds to a partition
of the configuration space 
into sets $R_a$, where each $a$ corresponds 
to a given state of mind for each human present.
And this means that in producing this partition,
we are concerned with the location of the 
particles in the brains of the humans. 
To refine $\pi_H$, we break each $R_a$ into
parts, according to the location of many other
particles that form all sorts of other 
things we may include and that have non-controversial
macroscopic descriptions for us. 
This could include all the other 
particles forming the bodies of these humans, 
forming other animals, like cats (alive and 
dead), computers, books (including all the 
letters and digits printed on each page), 
the moon (even if no
one is looking at it), ... 

The effective state of the universe based
on our perceptions is the vector $\Phi
= \Psi_{\pi_H}$. 
And when we compute Born-rule probabilities
we are using the best information we have to 
approximate $\Phi$ (or at least how $\Phi$ looks 
inside our lab). 
When we consider a $\pi$ that refines $\pi_H$, as 
in the previous paragraph, we may wonder if 
$\Psi_\pi$ would not be substantially different 
from $\Phi$. But we should not worry about it 
when the refinement of $\pi_H$ is, as above, 
based on ``macroscopic descriptions'', 
once again because of decoherence. 
The point is that we 
are aware of the phenomenon of environmental 
decoherence, and have proposed ways in which it
happens in our universe, based on our knowledge 
about the state of the universe that we perceive,
namely $\Phi$.
From this knowledge, we see that 
$\Phi$ includes a rich enough environment to assure
enough decoherence affecting cats, books, 
the moon, etc, that 
$\Phi \in \Hp''=\Hp$, as explained in 
Subsection~\ref{subsec:decoherence}. 
Therefore, since $\Hp \subset \H_{\pi_H}$, we 
have 
$\Psi_\pi = p_\pi \Psi = p_\pi p_{\pi_H} \Psi 
= p_\pi \Psi_{\pi_H} = p_\pi \Phi = \Phi$.

It is interesting to compare a typical event
$A \in \Sigma_H$ with its refinement $A' \in \Sigma$, 
corresponding to a $\pi$ that refines $\pi_H$ 
in the fashion described above. (For the definition 
of $A'$ see Section~\ref{sec:refinements}. 
The intuitive meaning of $A'$ is 
that it provides 
all the ways in which $A$ could 
happen in terms of 
descriptions based on $\pi$.) 
$A$ could, for instance, be the event that 
``at time $t_1$ Jane saw three moons of Jupiter,
and between times $t_2$ and $t_3$ Hui heard 
a meow sound''. In comparison, depending on 
what $\pi$ is, $A'$ could 
also include a description of the 
telescopes that Jane could have used, the hats she 
possibly had on,
the possible expressions
on her face, ..., and the cat that 
produced the sound that Hui heard, or the 
person who was imitating the sound of a cat, or ....
Using Theorem~\ref{thm:refinement} and the
fact that $p_\pi \Phi = \Phi$, we 
have $\P_\Phi(A') = ||p_{A'}\hat{\Phi}||^2 
= ||p_{A} p_{\pi} \hat{\Phi}||^2
= ||p_{A} \hat{\Phi}||^2
= \P_\Phi(A)$.
So that $A'$ is part of reality if and only
if $A$ is. 
We can also consider $B \subset A'$ that specifies 
a certain telescope, a certain hat, and a 
certain cat. In the view presented here, 
these are parts of the reality derived 
from $\Psi$, provided that 
$p_{B} \Psi \not= 0$, which from 
(\ref{axiomPno}) and what we saw in the last 
paragraph amounts to 
$p_{B} \Phi \not= 0$, or equivalently
$\P_\Phi(B) > 0$. Now, if in $C \subset A'$ 
Jane was looking at Jupiter, at time $t_1$,
with naked eyes, then the laws of physics
would entail $p_C \Psi = p_C \Phi = 0$ and, 
in particular, we would have 
$\P_\Phi (C) = 0$. And as much as Jane had 
branched in her life, this event would not 
be part of her reality along any branch. 
The point of all of this is that   
human perceptions and their causes 
``out there in the world'' are,
in the view presented here, 
possible parts of reality, and physics
laws restrict them and relate them in the 
proper way. 
(In this paragraph we abused notation in the
following way. Theorem~\ref{thm:basic}
provides a probability measure $\P_\Phi$, 
on $(\Omega_H, \Sigma_H)$
associated to $\pi_H$ and a different 
one, on $(\Omega,\Sigma)$, 
associated to its refinement $\pi$.
We should have distinguished them in 
the notation used, 
but did not do it, since when we write
$\P_\Phi(A)$, with $A \in \Sigma_H$, 
or $\P_\Phi (C)$, with $C \in \Sigma$, 
it should be clear which probability 
measure we mean. We will continue to 
abuse notation in this way, when no confusion
is possible.)

Of course, there should also be refinements of 
$\pi_H$
that specify the mental state of other 
animals. The only reason we did not include this 
feature 
in $\pi_H$ itself was lack of necessity, when 
our task was to account for human experience. 

It is also natural to consider coarsenings of
$\pi_H$, or of 
the refinements of $\pi_H$ discussed above.
Those can naturally be obtained
by focusing on a subset of the 
humans and then lumping together the 
$R_a$ according to the state of mind of 
the humans in this subset, regardless 
of the state of mind of the 
other humans. 
For instance, at the risk of being called
solipsistic, an individual may consider only 
the states of his or her mind in defining 
$\pi$, and 
perhaps refine it to include only 
certain aspects of 
the world of his or her interest. 
In another example, as we make predictions
for our foreseeable future, we may cap 
the number of humans that we distinguish
in $\pi$
to a large 
but finite upper bound, lumping together in 
one $R_a$ all the configurations with 
a larger number of humans. Assuming that each human 
brain can only encode finitely many 
different computational states, we see that 
the corresponding $\I$ is now finite. 

Again we should wonder if $\Psi_\pi$ could 
be substantially different from $\Phi$ 
in case, say,
that only the state of mind of some of 
the humans
is used in the definition of $\pi$.
Here the
equality
$\Psi_\pi = \Phi$ seems to be justified by invoking the assumption of
``homogeneity of the scales on 
which decoherence operates''.
As we understand it, 
decoherence affects our brains by means 
of phenomena that operate in homogeneous 
fashion in a very large scale, 
including the electromagnetic
background radiation that 
fills the universe.
It is true that 
the phenomena that we are aware of 
as producing decoherence are accounted 
for by $\Phi$, and we 
do not know anything about 
$\Psi - \Phi$. But it seems reasonable to think 
that that component of $\Psi$ 
shares this sort of large 
scale homogeneity feature, since it is 
part of 
the same natural phenomenon, namely our 
universe. 
This assumption implies that the 
component of $\Psi$ on which the brain of 
each one of us decohers should be the
same one. And therefore it should be 
the common $\Phi = \Psi_{\pi_H}$ 
that we infer from our perceptions. 

We can now summarize 
what we propose for 
our universe, based on the ideas 
presented above. 
We suppose that it 
is well modeled by a 
particle model of the kind described in 
Subsection~\ref{subsec:particles}, with 
the appropriate particles. Those, as far as 
we currently understand it, are the photons, 
the quarks, the leptons, etc, from what 
is called the ``standard model''. The 
corresponding Hilbert space $\H$ is the 
appropriate Fock space.  
The nature of the Hamiltonian and of the 
Heisenberg-picture 
state vector $\Psi \in \H$
are such that they allow for an 
analyser $\pi_C$, which is a substantial
refinement of $\pi_H$ based on the 
objects that we describe in classical 
physics, and has the 
properties that yield what we call 
our ``classical world(s)'',
or the ``realm of classical physics''
(hence the ``C'' in $\pi_C$).
These properties include the assumption 
that $\Psi_\pi = \Phi$ is the same for 
a wide range of coarsenings $\pi$ 
of $\pi_C$. 
The idea is that $\Phi$ provides enough 
decoherance, and $\Psi-\Phi$ does not 
add to it across the relevant scales, 
to assure this constancy of $\Psi_\pi$.  
Here $\pi$ could be as fine as $\pi_C$ 
itself, or as coarse as only describing 
a piece of dust, say (anything that is 
still well described by classical physics 
would fit here). Other examples of
allowed $\pi$
would be $\pi_H$ and the other analysers 
mentioned above in this subsection, 
or the $\pi$ associated to the computational 
states of a computer, from the previous 
subsection. 
How fine can the 
partition corresponding to $\pi_C$ be? 
This is the question of understanding how
small something can be to still be 
the subject of enough decoherence in $\Phi$
to fit in the definition of $\pi_C$
(and hence be called ``macroscopic''). 
Reflecting on 
this question makes clear that there
is a certain fuzziness in the definition of
$\pi_C$, and the answer may depend on the 
cosmological time scale considered, 
for the reasons presented in 
Subsection~\ref{subsec:decoherence}. 
This question is 
nothing but the question of 
determining the limits of the realm 
of classical 
physics, which certainly has a fuzzy 
boundary, which can and should 
be investigated experimentally.
Another question is what we can say about
$\Psi - \Phi$. The answer suggested by all 
the considerations so far is {\it nothing}. 
This vector could be 0, 
it could be comparable to $\Phi$ in norm, 
or it could be 
much larger than $\Phi$. In any case 
it would not affect our experiences in the 
theory proposed here, based on the 
Ontological Postulate from 
Subsection~\ref{subsec:ontology}, and
the idea that our mental processes 
correspond to $\pi_H$-patterns of $\Psi$,
or equivalently, $\pi_C$-patterns of $\Psi$.
(To see this last equivalence, let $A \in \Sigma_H$ and $A'$ be its refinement to 
$\pi_C$. Then Theorem~\ref{thm:refinement}
tells us that $p_{A'} \Psi = p_A p_{\pi_C}
\Psi = p_A \Phi = p_A \Psi$, so that 
$A$ is a $\pi_H$-pattern of $\Psi$ iff
$A'$ is a $\pi_C$-pattern of $\Psi$.
Note that, as in the example involving 
Jane and Hui, here we also have 
$\P_\Phi(A) = \P_\Phi(A')$.) 
Since $\Psi-\Phi$ has no effect on our experiences,
and is therefore inaccessible to us via 
experiments, 
it is tempting to assume that $\Psi = \Phi$, i.e., 
that $\Psi \in \H_{\pi_C}$. 
This assumption is 
nevertheless not needed and may be criticized 
as being anthropocentric. For this reason we take 
an agnostic position on whether $\Psi \in \H_{\pi_c}$, or not.  

The analyser $\pi_H$ and the other ones 
discussed so far in this subsection 
involve ``macroscopic''
objects (we are supposing that each 
computational bit
in our brains is also encoded by the state
of a large number of particles). 
The relevance of this assumption was  
emphasized repeatedly above, in connection
to the need of decoherence.  But in a refinement
of $\pi_H$, could we include microscopic 
phenomena too? For instance, if we are 
performing a double slit experiment with 
one electron, can we refine $\pi_C$, by
partitioning the sets $R_a$ according 
to the location of this electron while in 
flight? Nothing 
prevents us from doing it. We lose the 
assurance that $\Psi_\pi \not= 0$. But in
case we 
still have $\Psi_\pi \not= 0$, we can even 
associate probability $\P_{\Psi_\pi}(A) 
= ||p_A \Psi ||^2 / ||\Psi_\pi||^2$  
to an event $A$ involving the 
location of the electron.
And this probability will be positive 
if and only if the event $A$ is part of
reality. If $\Psi_\pi = 0$, then also 
$p_A \Psi = 0$, and $A$ is not part of 
reality. 
Now, it should be pointed out that these 
considerations have no operational 
meaning for us humans. 
When for an 
event $B \in \Sigma_H$ we compute 
$\P_\Phi(B)$, the meaning is that
$B$ is part of our perceptual 
reality if and only if this probability
is positive. So that we can predict that 
if this probability is 0, $B$ will not 
happen. But in the case of the event 
$A$ mentioned above, whether it is or is
not part of reality has no implication 
for our perceptions, unless those are 
also accounted for by an event 
$B \in \Sigma_H$. 
From a pragmatic point of view, 
this is in agreement with the quantum 
recipe of the textbooks, that tell us 
to only associate probabilities to outcomes
of experiments, not to events like the 
location of the electron in the double slit 
experiment, before it hits the screen. 
In the view presented here, 
it is not that such probabilities cannot 
be defined and related to the  
ontology of the theory. 
They can, when $\Psi_\pi \not= 0$.  
The issue is that  
they are related to aspects of that
ontology that are not amenable to
experimental scrutiny by us.  

We end this subsection with a 
further discussion and 
clarification of the role of some of 
the different
probability measures provided by 
Theorem~\ref{thm:basic}, part (b). 
Strictly speaking, for each analyser 
$\pi$ we have a distinct measurable 
space $(\Omega, \Sigma)$, and then, 
for each $\phi \in \Hp \backslash \{0\}$
we have a distinct 
probability measure $\P_\phi$
on this space. This is a very large set
of probability measures! Now, the proposal 
in the previous subsection and in this one
is that the most relevant analysers, from 
our perspective, are 
$\pi_H$ and its refinement $\pi_C$, 
that both share $\Phi = \Psi_\pi$ and 
that the relevant probabilities are 
$\P_\Phi(A)$, either for $A \in \Sigma_H$,
or $A \in \Sigma_C$ (abusing notation as
explained before). The probabilities
$\P_\Phi(A)$, for $A \in \Sigma_C$ are the 
ones that the quantum mechanics textbooks 
tell us to compute, to make predictions. 
(In doing it, they sometimes state that a
microscopic system 
must interact with a macroscopic measuring 
device, for a ``potentiality'' to become 
a ``reality''. The role of decoherence 
removes the need for such mysterious 
statements.)
For instance,
in a double slit experiment with a single 
electron, in which we are using an old
fashioned photographic plate as the screen, 
$A_i$ may be the event in $\Sigma_C$ that a
pixel $i$ is sensitized. $B_i$ may be the
event in $\Sigma_H$ that Carla sees the 
pixel $i$ as a white dot, as she looks at 
the plate after developing it. 
And $B_i'$ will denote the refinement of 
$B_i$ to $\Sigma_C$. The textbooks 
tell us how to compute $\P_\Phi(A_i)$
(precisely in the same way that 
Theorem~\ref{thm:basic} does), and 
our understanding of what happens when one 
looks at a developed photographic plate 
tells us that 
$B_i' = A_i$, and therefore  
$\P_\Phi(A_i) = \P_\Phi(B_i') = 
\P_\Phi(B_i)$. (For the justification of 
the last equality, one can use
Theorem~\ref{thm:refinement}, as in the 
example involving Jane and Hui.) 
Suppose now that $\pi$ is a refinement 
of $\pi_C$ based on the position of the 
electron at a certain time $t$, before 
it hits the screen. And let $A_i''$ denote
the corresponding refinement of $A_i \in 
\Sigma_C$
to the sigma-algebra $\Sigma$ associated to 
$\pi$. Lack of decoherence of the 
electron's location in $\Phi$, 
i.e., lack of recording of the electron's 
position at time $t$ in the environment provided by $\Phi$, 
means that $\Phi \not\in \Hp'' = \Hp$
(see Subsection~\ref{subsec:decoherence}).
And this implies that 
$\Psi_\pi = \
p_\pi \Psi = p_\pi p_{\pi_C} \Psi =
p_\pi \Phi =
\Phi_\pi \not= \Phi$. 
In case $\Phi_\pi \not= 0$, the probability 
$\P_{\Psi_\pi}(A''_i) = 
\P_{\Phi_\pi}(A''_i)$ 
is well defined, 
but is not related to the 
relevant probability $\P_{\Phi}(A_i)
= \P_\Phi(B_i)$
in any simple way.

\subsection{The relevance of 
$\{p_A : A \in \Sigma \}$ being a 
projection valued measure for the 
validity of our logical reasoning 
about our perceptions, and for the 
computational aspect of life}
\label{subsec:logic-pvm}

In the theory of 
Subsection~\ref{subsec:ontology}, 
our perceptions are events $A \in \Sigma_H$
that satisfy $p_A \Psi \not= 0$. 
The fact that 
$\{p_A : A \in \Sigma_H\}$ is a p.v.m. 
explains then the validity of our use of 
some basic rules of logic in thinking 
about these perceptions. Below go
some basic instances:

If $A \subset B$ and 
we believe that 
$B$ will not be one of 
our perceptions (along any branch of our 
existence), then we reason that 
also $A$ will not be one of our perceptions. 
And indeed, (PVM7) implies that if 
$p_B \Psi = 0$, then $p_A \Psi = 0$.

If we believe 
that each one of the disjoint 
$A_1, ..., A_n$ will not be
one of our perceptions, then we reason 
that also $A = \cup_{i=1}^n A_i$ will not be
one of our perceptions. And indeed, 
(PVM4) implies that if $p_{A_i} \Psi = 0$,
$i=1, ..., n$, then 
$p_A \Psi = 0$.

If we believe 
that one among $A_1, ..., A_n$ will not be
one of our perceptions, then we reason 
that also $A =\cap_{i=1}^n A_i$ will not be
one of our perceptions. And indeed, 
(PVM8) implies that if $p_{A_i} \Psi = 0$,
for some $i=1, ..., n$, then 
$p_A \Psi = 0$.

If we believe 
that each one of the disjoint 
$A_1, A_2, ...$ will not be
one of our perceptions, then we reason 
that also 
$A = \cup_{i=1}^\infty A_i$ will not be
one of our perceptions. And indeed, 
(PVM2) implies that if $p_{A_i} \Psi = 0$,
$i=1,2, ..., $, then 
$p_A \Psi = 0$.

More formally, the examples above and others
can be derived from the observation that 
the set $\{A \in \Sigma_H : 
p_A \Psi = 0\}$ is a sigma-ideal, i.e., 
a family of 
elements of $\Sigma_H$ that
has the following three properties: 
It contains $\emptyset$, is closed with 
respect to taking subsets and with respect 
to taking countable unions.

But we do make mistakes: If we believe 
that each one of the disjoint, but 
possibly uncountably many, 
$A_\alpha$, will not be
one of our perceptions, then we 
(sometimes) reason 
that also 
$A = \cup_{\alpha} A_\alpha$ will not be
one of our perceptions. This is not 
justified, and counter-examples are not
hard to find. For instance, consider 
an infinite
sequence of independent identical 
experiments that
may result in one of two outcomes, each with 
a Born probability that is positive. Each 
possible sequence of outcomes 
has Born probability 0, 
and therefore will not be one of our 
perceptions (along any branch). But the 
union of all the individual outcomes 
is $\Omega$, which has $p_\Omega \Psi =
p_\pi \Psi = \Psi_\pi \not = 0$. 

Such mistakes are of little consequence 
for the survival of a species that 
usually only needs to deal with finite 
sets of $A_i$ at a time. So the persistence 
of such mistakes, not having been eliminated 
by natural selection is not a surprise. 

It is very interesting to observe that the
structure of quantum mechanics, in a 
rich enough universe like ours, provides 
for the existence of patterns in the 
state vector that embed rules of logic
and therefore can 
instantiate classical computations.
Computations performed by DNA-based, 
or RNA-based wetware that are essential 
for life as we know it,  
animal brains and silicon based computers 
are possible thanks to this structure 
and this richness. (See also the concept
of IGUS in \cite{GMH}.) 

\subsection{
Reducing the Ontological Postulate 
to more intuitive statements} 
\label{subsec:P1P2P3}

The Ontological Postulate can be justified by its success.
It produces a theory that makes the same predictions
of Born's-rule-collapse quantum mechanics, without 
the collapses. But one may wonder if it can be reduced
to more intuitive statements. 

This can be done as we 
relate the theory proposed 
in Subsection~\ref{subsec:ontology} to 
the results in Sections 
\ref{sec:F} and \ref{sec:Farbitrary} and
the ideas from \cite{Sch1} and \cite{Sch2} 
presented in Subsection~\ref{subsec:superposition}.
Suppose that 
we accept the following two premises:
\begin{itemize}
    \item[(P1)] If $\Psi \in F_A$, then 
$A$ is not part of reality. 
     \item[(P2)] The one-sided 
     superposition principle.
\end{itemize}
Then, 
as explained in Subsection~\ref{subsec:superposition},
we conclude that when $\Psi \in \overline{F_A}$, 
$A$ should not be part of reality. Combining 
this with part (a) of 
Theorem~\ref{thm:Farbitrary} we have 
\begin{equation}
p_A \Psi = 0 \ \ \ \Longrightarrow \ \ \
\Psi \in \overline{F_A} \ \ \
\Longrightarrow \ \ \
\mbox{$A$ is not part of reality}. 
\label{reduction1}
\end{equation}
This reduces half of our postulate 
to (P1) and (P2) above. 

Now, suppose we go further and also 
accept a third premise:
\begin{itemize}
    \item[(P3)] If $A$ is not excluded from reality 
    by (P1) and (P2) above, then 
    $A$ is part of reality.  
\end{itemize}
Since $F_A \subset \overline{F_A}$ and 
$\overline{F_A}$ is closed with respect to taking 
superpositions, we have then 
\begin{equation}
p_A \Psi = 0 \ \ \ \Longleftrightarrow \ \ \
\Psi \in \overline{F_A} \ \ \
\Longleftrightarrow \ \ \
\mbox{$A$ is not part of reality}, 
\label{reduction2}
\end{equation}
where the leftmost 
implication to the left can currently
only be justified by 
Theorem~\ref{thm:F} 
if $S$ is countable 
(e.g., $S \subset \Q$,
or $S \subset \epsilon \N$), or 
by Theorem~\ref{thm:Farbitrary}
if $\I$ is finite. This 
reduces our postulate 
to (P1), (P2) 
and (P3), in these cases.

\subsection{A close relationship 
between Born's 
rule and the one-sided 
superposition principle}
\label{subsec:Born-superposition}

From 
(\ref{lemmaG}) in 
part (d) of Theorem~\ref{thm:basic}
we know that 
$$
p_A \Psi = 0 \ \ \ \Longleftrightarrow \ \ \
\Psi_\pi = 0 \ \ \mbox{or} \ \ \P_{\Psi_\pi}(A) = 0. 
$$
Therefore, if we accept the 
assumptions (P1) and (P2), from 
Subsection~\ref{subsec:P1P2P3},
then (\ref{reduction1}) implies 
$$
\P_{\Psi_\pi}(A) = 0 
\ \ \ \Longrightarrow \ \ \
\mbox{$A$ is not part of reality}.
$$
And if we also accept (P3) and assume $S$ countable
or $\I$ finite, 
then (\ref{reduction2}) implies
$$
\P_{\Psi_\pi}(A) > 0
\ \ \  \Longleftrightarrow \ \ \
\mbox{$A$ is part of reality}.
$$
(In the left-hand side, the assumption 
$\Psi_\pi \not = 0$ is implicit, since 
otherwise $\P_{\Psi_\pi}(A)$ would 
not be defined.)
These consideration show how closely related Born's 
rule is, in the context of non-collapse quantum
mechanics, to the one-sided superposition principle,
expanding on the thesis of \cite{Sch1} and
\cite{Sch2}.

It is natural to ask if one can have a 
version of non-collapse quantum mechanics
that does not satisfy Born's rule, 
but instead satisfies some other probability 
rule. This is the case 
in the context of
the particle models of 
Subsection~\ref{subsec:particles}, 
if nature 
chooses at each time $t \in \R$, 
independently of anything else, a point 
$x_t$ from the configuration space 
$\mathcal{C}$ with probability 
$\P'_{\Psi_\pi} (x_t \in \cup_i R_{a,i})$ 
proportional 
to $||p^t_a \hat\phi||^\alpha$,
$\phi = \Psi_\pi, $
with some $\alpha \not= 2$, where 
$\Psi$ is the 
Heisenberg-picture
state of the universe 
and $\pi$ is some special analyser, 
that satisfies $\Psi_\pi \not = 0$. 

\subsection{Choosing between pilot-wave theories and the minimalistic ontology of 
Subsection~\ref{subsec:ontology}}
\label{subsec:comparison}
As we observed in
Subsection~\ref{subsec:ontology},
the minimalistic ontology proposed there 
cannot be distinguished through experiments
from the alternative proposal that we live 
in an infinite-ensemble universe, 
in which each world in the ensemble is 
an independent realization of a process 
$(x_t)_{t\in S}$ that satisfies 
(\ref{Bornx}) with $\phi = \Psi_{\pi}$, 
where the partition of the 
configuration space into the sets $R_a$, 
$a \in \I$ and the set $S$ correspond to 
the analyser $\pi = \pi_H$. 
The same is true if $\pi$ is an  
appropriate 
refinement of $\pi_H$, as described in 
Subsection~\ref{subsec:ontology+}, 
including $\pi_C$. 

Experiments would also not distinguish 
our experiences under these proposals
from those in 
a universe with a single, or any 
finite number of independent realizations 
of such a process 
$(x_t)_{t\in S}$. 

Choosing among these theories 
seems to be a pure 
matter of personal taste. 

If the issue is only experimental 
adequacy, also pilot-wave theories that 
satisfy (\ref{equivariance}), 
but do not 
satisfy (\ref{Bornx}), are alternatives. 
But here there are already  
serious 
manifestations of
discontent in the literature: 
in Section 10.2 of \cite{Nelson1}
and Section 5 of \cite{Nelson2}, 
and in Section 5 of 
\cite{Bell}, because in such 
theories our memories and records 
do not have to correspond to our 
true past. 

One should
try to show that (under 
appropriate conditions on the 
Hamiltonian, perhaps) a process
$(x_t)_{t\in S}$ that satisfies 
(\ref{Bornx}) can have very nice properties,
including continuous paths interrupted by 
jumps at creation and annihilation 
of particles, Markovianity, being 
described by a differential equation, etc.

If this turns out to be the case, such
a pilot-wave theory would probably be 
very attractive. On the other hand,
for people who prefer less baggage in
a metaphysical theory and see abstraction 
as no obstacle (perhaps even an advantage)
the minimalistic ontology of 
Subsection~\ref{subsec:ontology} will 
probably still be a better choice. 

In the discussion above the choice between 
the minimalistic ontology of 
Subsection~\ref{subsec:ontology} and the
alternatives is only about what is considered
to be in the primary ontology of the theory.
Even in the minimalistic ontology, the 
processes $(x_t)_{t \in S}$ that satisfy 
(\ref{Bornx}) exist in the 
derived ontology 
(as mathematical constructs). And an infinite
ensemble of independent versions of such
processes exists as well in the same sense. 
And as we saw in 
Subsection~\ref{subsec:ontology}, such 
an infinite ensemble helps us understand 
the meaning of the Ontological Postulate 
introduced there. This postulate is
equivalent to the statement that the only 
events in $\Sigma$
that are realized are those that 
are realized in this infinite-ensemble of 
independent processes. Therefore one can 
think of these processes as aids to the 
visualization of the realities encoded 
in the primary ontology given simply by
$\H$, $(U_t)_{t \in T}$ and $\Psi$. 

A person choosing a metaphysics in which
a realization of a process $(x_t)_{t \in S}$
is also part of the primary ontology will
be faced with the question of why to 
prefer one single realization of such a 
process with this status rather than 
infinitely many independent ones. In 
other words, between a pilot-wave theory 
with a single realization and one with 
an infinite ensemble, why prefer one to 
the other? (The remaining choice of a 
finite number, larger than 1, 
of realizations will probably be discarded 
in comparison with those, for lack of 
motivation, or even on aesthetic grounds.)

In favor of a single realization one can 
argue that it is more economic and at 
the same time sufficient.
(But then, the minimalistic ontology is 
even more so.) 
Perhaps one would add a strong preference 
for having a theory without human 
branching. (But humans would still
be branching in the wave function, even if 
only one branch would be considered to be
real. And this raises the question of what
to make of the humans in the 
branches of the wave function that evolve
and behave like you and me, without being 
``real''.) 

In favor of an infinite
ensemble one can argue that it preserves 
the symmetry among the branches of the
wave function. For instance, at the end
of the infamous experiment involving a cat, 
there will be two branches that evolve 
quite differently (for the cat at least). 
Regardless of the way a single process 
$(x_t)_{t \in S}$ 
goes at the end of this experiment,
both branches evolve according to 
$(U_t)_{t \in T}$ 
in ways that encode coherent 
subsequent stories. Why would nature 
produce all of this and only realize
one of these stories? 

The debate above, between one or an 
infinite ensemble of realizations of 
$(x_t)_{t \in S}$ 
may be taken as an argument in 
favor of the minimalistic ontology. 
That ontology avoids the choice, 
by considering 
any number of such processes with the 
same status of derived realities. 
All very useful for our understanding 
of how the universe evolves, of our
place in it, and why we can use textbook
quantum mechanics to predict our future
and describe our past. 

One should not think that by being placed
in the derived ontology the 
pilot-wave processes 
$(x_t)_{t \in S}$ become less important.
Their existence, as mathematical objects,
shows that it is possible to have the 
particles of the universe move in physical 
space, between their creation and 
annihilation, in 
ways that are compatible with our
experiences and our records and memories 
of these experiences. This negates the 
very common statements according to 
which nothing like this could be done.
See for instance Chapters 1, 2 and 7 of 
\cite{Bricmont}, for an extensive 
criticism of such statements. 

The reader should have inferred from the 
discussion in this subsection and previous
ones what my own preferences for
interpretations of quantum mechanics are.
Among those discussed here, they are 
ordered as follows. The minimalistic
ontology first, followed by an
infinite-ensemble-pilot-wave model 
that satisfies (\ref{Bornx}), 
followed by a single-pilot-wave model  
that satisfies (\ref{Bornx}). 
The option of a pilot-wave model that 
satisfies (\ref{equivariance}), but does
not satisfy (\ref{Bornx}) does not 
seem plausible to me, for the reasons 
(admitting incorrect records and memories)
presented before. 

Readers are invited to come to their own 
conclusions and to possibly 
apply the theorems 
in this paper in different ways 
that may further shed light on issues in 
the interpretation of quantum mechanics. 
They are also invited to expand and elaborate
on the proposals in this paper, and to 
possibly settle mathematical issues left 
open here, as the conjecture raised after 
Theorem~\ref{thm:Farbitrary}, and the 
questions posed in 
Subsection~\ref{subsec:pilot-wave}.

\ 

\ 

\noindent {\bf Acknowledgements:}
It is a pleasure to thank Marek Biskup, 
Christopher de Firmian and Jim Ralston for enlightening conversations. Jim Ralston also 
deserves many thanks for carefully reading the paper 
and making several constructive suggestions. 
Thanks are also 
given to Jean Bricmont and Shelly 
Goldstein for enlightening conversations on 
pilot-wave theories, especially Bohmian 
mechanics. 

\ 

\bibliographystyle{unsrt}  


\end{document}